\documentclass[aps,eqsecnum,preprint,floats,epsf,epsfig,nofootinbib]{revtex4}
\usepackage{graphicx}
\usepackage{epsfig}

\begin{document}
\def\be{\begin{eqnarray}}
\def\en{\end{eqnarray}}
\def\non{\nonumber}
\def\la{\langle}
\def\ra{\rangle}
\def\pp{{\prime\prime}}
\def\nc{N_c^{\rm eff}}
\def\vp{\varepsilon}
\def\hep{\hat{\varepsilon}}
\def\a{{\cal A}}
\def\B{{\cal B}}
\def\c{{\cal C}}
\def\d{{\cal D}}
\def\e{{\cal E}}
\def\p{{\cal P}}
\def\t{{\cal T}}
\def\up{\uparrow}
\def\dw{\downarrow}
\def\vma{{_{V-A}}}
\def\vpa{{_{V+A}}}
\def\smp{{_{S-P}}}
\def\spp{{_{S+P}}}
\def\J{{J/\psi}}
\def\ov{\overline}
\def\Lqcd{{\Lambda_{\rm QCD}}}
\def\pr{{Phys. Rev.}~}
\def\prl{{ Phys. Rev. Lett.}~}
\def\pl{{ Phys. Lett.}~}
\def\np{{ Nucl. Phys.}~}
\def\zp{{ Z. Phys.}~}
\def\lsim{ {\ \lower-1.2pt\vbox{\hbox{\rlap{$<$}\lower5pt\vbox{\hbox{$\sim$}
}}}\ } }
\def\gsim{ {\ \lower-1.2pt\vbox{\hbox{\rlap{$>$}\lower5pt\vbox{\hbox{$\sim$}
}}}\ } }

\font\el=cmbx10 scaled \magstep2{\obeylines\hfill January, 2004}

\vskip 1.5 cm

\centerline{\large\bf Covariant Light-Front Approach for $s$-wave
and $p$-wave Mesons:}
 \centerline{\large\bf Its Application to Decay Constants and
 Form Factors}
\bigskip
\centerline{\bf Hai-Yang Cheng,$^1$ Chun-Khiang Chua$^1$ and
Chien-Wen Hwang$^2$}
\medskip
\centerline{$^1$ Institute of Physics, Academia Sinica}
\centerline{Taipei, Taiwan 115, Republic of China}
\medskip

\medskip
\centerline{$^2$ Department of Physics, National Kaohsiung Normal
University} \centerline{Kaohsiung, Taiwan 802, Republic of China}
\medskip

\bigskip
\bigskip
\centerline{\bf Abstract}
\bigskip
\small

We study the decay constants and form factors of the ground-state
$s$-wave and low-lying $p$-wave mesons within a covariant
light-front approach. Numerical results of the form factors for
transitions between a heavy pseudoscalar meson and an $s$-wave or
$p$-wave meson and their momentum dependence are presented in
detail. In particular, form factors for heavy-to-light and $B\to
D^{**}$ transitions, where $D^{**}$ denotes generically a $p$-wave
charmed meson, are compared with other model calculations. The
experimental measurements of the decays $B^-\to D^{**}\pi^-$ and
$B\to\ov DD^{**}_s$ are employed to test the decay constants of
$D_s^{**}$ and the $B\to D^{**}$ transition form factors. The
heavy quark limit behavior of the decay constants and form factors
is examined and it is found that the requirement of heavy quark
symmetry is satisfied. The universal Isgur-Wise (IW) functions,
one for $s$-wave to $s$-wave and two for $s$-wave to $p$-wave
transitions, are obtained. The values of IW functions at zero
recoil and their slope parameters can be used to test the Bjorken
and Uraltsev sum rules.

%
%
%
%





\eject
\section{Introduction}

Mesonic weak transition form factors and decay constants are two
of the most important ingredients in the study of hadronic weak
decays of mesons. There exist many different model calculations.
The light-front quark model \cite{Ter,Chung} is the only
relativistic quark model in which a consistent and fully
relativistic treatment of quark spins and the center-of-mass
motion can be carried out. This model has many advantages. For
example, the light-front wave function is manifestly Lorentz
invariant as it is expressed in terms of the momentum fraction
variables  in analog to the parton distributions in the infinite
momentum frame. Moreover, hadron spin can also be correctly
constructed using the so-called Melosh rotation. This model is
very suitable to study hadronic form factors. Especially, as the
recoil momentum increases (corresponding to a decreasing $q^2$),
we have to start considering relativistic effects seriously. In
particular, at the maximum recoil point $q^2=0$ where the
final-state meson could be highly relativistic, there is no reason
to expect that the non-relativistic quark model is still
applicable.

The relativistic quark model in the light-front approach has been
employed to obtain decay constants and weak form
factors~\cite{Jaus90,Jaus91,Ji92,Jaus96,Cheng97}. There exist,
however, some ambiguities and even some inconsistencies in
extracting the physical quantities. In the light-front quark model
formulation one often picks up a specific Lorentz frame (e.g. the
purely longitudinal frame $q_\bot=0$, or the purely transverse
frame $q^+=q^0+q^3=0$) and then calculates a particular component
(the ``plus" component) of the associated current matrix element.
Due to the lack of relativistic covariance, the results may not be
unique and may even cause some inconsistencies. For example, it
has been pointed out in \cite{Cheng97} that in the $q_\bot=0$
frame, the so-called $Z$-diagram contributions must be
incorporated in the form-factor calculations in order to maintain
covariance. Another issue is that the usual recipe of taking only
the plus component of the current matrix elements will miss the
zero-mode contributions and render the matrix element
non-covariant. A well known example is the electromagnetic form
factor $F_2(q^2)$ of the vector meson (see e.g. \cite{BCJ02}). In
other words, the familiar expression of $f_V$, for example, in the
conventional light-front approach \cite{Jaus91} is not trustworthy
due to the lack of the zero-mode contributions. As a consequence,
it is desirable to construct a covariant light-front model that
can provide a systematical way of exploring the zero-mode effects.
Such a covariant model has been constructed in \cite{CCHZ} for
heavy mesons within the framework of heavy quark effective theory.

Without appealing to the heavy quark limit, a covariant approach
of the light-front model for the usual pseudoscalar and vector
mesons has been put forward by Jaus \cite{Jaus99} (for a different
approach, see \cite{BCJ03}). The starting point of the covariant
approach is to consider the corresponding covariant Feynman
amplitudes in meson transitions. Then one can pass to the
light-front approach by using the light-front decomposition of the
internal momentum in covariant Feynman momentum loop integrals and
integrating out the $p^-=p^0-p^3$ component~\cite{CM69}. At this
stage one can then apply some well-studied vertex functions in the
conventional light-front approach after $p^-$ integration. It is
pointed out by Jaus that in going from the manifestly covariant
Feynman integral to the light-front one, the latter is no longer
covariant as it receives additional spurious contributions
proportional to the lightlike vector
$\tilde\omega^\mu=(1,0,0,-1)$. This spurious contribution is
cancelled after correctly performing the integration, namely, by
the inclusion of the zero mode contribution~\cite{zeromode}, so
that the result is guaranteed to be covariant. Before proceeding,
it is worth mentioning that in the literature there is a
controversy about the zero mode contributions to the vector decay
constant $f_V$ and the form factor $A_1(q^2)$ in the pseudoscalar
to vector transition: While Jaus \cite{Jaus99,Jaus03} claimed that
there are zero effects in the aforementioned two quantities,
Bakker, Choi and Ji \cite{BCJ02,BCJ03} argued that both $f_V$ and
$A_1(q^2)$ are free of zero-mode contributions. This issue will be
addressed in Sec.~III.B.

The main purposes of this work are twofold: First, we wish to
extend the covariant analysis of the light-front model in
\cite{Jaus99} to even-parity, $p$-wave mesons. Second, the
momentum dependence of the form factors is parametrized in a
simple three-parameter form so that the reader is ready to use our
numerical results as the analytic expressions of various form
factors in the covariant light-front model are usually complicated
(see Sec. III). Interest in even-parity charmed mesons has been
revived by recent discoveries of two narrow resonances: the $0^+$
state $D_{s0}^*(2317)$ \cite{BaBar} and the $P^{1/2}_1$ state
$D_{s1}(2460)$ \cite{CLEO}, and two broad resonances,
$D_0^*(2308)$ and $D_1(2427)$ \cite{BelleD}.\footnote{We follow
the naming scheme of the Particle Data Group \cite{PDG} to add a
superscript ``*" to the states if the spin-parity is in the
``normal" sense, $J^P=0^+,1^-,2^+,\cdots$.} Furthermore, the
hadronic $B$ decays such as $B\to D^{**}\pi$ and $B\to D_s^{**}\ov
D$ have been recently observed, where $D^{**}$ denotes a $p$-wave
charmed meson. A theoretical study of them requires the
information of the $B\to D^{**}$ form factors and the decay
constants of $D^{**}$ and $D_s^{**}$. In the meantime, three body
decays of $B$ mesons have been recently studied at the $B$
factories: BaBar and Belle. The Dalitz plot analysis allows one to
see the structure of exclusive quasi-two-body intermediate states
in the three-body signals. The $p$-wave resonances observed in
three-body decays begin to emerge. Theoretically, the
Isgur-Scora-Grinstein-Wise (ISGW) quark model \cite{ISGW} is so
far the only model in the literature that can provide a
systematical estimate of the transition of a ground-state $s$-wave
meson to a low-lying $p$-wave meson. However, this model and, in
fact, many other models in $P\to P,V$ ($P$: pseudoscalar meson,
$V$: vector meson) calculations, are based on the non-relativistic
constituent quark picture. As noted in passing, the final-state
meson at the maximum recoil point $q^2=0$ or in heavy-to-light
transitions could be highly relativistic. It is thus important to
consider a relativistic approach.

It has been realized that the zero mode contributions can be
interpreted as residues of virtual pair creation processes in the
$q^+\to 0$ limit~\cite{deMelo98}. In \cite{Jaus99}, the
calculation of the zero mode contribution is obtained in a frame
where the momentum transfer $q^+$ vanishes. Because of this
($q^+=0$) condition, form factors are known only for spacelike
momentum transfer $q^2=-q^2_\bot\leq 0$. One needs to analytically
continue them to the timelike region~\cite{Jaus96}, where the
physical decay processes are relevant. Recently, it has been shown
that within a specific model, form factors obtained directly from
the timelike region (with $q^+>0$) are identical to those obtained
by the analytic continuation from the spacelike
region~\cite{BCJ03}.

There are some theoretical constraints implied by heavy quark
symmetry (HQS) in the case of heavy-to-heavy transitions and
heavy-to-vacuum decays~\cite{IW89}. It is important to check if
the calculated form factors and decay constants do satisfy these
constraints. Furthermore, under HQS the number of the independent
form factors is reduced and they are related to some universal
Isgur-Wise (IW) functions. In this work, we shall follow
\cite{CCHZ} to evaluate the form factors and decay constants in a
covariant light-front formulism within the framework of heavy
quark effective theory. It is found that the resultant decay
constants and form factors do agree with those obtained from the
covariant light-front approach and then extended to the heavy
quark limit. The relevant IW functions, namely, $\xi$,
$\tau_{1/2}$ and $\tau_{3/2}$ are obtained. One can then study
some properties of these IW functions, including the slopes and
sum rules~\cite{Uraltsev,Bjorken}.

The paper is organized as follows. In Sec.~II, we give the
calculations for the decay constants of $s$-wave and $p$-wave
mesons in a covariant light-front model. The calculation for
$s$-wave meson transitions has been done by Jaus~\cite{Jaus99}. We
extend it to the $p$-wave meson case. In Sec.~III, $P\to
P$,$V$,$S$,$A$,$T$ ($S$,$A$,$T$ standing for scalar, axial-vector
and tensor mesons, respectively) transitions are considered. It is
interesting to notice that the analytic forms of $P\to S,A$
transitions are similar to that of $P\to P,V$ transitions,
respectively, while the $P\to T$ calculation needs formulas beyond
\cite{Jaus99}. We provide numerical results for $B$ and $D$ decay
form factors and their $q^2$ dependence. These results are then
compared to the other model calculations. In Sec.~IV, properties
of the decay constants and form factors in the heavy quark limit
are studied. The universal Isgur-Wise functions, one for $s$-wave
to $s$-wave and two for $s$-wave to $p$-wave transitions, are
obtained. Their values at zero recoil and their slope parameters
can be used to test the sum rules derived by Bjorken
\cite{Bjorken} and by Uraltsev \cite{Uraltsev}. Conclusion is
given in Sec~V followed by two Appendices devoted to the
derivations of conventional light-front vertex functions and some
useful formulas.

\section{Formalism of a covariant light-front model}

\begin{figure}[t!]
\centerline{
            {\epsfxsize2.1 in \epsffile{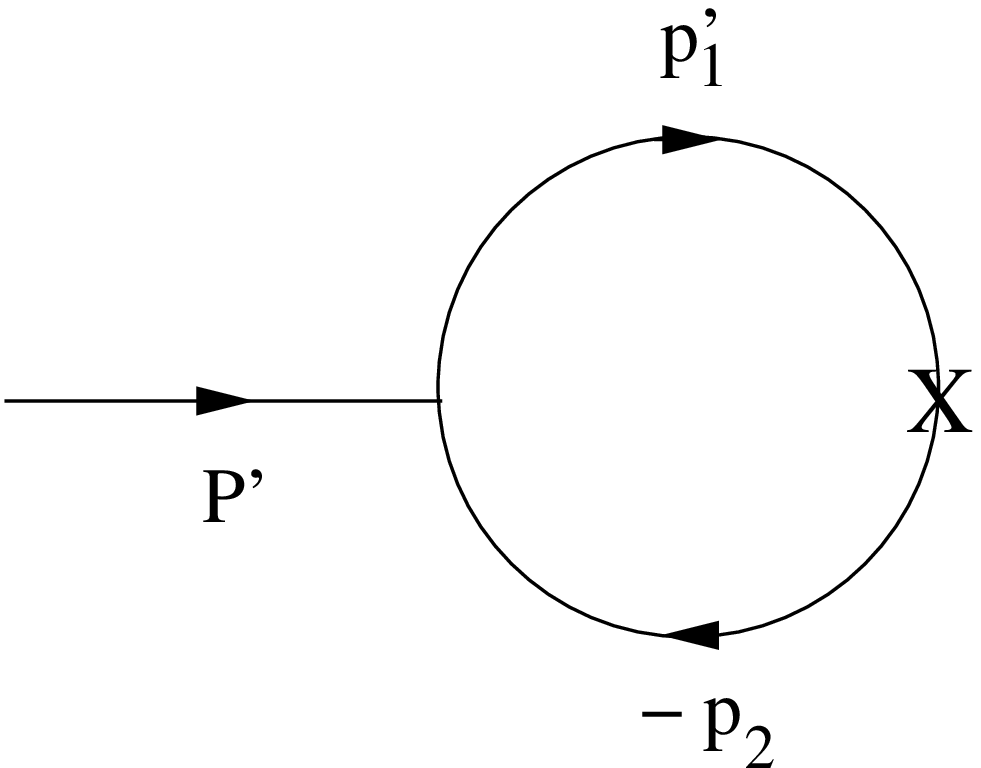}}
            \hspace{1cm}
            {\epsfxsize3 in \epsffile{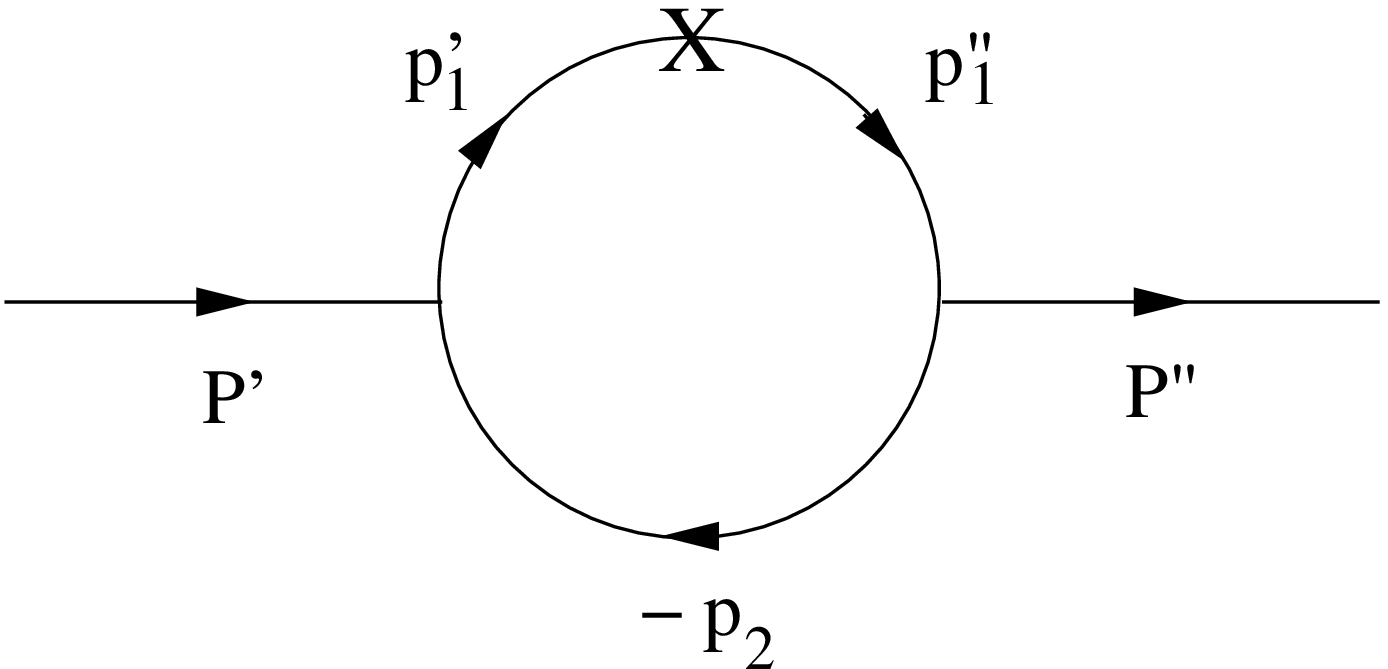}}}
\centerline{\,\,\,\,\,(a)\hspace{6.2cm}(b)} \vskip0.2cm
\caption{Feynman diagrams for (a) meson decay and (b) meson
transition amplitudes, where $P^{\prime(\pp)}$ is the incoming
(outgoing) meson momentum, $p^{\prime(\pp)}_1$ is the quark
momentum, $p_2$ is the anti-quark momentum and $X$ denotes the
corresponding $V-A$ current vertex.}\label{fig:feyn}
\end{figure}

\subsection{Formalism}
In the conventional light-front framework, the constituent quarks
of the meson are required to be on their mass shells (see Appendix
A for an introduction) and various physical quantities are
extracted from the plus component of the corresponding current
matrix elements. However, this procedure will miss the zero-mode
effects and render the matrix elements non-covariant. Jaus
\cite{Jaus99} has proposed a covariant light-front approach that
permits a systematical way of dealing with the zero mode
contributions. Physical quantities such as the decay constants and
form factors can be calculated in terms of Feynman momentum loop
integrals which are manifestly covariant. This of course means
that the constituent quarks of the bound state are off-shell. In
principle, this covariant approach will be useful if the vertex
functions can be determined by solving the QCD bound state
equation. In practice, we would have to be contended with the
phenomenological vertex functions such as those employed in the
conventional light-front model. Therefore, using the light-front
decomposition of the Feynman loop momentum, say $p_\mu$, and
integrating out the minus component of the loop momentum $p^-$,
one goes from the covariant calculation to the light-front one.
Moreover, the antiquark is forced to be on its mass shell after
$p^-$ integration. Consequently, one can replace the covariant
vertex functions by the phenomenological light-front ones.

As stated in passing, in going from the manifestly covariant
Feynman integral to the light-front one, the latter is no longer
covariant as it can receive additional spurious contributions
proportional to the lightlike four vector $\tilde\omega$. The
undesired spurious contributions can be eliminated by the
inclusion of the zero mode contribution which amounts to
performing the $p^-$ integration in a proper way in this approach.
The advantage of this covariant light-front framework is that it
allows a systematical way of handling the zero mode contributions
and hence permits to obtain covariant matrix elements.

To begin with, we consider decay and transition amplitudes given
by one-loop diagrams as shown in Fig.~\ref{fig:feyn} for the decay
constants and form factors of ground-state $s$-wave mesons and
low-lying $p$-wave mesons.  We follow the approach of
\cite{Jaus99} and use the same notation. The incoming (outgoing)
meson has the momentum $P^{\prime(\pp)}=p_1^{\prime(\pp)}+p_2$,
where $p_1^{\prime(\pp)}$ and $p_2$ are the momenta of the
off-shell quark and antiquark, respectively, with masses
$m_1^{\prime(\pp)}$ and $m_2$. These momenta can be expressed in
terms of the internal variables $(x_i, p_\bot^\prime)$,
 \be
 p_{1,2}^{\prime+}=x_{1,2} P^{\prime +},\qquad
 p^\prime_{1,2\bot}=x_{1,2} P^\prime_\bot\pm p^\prime_\bot,
 \en
with $x_1+x_2=1$. Note that we use $P^{\prime}=(P^{\prime +},
P^{\prime -}, P^\prime_\bot)$, where $P^{\prime\pm}=P^{\prime0}\pm
P^{\prime3}$, so that $P^{\prime 2}=P^{\prime +}P^{\prime
-}-P^{\prime 2}_\bot$.
In the covariant light-front approach, total four momentum is
conserved at each vertex where quarks and antiquarks are
off-shell. These differ from the conventional light-front approach
(see, for example ~\cite{Jaus91,Cheng97}) where the plus and
transverse components of momentum are conserved, and quarks as
well as antiquarks are on-shell.
It is useful to define some internal quantities analogous to
those defined in Appendix A for on-shell quarks:

\begin{eqnarray} \label{eq:internalQ}
 M^{\prime2}_0
          &=&(e^\prime_1+e_2)^2=\frac{p^{\prime2}_\bot+m_1^{\prime2}}
                {x_1}+\frac{p^{\prime2}_{\bot}+m_2^2}{x_2},\quad\quad
                \widetilde M^\prime_0=\sqrt{M_0^{\prime2}-(m^\prime_1-m_2)^2},
 \nonumber\\
 e^{(\prime)}_i
          &=&\sqrt{m^{(\prime)2}_i+p^{\prime2}_\bot+p^{\prime2}_z},\quad\qquad
 p^\prime_z=\frac{x_2 M^\prime_0}{2}-\frac{m_2^2+p^{\prime2}_\bot}{2 x_2 M^\prime_0}.
 \end{eqnarray}
Here $M^{\prime2}_0$ can be interpreted as the kinetic invariant
mass squared of the incoming $q\bar q$ system, and $e_i$ the
energy of the quark $i$.

\begin{table}[b]
\caption{\label{tab:feyn} Feynman rules for the vertices
($i\Gamma^\prime_M$) of the incoming mesons-quark-antiquark, where
$p^\prime_1$ and $p_2$ are the quark and antiquark momenta,
respectively. Under the contour integrals to be discussed below,
$H^\prime_M$ and $W^\prime_M$ are reduced to $h^\prime_M$ and
$w^\prime_M$, respectively, whose expressions are given by
Eq.~(\ref{eq:h}). Note that for outgoing mesons, we shall use
$i(\gamma_0\Gamma^{\prime\dagger}_M\gamma_0)$ for the
corresponding vertices.}
\begin{tabular}{|c| c|}
\hline
 $M\,(^{2S+1}L_J) $
      &$i\Gamma^\prime_M$
      \\
      \hline
 pseudoscalar ($^1S_0$)
      &$H^\prime_P\gamma_5$
      \\
 vector ($^3S_1$)
      &$i H^\prime_V [\gamma_\mu-\frac{1}{W^\prime_V}(p^\prime_1-p_2)_\mu]$
      \\
 scalar ($^3P_0$)
      &$-i H^\prime_S$
      \\
 axial ($^3 P_1$)
      &$-i H^\prime_{^3\!A}[\gamma_\mu+\frac{1}{W^\prime_{^3\!A}}(p^\prime_1-p_2)_\mu]\gamma_5$
      \\
 axial ($^1 P_1$)
      &$-i H^\prime_{^1\!A} [\frac{1}{W^\prime_{^1\!A}}(p^\prime_1-p_2)_\mu]\gamma_5$
      \\
 tensor ($^3P_2$)
      &$i\frac{1}{2} H^\prime_T [\gamma_\mu-\frac{1}{W^\prime_V}(p^\prime_1-p_2)_\mu](p^\prime_1-p_2)_\nu$
      \\
\hline
\end{tabular}
\end{table}

It has been shown in \cite{CM69} that one can pass to the
light-front approach by integrating out the $p^-$ component of the
internal momentum in covariant Feynman momentum loop integrals.
We need Feynman rules for the meson-quark-antiquark vertices to
calculate the amplitudes shown in Fig.~1. These Feynman rules for
vertices ($i\Gamma^\prime_M$) of ground-state $s$-wave mesons and
low-lying $p$-wave mesons are summarized in Table~\ref{tab:feyn}.
As we shall see later, the integration of the minus component of
the internal momentum in Fig.~1 will force the antiquark to be on
its mass shell. The specific form of the covariant vertex
functions for on-shell quarks can be determined by comparing to
the conventional
vertex functions as shown in Appendix A. 
%
Next, we shall use the decay constants as an example to illustrate
a typical calculation in the covariant light-front approach.

\subsection{Decay constants}

The decay constants for $J=0,1$ mesons are defined by the matrix
elements
 \be \label{eq:AM}
  \la 0|A_\mu|P(P^\prime)\ra &\equiv& {\cal A}^{P}_\mu=i  f_P P^\prime_\mu ,\qquad
      \la 0|V_\mu|S(P^\prime)\ra\equiv {\cal A}^{S}_\mu= f_S P^\prime_\mu ,
  \\
  \la 0|V_\mu|V(P^\prime,\vp')\ra &\equiv& {\cal A}^{V}_\mu=M^\prime_V
  f_V\vp^\prime_\mu,\quad
      \la 0|A_\mu|\,^{3(1)}\!A(P^\prime,\vp')\ra \equiv {\cal A}^{^3\!A(^1\!A)}_\mu
      =M^\prime_{^3\!A(^1\!A)} f_{^3\!A(^1\!A)}\varepsilon^\prime_\mu, \non
 \en
where the $^{2S+1} L_J= {}^1S_0$, $^3P_0$, $^3S_1$, $^3P_1$,
$^1P_1$ and $^3P_2$ states of $q_1^\prime \bar q_2$ mesons are
denoted by $P$, $S$, $V$, $^3\!A$, $^1\!A$ and $T$, respectively.
Note that a $^3P_2$ state cannot be produced by a current. It is
useful to note that in the SU(N)-flavor limit ($m_1^\prime=m_2$)
we should have vanishing $f_S$ and $f_{^1\!A}$. The former can be
seen by applying equations of motion to the matrix element of the
scalar resonance in Eq. (\ref{eq:AM}) to obtain
 \be \label{eq:Seom}
 m_S^2f_S=\,i(m'_1-m_2)\la 0|\bar q_1q_2|S\ra.
 \en
The latter is based on the argument that the light $^3P_1$ and
$^1P_1$ states transfer under charge conjugation as
 \be
 M_a^b(^3P_1) \to M_b^a(^3P_1), \qquad M_a^b(^1P_1) \to
 -M_b^a(^1P_1),~~~(a=1,2,3),
 \en
where the light axial-vector mesons are represented by a $3\times
3$ matrix. Since the weak axial-vector current transfers as
$(A_\mu)_a^b\to (A_\mu)_b^a$ under charge conjugation, it is clear
that the decay constant of the $^1P_1$ meson vanishes in the SU(3)
limit \cite{Suzuki}. This argument can be generalized to heavy
axial-vector mesons. In fact, under similar charge conjugation
argument [$(V_\mu)_a^b\to -(V_\mu)_b^a$, $M_a^b(^3P_0) \to
M_b^a(^3P_0)$] one can also prove the vanishing of $f_S$ in the
SU(N) limit.

Furthermore, in the heavy quark limit ($m_1^\prime\to\infty$), the
heavy quark spin $s_Q$ decouples from the other degrees of freedom
so that $s_Q$ and the total angular momentum of the light
antiquark $j$ are separately good quantum numbers. Hence, it is
more convenient to use the $L^j_J=P^{3/2}_2$, $P^{3/2}_1$,
$P^{1/2}_1$ and $P^{1/2}_0$ basis. It is obvious that the first
and the last of these states are $^3P_2$ and $^3P_0$,
respectively, while \cite{IW91}
\begin{equation} \label{eq:Phalf}
\left|P^{3/2}_1\right\ra=\sqrt{\frac{2}{3}}\,\left|^1P_1\right\ra
+{1\over \sqrt{3}}\,\left|^3P_1\right\ra,\qquad
\left|P^{1/2}_1\right\ra={1\over \sqrt{3}}\,\left|^1P_1\right\ra
-\sqrt{\frac{2}{3}}\,\left|^3P_1\right\ra.
\end{equation}
Heavy quark symmetry (HQS) requires (see Sec.
IV)~\cite{IW89,HQfrules}
\begin{equation} \label{eq:HQSf}
 f_V=f_P,\qquad
 f_{A^{1/2}}=f_S,\qquad
 f_{A^{3/2}}=0,
\end{equation}
where we have denoted the $P^{1/2}_1$ and $P^{3/2}_1$ states by
$A^{1/2}$ and $A^{3/2}$, respectively.
These relations in the above equation can be understood from the
fact that $(S^{1/2}_0,S^{1/2}_1)$, $(P_0^{1/2},P_1^{1/2})$ and
$(P_1^{3/2},P_2^{3/2})$ form three doublets in the HQ limit and
that the tensor meson cannot be induced from the $V-A$ current. It
is important to check if the calculated decay constants satisfy
the non-trivial SU(N)-flavor and HQS relations.

We now follow \cite{Jaus99} to evaluate meson decay constants. The
matrix element for the annihilation of a pseudoscalar state via
axial currents can be easily written down and it has the
expression
 \be
 {\cal A}^P_\mu&=&-i^2\frac{N_c}{(2\pi)^4}\int d^4 p_1^\prime
                     \frac{H_P^\prime}{N_1^\prime N_2} s_\mu^P,
\label{eq:AP}
 \en
 where
 \be
     s^P_\mu&\equiv&{\rm Tr}[\gamma_\mu\gamma_5(\not \!p^\prime_1+m_1^\prime)\gamma_5
                            (-\not\!p_2+m_2)]
 \nonumber\\
               &=&-4[m_1^\prime P^\prime_\mu+(m_2-m_1^\prime)p^\prime_{1\mu}],
 \label{eq:AP1}
 \en
$N_1^\prime=p_1^{\prime2}-m_1^{\prime2}+i\epsilon$ and
$N_2=p_2^2-m_2^2+i\epsilon$. We need to integrate out
$p_1^{\prime-}$ in ${\cal A}_\mu^P$. As stressed in \cite{Jaus99},
if it is assumed that the vertex function $H'$ has no pole in the
upper complex $p_1^{\prime -}$ plane, then the covariant
calculation of meson properties and the calculation of the
light-front formulism will give identical results at the one-loop
level. Therefore, by closing the contour in the upper complex
$p_1^{\prime-}$ plane and assuming that $H_P^\prime$ is analytic
within the contour, the integration picks up a residue at
$p_2=\hat p_2$, where $\hat p^2_2=m_2^2$. The other momentum is
given by momentum conservation, $\hat p_1^\prime=P^\prime-\hat
p_2$. Consequently, one has the following replacements:
 \be
 N_1^\prime      &\to&\hat N_1^\prime
           =\hat p^{\prime2}_1-m_1^{\prime2}
           =x_1(M^{\prime2}-M_0^{\prime2}), \non\\
H^\prime_M
       &\to&\hat H^\prime_M
           =H^\prime_M(\hat p^{\prime 2}_1,\hat p^2_2)
           \equiv h^\prime_M,
\non\\
W^\prime_M
       &\to&\hat W^\prime_M
           =W^\prime_M(\hat p^{\prime 2}_1,\hat p^2_2)
           \equiv w^\prime_M,
\non\\
\int \frac{d^4p_1^\prime}{N^\prime_1 N_2} H^\prime_M s^M
       &\to& -i \pi \int \frac{d x_2 d^2p^\prime_\bot}
                              {x_2\hat N^\prime_1} h^\prime_M \hat s^M,
 \label{eq:contourA}
 \en
in a generic one-loop vacuum to particle $M$ amplitude ${\cal
A}^M_\mu$. In this work the explicit forms of $h^\prime_M$ and
$w^\prime_M$ are given by (see Appendix A)
\begin{eqnarray} \label{eq:vertex}
 h^\prime_P&=&h^\prime_V
                  =(M^{\prime2}-M_0^{\prime2})\sqrt{\frac{x_1 x_2}{N_c}}
                    \frac{1}{\sqrt{2}\widetilde M^\prime_0}\varphi^\prime,
 \nonumber\\
 h^\prime_S &=&\sqrt{\frac{2}{3}}h^\prime_{^3\!A}
                  =(M^{\prime2}-M_0^{\prime2})\sqrt{\frac{x_1 x_2}{N_c}}
                    \frac{1}{\sqrt{2}\widetilde M^\prime_0}\frac{\widetilde
                     M^{\prime
                     2}_0}{2\sqrt{3}M^\prime_0}\varphi^\prime_p,
       \nonumber\\
 h^\prime_{^1\!A}&=& h^\prime_T =(M^{2\prime}-M_0^{\prime 2})\sqrt{\frac{x_1
 x_2}{N_c}}\frac{1}{\sqrt{2}\widetilde M^\prime_0}\varphi'_p\, ,
 \non\\
 w^\prime_V&=&M^\prime_0+m^\prime_1+m_2,\quad
 w^\prime_{^3\!A}=\frac{\widetilde{M}'^2_0}{m^\prime_1-m_2},\quad
 w^\prime_{^1\!A}=2\,,
 \label{eq:h}
\end{eqnarray}
where $\varphi'$ and $\varphi'_p$ are the light-front momentum
distribution amplitudes for $s$-wave and $p$-wave mesons,
respectively. There are several popular phenomenological
light-front wave functions that have been employed to describe
various hadronic structures in the literature. In the present
work, we shall use the Gaussian-type wave function \cite{Gauss}
\begin{eqnarray} \label{eq:Gauss}
 \varphi^\prime
    &=&\varphi^\prime(x_2,p^\prime_\perp)
             =4 \left({\pi\over{\beta^{\prime2}}}\right)^{3\over{4}}
               \sqrt{{dp^\prime_z\over{dx_2}}}~{\rm exp}
               \left(-{p^{\prime2}_z+p^{\prime2}_\bot\over{2 \beta^{\prime2}}}\right),
\nonumber\\
 \varphi^\prime_p
    &=&\varphi^\prime_p(x_2,p^\prime_\perp)=\sqrt{2\over{\beta^{\prime2}}}~\varphi^\prime,\quad\qquad
         \frac{dp^\prime_z}{dx_2}=\frac{e^\prime_1 e_2}{x_1 x_2 M^\prime_0}.
 \label{eq:wavefn}
\end{eqnarray}
The parameter $\beta'$ is expected to be of order $\Lambda_{\rm
QCD}$. The derivation of these vertex functions is shown in
Appendix~A.

The matrix element ${\cal A}^P_\mu$ can be evaluated readily by
using above equations. However, ${\cal A}^P_\mu$ obtained in this
way contains a spurious contribution  proportional to
$\tilde\omega^\mu=(\tilde\omega^-,\tilde\omega^+,\tilde\omega_\bot)=(2,0,0_\bot)$.
It arises from the momentum decomposition of $\hat
p_1^{\prime\mu}$
 \be
 \hat p_1^{\prime\mu} &=& (P'-\hat p_2)^\mu \non \\
 &=& x_1 P^{\prime\mu} +(0,0,\vec
p^\prime_\bot)^\mu+{1\over 2}\left(x_2P^{\prime
-}-{\vec{p}_{2\bot}^2+m_2^2\over
x_2P^{\prime+}}\right)\,\tilde\omega^\mu.
 \en
In fact, after the integration, $p^\prime_1$ can be expressed in
terms of two external vectors, $P^\prime$ and $\tilde\omega$.
Therefore, in the integrand of ${\cal A}^M_\mu$, one has
 \be
 p^\prime_{1\mu}&\doteq&\frac{\tilde\omega\cdot p^\prime_1}{\tilde\omega\cdot P^\prime}
 P^\prime_\mu+\frac{1}{\tilde\omega\cdot P^\prime}\tilde\omega_\mu\left(P^\prime\cdot
                         p_1^\prime-\frac{\tilde\omega\cdot p^\prime_1}{\tilde\omega\cdot
                         P^\prime}P^{\prime2}\right)
 \nonumber\\
                &\doteq&x_1 P^\prime_\mu+\frac{1}{2\,\tilde\omega\cdot P^\prime}\tilde\omega_\mu[-N_2+
                        N_1^\prime+m_1^{\prime2}-m_2^2+(1-2x_1)M^{\prime2}].
 \en
The symbol $\doteq$ in the above equation reminds us that it is
true only in the ${\cal A}^M$ integration.
There is one missing piece in the contour integration, namely, the
contribution of the zero mode from the $p_1^{\prime +}=0$
region~\cite{zeromode}. The appearance of $N_2$ in the numerator
as shown in the above equation (\ref{eq:AP}) also prompts an extra
care in performing the $p_1^{\prime-}$ contour integration. It is
interesting that this zero mode contribution provides a cue for
the spurious term in ${\cal A}^M_\mu$. As shown in \cite{Jaus99},
the inclusion of the zero mode contribution in ${\cal A}^{M}_\mu$
matrix elements in practice amounts to the replacements
 \be
 \hat p_1^\prime\to x_1 P^\prime,\quad
 \hat N_2\to \hat
 N_1^\prime+m_1^{\prime2}-m_2^2+(1-2x_1)M^{\prime2},
 \label{eq:p1A}
 \en
in the $\hat S^M$ under the integration. By virtue of
Eqs.~(\ref{eq:AM}), (\ref{eq:contourA}) and (\ref{eq:p1A}), we
obtain \cite{Jaus99}
 \be
 f_P=\frac{N_c}{16\pi^3}\int dx_2 d^2p^\prime_\bot \frac{h^\prime_P}{x_1
 x_2 (M^{\prime2}-M^{\prime2}_0)}4(m_1^\prime x_2+m_2 x_1).
 \label{eq:fP}
 \en
It should be stressed that $f_P$ itself is free of zero mode
contributions as its derivation does not involve the replacement
of $\hat N_2$ (see also Sec. III.B). With the explicit form of
$h'_P$ shown in Eq.~(\ref{eq:h}), the familiar expression of $f_P$
in the conventional light-front approach~\cite{Jaus91,Cheng97},
namely,
 \be \label{eq:fP0}
  f_P=2\frac{\sqrt{2N_c}}{16\pi^3}\int dx_2 d^2p^\prime_\bot \frac{1}{\sqrt{x_1
 x_2} \widetilde M'_0}\,(m_1^\prime x_2+m_2
 x_1)\,\varphi^\prime(x_2,p^\prime_\perp),
 \en
is reproduced.

The decay constant of a scalar meson can be obtained in a similar
manner. By using the corresponding Feynman rules shown in
Table~\ref{tab:feyn}, we have
 \be
 {\cal A}^S_\mu&=&-i^2\frac{N_c}{(2\pi)^4}\int d^4 p_1^\prime
 \frac{H_S^\prime}{N_1^\prime N_2}
 {\rm Tr}[\gamma_\mu(\not \!p^\prime_1+m_1^\prime)(-i)
                            (-\not\!p_2+m_2)].
 \en
Note that the trace ($\equiv s^S_\mu$) in the above equation is
related to $s^P_\mu$ in Eq.~(\ref{eq:AP}) by the replacement of
$m_2\to-m_2$ and by adding an overall factor of $-i$. Likewise, by
using Eqs.~(\ref{eq:AM}), (\ref{eq:contourA}) and (\ref{eq:p1A}),
it follows that
 \be
 f_S=\frac{N_c}{16\pi^3}\int dx_2 d^2p^\prime_\bot \frac{h^\prime_S}{x_1
 x_2 (M^{\prime2}-M^{\prime2}_0)}4(m_1^\prime x_2-m_2 x_1).
 \label{eq:fS}
 \en
For $m_1^\prime=m_2$, the meson wave function is symmetric with
respect to $x_1$ and $x_2$, and hence $f_S=0$, as it should be.

We now turn to the decay constants of vector and axial-vector
mesons. The decay amplitude for a vector meson is given by
 \be \label{eq:AV}
 {\cal A}^V_\mu=-i^2\frac{N_c}{(2\pi)^4}\int d^4 p^\prime_1 \frac{i
 H^\prime_V}{N_1^\prime N_2}
 {\rm Tr}\left\{\gamma_\mu(\not \!p^\prime_1+m_1^\prime)
          \left[\gamma_\nu-\frac{(p_1^\prime-p_2)_\nu}{W^\prime_V}\right]
                            (-\not\!p_2+m_2)\right\}\vp^{\prime\nu}.
 \en
We consider the case with the transverse polarization
 \be
 \vp(\pm)=\left(\frac{2}{P^{\prime+}}\varepsilon_\bot\cdot
 P^\prime_\bot,0,\varepsilon_\bot\right),\qquad
 \varepsilon_\bot=\mp\frac{1}{\sqrt2}(1,\pm i).
 \en
Contracting ${\cal A}^V_\mu$ with $\vp^*(\pm)$ and applying
Eqs.~(\ref{eq:AM}), (\ref{eq:contourA}) and (\ref{eq:p1A}) lead to
~\cite{Jaus99}\footnote{ When ${\cal A}^V_\mu$ is contracted with
the longitudinal polarization vector $\vp^\mu(0)$, $f_V$ will
receive additional contributions characterized by the $B$
functions defined in Appendix B (see Eq. (3.5) of \cite{Jaus03})
which give about 10\% corrections to $f_V$ for the vertex function
$h'_V$ used in Eq. (\ref{eq:h}). It is not clear to us why the
result of $f_V$ depends on the polarization vector. Note that the
new residual contributions are absent in the approach of
\cite{CDKM} in which a different scheme has been developed to
identify the zero mode contributions to the decay constants and
form factors.}
 \be
 f_V&=&\frac{N_c}{4\pi^3M^\prime}\int dx_2 d^2p^\prime_\bot \frac{h^\prime_V}{x_1
      x_2 (M^{\prime2}-M^{\prime2}_0)}
 \nonumber\\
      &&\qquad\qquad\times\left[x_1 M^{\prime2}_0-m_1^\prime (m_1^\prime-m_2)-p^{\prime2}_\bot
            +\frac{m_1^\prime+m_2}{w^\prime_V}\,p^{\prime2}_\bot \right].
 \label{eq:fV}
 \en
We wish to stress that the vector decay constant obtained in the
conventional light-front model \cite{Jaus91} does not coincide
with the above result (\ref{eq:fV}) owing to the missing zero mode
contribution, whose presence is evidenced by its involvement of
$\hat N_2$ \cite{Jaus99,Jaus03}. Since ${\cal A}^{^3\!A}_\mu$
(${\cal A}^{^1\!A}_\mu$) is related to ${\cal A}^V_\mu$ by a
suitable replacement of $H^\prime_V\to -H^\prime_{^3\!A(^1\!A)}$
and $m_2\to -m_2$, $W^\prime_V \to -W^\prime_{^3\!A(^1\!A)}$ in
the trace (only the $1/W^\prime$ terms being kept in the $^1\!A$
case), this allows us to readily obtain
 \be
 f_{^3\!A}&=&-\frac{N_c}{4\pi^3M^\prime}\int dx_2 d^2p^\prime_\bot
            \frac{h^\prime_{^3\!A}}{x_1 x_2 (M^{\prime2}-M^{\prime2}_0)}
 \nonumber\\
      &&\qquad\qquad\times\left[x_1 M^{\prime2}_0-m_1^\prime (m_1^\prime+m_2)-p^{\prime2}_\bot
            -\frac{m_1^\prime-m_2}{w^\prime_{^3\!A}}\,p^{\prime2}_\bot \right],
 \nonumber\\
  f_{^1\!A}&=&\frac{N_c}{4\pi^3 M^\prime}\int dx_2 d^2p^\prime_\bot
            \frac{h^\prime_{^1\!A}}{x_1 x_2 (M^{\prime2}-M^{\prime2}_0)}
      \left(\frac{m_1^\prime-m_2}{w^\prime_{^1\!A}}\,p^{\prime2}_\bot \right).
 \label{eq:fA}
 \en
It is clear that $f_{^1\!A}=0$ for $m^\prime_1=m_2$. The
SU(N)-flavor constraints on $f_S$ and $f_{^1\!A}$ are thus
satisfied. The HQS relations on decay constants will be discussed
in Section IV.

\begin{table}[b!]
\caption{\label{tab:beta} The input parameter $\beta$ (in units of
GeV) in the Gaussian-type wave function (\ref{eq:wavefn}).}
\begin{ruledtabular}
\begin{tabular}{|c|ccccc|}
$^{2S+1} L_J$
          & $\beta_{u\bar d}$
          & $\beta_{s\bar u}$
          & $\beta_{c\bar u}$
          & $\beta_{c\bar s}$
          & $\beta_{b\bar u}$
          \\
\hline $^1S_0$
          & $0.3102$
          & $0.3864$
          & $0.4496$
          & $0.4945$
          & $0.5329$
          \\
$^3S_1$
          & $0.2632$
          & $0.2727$
          & $0.3814$
          & $0.3932$
          & $0.4764$
          \\
$^3P_0$
          & $\beta_{a_1}$
          & $\beta_{K(^3P_1)}$
          & $0.3305$
          & $0.3376$
          & $0.4253$
          \\
$^3P_1$
          & $0.2983$
          & $0.303$
          & $0.3305$
          & $0.3376$
          & $0.4253$
          \\
$^1P_1$
          & $\beta_{a_1}$
          & $\beta_{K(^3P_1)}$
          & $0.3305$
          & $0.3376$
          & $0.4253$
          \\
\end{tabular}
\end{ruledtabular}
\end{table}

\begin{table}[t!]
\caption{\label{tab:f} Mesonic decay constants (in units of MeV)
obtained by using Eqs.~(\ref{eq:fP}), (\ref{eq:fS}), (\ref{eq:fV})
and (\ref{eq:fA}). Those in parentheses are taken as inputs to
determine the corresponding $\beta$'s shown in
Table~\ref{tab:beta}. The decay constant $f_{K_1(1270)}=175$ MeV
is also used as an input (see the text for detail). }
\begin{ruledtabular}
\begin{tabular}{|c|ccccc|}
$^{2S+1} L_J$
          & $f_{u\bar d}$
          & $f_{s\bar u}$
          & $f_{c\bar u}$
          & $f_{c\bar s}$
          & $f_{b\bar u}$
          \\
\hline $^1S_0$
          & $(131)$
          & $(160)$
          & $(200)$
          & $(230)$
          & $(180)$
          \\
$^3S_1$
          & $(216)$
          & $(210)$
          & $(220)$
          & $(230)$
          & $(180)$
          \\
$^3P_0$
          & $0$
          & $21$
          & $86$
          & $71$
          & $112$
          \\
$^3P_1$
          & $(-203)$
          & $-186$
          & $-127$
          & $-121$
          & $-123$
          \\
$^1P_1$
          & $0$
          & $11$
          & $45$
          & $38$
          & $68$
          \\
\hline $P^{1/2}_1$
          & --
          & --
          & $130$
          & $122$
          & $140$
          \\
$P^{3/2}_1$
          & --
          & --
          & $-36$
          & $-38$
          & $-15$
          \\
\end{tabular}
\end{ruledtabular}
\end{table}

In order to have a numerical study for decay constants, we need to
specify the constituent quark masses and the parameter $\beta$
appearing in the Gaussian-type wave function (\ref{eq:wavefn}).
For constituent quark masses we
use~\cite{Jaus96,Cheng97,Hwang02,Jaus99}
 \be \label{eq:quarkmass}
m_{u,d}=0.26\,{\rm GeV},\qquad m_s=0.37\,{\rm GeV},\qquad
m_c=1.40\,{\rm GeV},\qquad m_b=4.64\,{\rm GeV}.
 \en
As we shall see in Sec. III, the masses of strange and charmed
quarks are constrained from the measured form-factor ratios in
semileptonic $D\to K^*\ell\bar\nu$ decays. Shown in
Tables~\ref{tab:beta} and \ref{tab:f} are the input parameter
$\beta$ and decay constants, respectively. In Table \ref{tab:f}
the decay constants in parentheses are used to determine $\beta$.
For the purpose of an estimation, for $p$-wave mesons in $D$,
$D_s$ and $B$ systems we shall use the $\beta$ parameters obtained
in the ISGW2 model~\cite{ISGW2}, the improved version of the ISGW
model, up to some simple scaling. Several remarks are in order:
(i) The values of the parameter $\beta_V$ presented in Table
\ref{tab:beta} are slightly smaller than the ones obtained in the
earlier literature. For example, $\beta_\rho=0.26$,
$\beta_{K^*}=0.27$ and $\beta_{D^*}=0.38$ are obtained here using
the Gaussian-type wave function, while the corresponding values
are 0.30,\,0.31,\,0.46 in~\cite{Cheng97}. This is because we have
utilized the correct light-front expression for the vector decay
constant $f_V$ [cf. Eq. (\ref{eq:fV})]. It is interesting to
notice that $\beta_V$ in the ISGW2 model also has a similar
reduction due to hyperfine interactions, which have been neglected
in the original ISGW model in the mass spectrum calculation. (ii)
The $\beta$ parameters for $p$-wave states of $D$, $D_s$ and $B$
systems are the smallest when compared to $\beta_{P,V}$. (iii) The
decay constants of $^3P_1$ and $P_1^{3/2}$ states have opposite
signs to that of $^1P_1$ or $P_1^{1/2}$ as can be easily seen from
Eq. (\ref{eq:Phalf}).

In principle, the parameter $\beta$ for $p$-wave mesons can be
determined from the study of the meson spectroscopy. Although we
have not explored this issue in this work, it is important to keep
in mind that $\beta$'s are closely related to meson masses. In
Table \ref{tab:f} we have employed $|f_{a_1}|=203$ MeV and
$f_{D_s^*}=f_{D_s}$ as inputs. It is generally argued that
$a_1(1260)$ should have a similar decay constant as the $\rho$
meson.  Presumably, $f_{a_1}$ can be extracted from the decay
$\tau\to a_1(1260)\nu_\tau$. Though this decay is not shown in the
Particle Data Group (PDG) \cite{PDG}, an experimental value of
$|f_{a_1}|=203\pm 18$ MeV is nevertheless quoted in
\cite{Bloch}.\footnote{The decay constant of $a_1$ can be tested
in the decay $B^+\to \bar D^0a_1^+$ which receives the main
contribution from the color-allowed amplitude proportional to
$f_{a_1}F^{BD}(m_{a_1}^2)$.} Contrary to the non-strange charmed
meson case where $D^*$ has a slightly larger decay constant than
$D$, the recent measurements of $B\to D_s^{(*)}D^{(*)}$
\cite{PDG,BaBarDs} indicate that the decay constants of $D_s^*$
and $D_s$ are similar. Hence we shall take $f_{D_s^*}=f_{D_s}$. As
for the decay constant of $B^*$, a recent lattice calculation
yields $f_{B^*}/f_B=1.01\pm0.01^{+0.04}_{-0.01}$ \cite{Bernard}.
Therefore we will set $f_{B^*}=f_B$ in Table \ref{tab:f}.

It is clear from Eq. (\ref{eq:Seom}) that the decay constant of
light scalar resonances is largely suppressed relative to that of
the pseudoscalar mesons owing to the small mass difference between
the constituent quark masses. However, as shown in Table
\ref{tab:f}, this suppression becomes less restrictive for heavy
scalar mesons because of heavy and light quark mass imbalance.
Note that what is the underlying quark structure of light scalar
resonances is still controversial. While it has been widely
advocated that the light scalar nonet formed by $\sigma(600)$,
$\kappa(800)$, $f_0(980)$ and $a_0(980)$ can be identified
primarily as four-quark states, it is generally believed that the
nonet states $f_0(1370)$, $a_0(1450)$, $K_0^*(1430)$ and
$f_0(1500)/f_0(1710)$ are the conventional $q\bar q'$ states (for
a review, see e.g. \cite{Close}). Therefore, the prediction of
$f_S=21$ MeV for the scalar meson in the $s\bar u$ content (see
Table \ref{tab:f}) is most likely designated for the $K_0^*(1430)$
state. Notice that this prediction is slightly smaller than the
result of 42 MeV obtained in \cite{Maltman} based on the
finite-energy sum rules, and far less than the estimate of
$(70\pm10)$ MeV  in \cite{Chernyak}. It is worth remarking that
even if the light scalar mesons are made from 4 quarks, the decay
constants of the neutral scalars $\sigma(600)$, $f_0(980)$ and
$a_0^0(980)$ must vanish owing to charge conjugation invariance.

In principle, the decay constant of the scalar strange charmed
meson $D^*_{s0}$ can be determined from the hadronic decay $B\to
\ov DD_{s0}^*$ since it proceeds only via external $W$-emission.
Indeed, a recent measurement of the $D\bar D_{s0}^*$ production in
$B$ decays by Belle \cite{BelleDs-1} indicates a $f_{D_{s0}^*}$ of
order 60 MeV \cite{Cheng:2003id} which is close to the expectation
of 71 MeV (see Table \ref{tab:f}). In Sec. III.E we will discuss
more about $\ov DD_s^{**}$ productions in $B$ decays. The
smallness of the decay constant $f_{D_{s0}^*}$ relative to
$f_{D_s}$ can be seen from Eqs. (\ref{eq:fP}) and (\ref{eq:fS})
that
 \be
 f_{D_s(D_{s0}^*)}\propto \int dx_2\cdots[m_c x_2\pm m_s(1-x_2)].
 \en
Since the momentum fraction $x_2$ of the strange quark in the
$D_s(D_{s0}^*)$ meson is small, its effect being constructive in
$D_s$ case and destructive in $D_{s0}^*$ is sizable and explains
why $f_{D_{s0}^*}/f_{D_s}\sim 0.3$.

Except for $a_1$ and $b_1$ mesons which cannot have mixing because
of the opposite $C$-parities, physical strange axial-vector mesons
are the mixture of $^3P_1$ and $^1P_1$ states, while the heavy
axial-vector resonances are the mixture of $P_1^{1/2}$ and
$P_1^{3/2}$. For example, $K_1(1270)$ and $K_1(1400)$ are the
mixture of $K_{^3P_1}$ and $K_{^1\!P_1}$ (denoted by $K_{1A}$ and
$K_{1B}$, respectively, by PDG \cite{PDG}) owing to the mass
difference of the strange and non-strange light quarks:
 \be \label{eq:K1mixing}
 K_1(1270)=K_{^3\!P_1} \sin\theta+K_{^1\!P_1}\cos\theta,
 \nonumber\\
 K_1(1400)=K_{^3\!P_1} \cos\theta-K_{^1\!P_1}\sin\theta,
 \en
with $\theta\approx -58^\circ$ as implied from the study of $D\to
K_1(1270)\pi,~K_1(1400)\pi$ decays \cite{Cheng:2003bn}. We use
$f_{K_1(1270)}=175$~MeV~\cite{Cheng:2003bn} to fix
$\beta_{K(^3P_1)}\simeq\beta_{K(^1P_1)}=0.303$~GeV and obtain
$f_{K_1(1400)}=-87$~MeV. Note that these $\beta_{K(^3P_1)}$,
$\beta_{K(^1P_1)}$ are close to $\beta_{K^*}$. For the masses of
$K_{^1\!P_1}$ and $K_{^3\!P_1}$, we follow \cite{Suzuki} to
determine them from the mass relations
$2m_{K_{^1\!P_1}}^2=m_{b_1(1232)}^2+m_{h_1(1380)}^2$ and
$m_{K_{^3\!P_1}}^2=m_{K_1(1270)}^2+m_{K_1(1400)}^2-m_{K_{^1\!P_1}}^2$.
For $D$ and $B$ systems, it is clear from Table~\ref{tab:f} that
$|f_{A^{3/2}}|\ll f_S <  f_{A^{1/2}}$, in accordance with the
expectation from HQS [cf. Eq. (\ref{eq:HQSf})].

\section{Covariant model analysis of form factors}

In this section we first review the analysis of the form factors
for $s$-wave mesons within the framework of the covariant
light-front quark model \cite{Jaus99} and then extend it to the
$p$-wave meson case followed by numerical results and discussion.

\subsection{Form factors}

Form factors for $P\to P,V$ transitions are defined by
  \be
\la P(P^\pp)|V_\mu|P(P^\prime)\ra
          &=& P_\mu f_+(q^2) + q_\mu f_-(q^2),
\non\\
 \la V(P^\pp,\vp^\pp)|V_\mu|P(P^\prime)\ra
          &=&\epsilon_{\mu\nu\alpha \beta}\,\vp^{\pp*\nu}P^\alpha q^\beta\, g({q^2}),
\non\\
\la V(P^\pp,\vp^\pp)|A_\mu|P(P^\prime)\ra
          &=&-i\left\{\varepsilon_\mu^{\pp*} f({q^2})
              +\vp^{*\pp}\cdot P \left[P_\mu a_+({q^2})+q_\mu a_-({q^2})\right]\right\},
 \label{eq:ffs}
 \en
where $P=P^\prime+P^\pp$, $q=P^\prime-P^\pp$ and the convention
$\epsilon_{0123}=1$ is adopted. These form factors are related to
the commonly used Bauer-Stech-Wirbel (BSW) form factors \cite{BSW}
via
 \be \label{eq:ffsdimless}
 F_1^{PP}(q^2)&=&f_+(q^2),\quad
                 F_0^{PP}(q^2)=f_+(q^2)+\frac{q^2}{q\cdot P} f_-(q^2),
 \non\\
 V^{PV}(q^2)&=&-(M^\prime+M^\pp)\, g(q^2),\quad
                 A_1^{PV}(q^2)=-\frac{f(q^2)}{M^\prime+M^\pp},
 \non\\
 A_2^{PV}(q^2)&=&(M^\prime+M^\pp)\, a_+(q^2),\quad
                 A_3^{PV}(q^2)-A_0^{PV}(q^2)=\frac{q^2}{2 M^\pp}\, a_-(q^2),
 \en
 where the latter form factors are defined by \cite{BSW}
  \be \label{eq:ffPPV}
 \la P(P^\pp)|V_\mu|P(P^\prime)\ra &=& \left(P_\mu-{M^{\prime 2}-M^{\pp 2}\over q^2}\,q_ \mu\right)
F_1^{PP}(q^2)+{M^{\prime 2}-M^{\pp 2}\over q^2}\,q_\mu\,F_0^{PP}(q^2), \non \\
 \la V(P^\pp,\vp^\pp)|V_\mu|P(P^\prime)\ra &=& -{1\over
M'+M^\pp}\,\epsilon_{\mu\nu\alpha \beta}\vp^{\pp*\nu}P^\alpha
q^\beta  V^{PV}(q^2),   \non \\
 \la V(P^\pp,\vp^\pp)|A_\mu|P(P^\prime)\ra &=& i\Big\{
(M'+M^\pp)\vp^{\pp*}_\mu A_1^{PV}(q^2)-{\vp^{\pp*}\cdot P\over
M'+M^\pp}\,P_\mu A_2^{PV}(q^2)    \non \\
&& -2M^\pp\,{\vp^{\pp*}\cdot P\over
q^2}\,q_\mu\big[A_3^{PV}(q^2)-A_0^{PV}(q^2)\big]\Big\},
 \en
with $F_1^{PP}(0)=F_0^{PP}(0)$, $A_3^{PV}(0)=A_0^{PV}(0)$, and
 \be
A_3^{PV}(q^2)=\,{M'+M^\pp\over
2M^\pp}\,A_1^{PV}(q^2)-{M'-M^\pp\over 2M^\pp}\,A_2^{PV}(q^2).
 \en
The general expressions for $P$ to low-lying $p$-wave meson
transitions are given by \cite{ISGW}
 \be \label{ISGWform}
 \la S(P^\pp)|A_\mu|P(P^\prime)\ra &=& i\Big[ u_+(q^2)P_\mu+u_-(q^2)q_\mu
 \Big], \non \\
 \la A^{1/ 2}(P^\pp,\vp^\pp)|V_\mu|P(P^\prime)\ra
           &=& i\left\{\ell_{1/2}(q^2)\vp_\mu^{\pp*}+\vp^{\pp*}\cdot
                  P[P_\mu c_+^{1/2}(q^2)+q_\mu c_-^{1/2}(q^2)]\right\},
 \non \\
 \la A^{1/2}(P^\pp,\vp^\pp)|A_\mu|P(P^\prime)\ra
           &=& -q_{1/2}(q^2)\epsilon_{\mu\nu\alpha\beta}\vp^{\pp*\nu}P^\alpha
q^\beta,
 \non \\
 \la A^{3/ 2}(P^\pp,\vp^\pp)|V_\mu|P(P^\prime)\ra
           &=& i\left\{\ell_{3/2}(q^2)\vp_\mu^{\pp*}+\vp^{\pp*}\cdot
               P[P_\mu c_+^{3/2}(q^2)+q_\mu c_-^{3/2}(q^2)]\right\},
 \non \\
  \la A^{3/2}(P^\pp,\vp^\pp)|A_\mu|P(P^\prime)\ra
           &=& -q_{3/2}(q^2)\epsilon_{\mu\nu\alpha\beta}\vp^{\pp*\nu}P^\alpha
q^\beta,
 \non \\
 \la T(P^\pp,\vp^\pp)|V_\mu|P(P^\prime)\ra
           &=& h(q^2)\epsilon_{\mu\nu\alpha\beta}\vp^{\pp*\nu\lambda}
                    P_\lambda P^\alpha q^\beta,
 \non \\
 \la T(P^\pp,\vp^\pp)|A_\mu|P(P^\prime)\ra
           &=& -i\Big\{k(q^2)\vp^{\pp*}_{\mu\nu}P^{\nu}+
                \vp^{\pp*}_{\alpha\beta}P^{\alpha} P^{\beta}
                [P_\mu b_+(q^2)+q_\mu b_-(q^2)]\Big\}.
 \label{eq:ffp}
  \en
The form factors $\ell_{1/2(3/2)},c_+^{1/2(3/2)},c_-^{1/2(3/2)}$
and $q_{1/2(3/2)}$ are defined for the transitions to the heavy
$P^{1/2}_1$ ($P^{3/2}_1$) state. For transitions to light
axial-vector mesons, it is more appropriate to employ the $L-S$
coupled states $^1P_1$ and $^3P_1$ denoted by the particles
$^1\!A$ and $^3\!A$ in our notation. The relation between
$P^{1/2}_1,P^{3/2}_1$ and $^1P_1,\,^3P_1$ states is given by Eq.
(\ref{eq:Phalf}). The corresponding form factors
$\ell_{^1\!A(^3\!A)},c_+^{^1\!A(^3\!A)},c_-^{^1\!A(^3\!A)}$ and
$q_{^1\!A(^3\!A)}$ for $P\to \,^1\!A$ ($^3\!A$) transitions can be
defined in an analogous way.\footnote{The form factors
$\ell_{^1\!A(^3\!A)},c_+^{^1\!A(^3\!A)},c_-^{^1\!A(^3\!A)}$ and
$q_{^1\!A(^3\!A)}$ are dubbed as $\ell(v),c_+(s_+),c_-(s_-)$ and
$q(r)$, respectively, in the ISGW model \cite{ISGW}.}

Note that only the form factors $u_+(q^2),u_-(q^2)$ and $k(q^2)$
in the above parametrization are dimensionless. It is thus
convenient to define dimensionless form factors by\footnote{The
definition here for dimensionless $P\to A$ transition form factors
differs than Eq. (3.17) of \cite{Cheng:2003id} where the
coefficients $(m_P\pm m_A)$ are replaced by $(m_P\mp m_A)$. It
will become clear in Sec. IV that this definition will lead to HQS
relations for $B\to D_0^*,D_1$ transitions [cf. Eq.
(\ref{eq:HQStau1half})] similar to that for $B\to D,D^*$ ones.}
 \be \label{eq:ffpdimless}
 \la S(P^\pp)|A_\mu|P(P^\prime)\ra &=&
 -i\left[\left(P_\mu-{M^{\prime 2}-M^{\pp 2}\over q^2}\,q_
 \mu\right) F_1^{PS}(q^2)+{M^{\prime 2}-M^{\pp 2}\over q^2}
 \,q_\mu\,F_0^{PS}(q^2)\right], \non \\
 \la A(P^\pp,\vp^\pp)|V_\mu|P(P^\prime)\ra &=&
-i\Bigg\{(m_P-m_A) \vp^*_\mu V_1^{PA}(q^2)  - {\vp^*\cdot
P'\over m_P-m_A}P_\mu V_2^{PA}(q^2) \non \\
&-& 2m_A {\vp^*\cdot P'\over
q^2}q_\mu\left[V_3^{PA}(q^2)-V_0^{PA}(q^2)\right]\Bigg\},
\non \\
   \la A(P^\pp,\vp^\pp)|A_\mu|P(P^\prime)\ra &=& -{1\over
  m_P-m_A}\,\epsilon_{\mu\nu\rho\sigma}\vp^{*\nu}P^\rho
  q^{\sigma}A^{PA}(q^2),
 \en
with
 \be V_3^{PA}(q^2)=\,{m_P-m_A\over 2m_A}\,V_1^{PA}(q^2)-{m_P+m_A\over
2m_A}\,V_2^{PA}(q^2),
 \en
and $V_3^{PA}(0)=V_0^{PA}(0)$. They are related to the form
factors in (\ref{eq:ffPPV}) via
 \be
 F^{PS}_1(q^2)&=&-u_+(q^2),\quad
                 F^{PS}_0(q^2)=-u_+(q^2)-\frac{q^2}{q\cdot P} u_-(q^2),
 \non\\
 A^{PA}(q^2)&=&-(M^\prime-M^\pp)\, q(q^2),\quad
                 V^{PA}_1(q^2)=-\frac{\ell(q^2)}{M^\prime-M^\pp},
 \non\\
 V^{PA}_2(q^2)&=&(M^\prime-M^\pp)\, c_+(q^2),\quad
                 V^{PA}_3(q^2)-V^{PA}_0(q^2)=\frac{q^2}{2 M^\pp}\, c_-(q^2).
 \en
In above equations, the axial-vector meson $A$ stands for
$A^{1/2}$ or $A^{3/2}$. Besides the dimensionless form factors,
this parametrization has the advantage that the $q^2$ dependence
of the form factors is governed by the resonances of the same
spin, for instance, the momentum dependence of $F_0(q^2)$ is
determined by scalar resonances.

To obtain the $P\to M$ transition form factors with $M$ being a
ground-state $s$-wave meson or a low-lying $p$-wave meson, we
shall consider the matrix elements
 \be
 \la M(P^\pp)|V_\mu-A_\mu|P(P^\prime)\ra\equiv {\cal B}^{PM}_\mu,
 \label{eq:BPM}
 \en
where the corresponding Feynman diagram is shown in Fig.~1(b). We
follow \cite{Jaus99} to obtain $P\to P,V$ form factors before
extending the formalism to the $p$-wave meson case. As we shall
see, the $P\to S\,(A)$ transition form factors can be easily
obtained by some suitable modifications on $P\to P\,(V)$ ones, and
we need some extension of the analysis in \cite{Jaus99} to the
$P\to T$ case.

For the case of $M=P$, it is straightforward to obtain
 \be
 {\cal B}^{PP}_\mu=-i^3\frac{N_c}{(2\pi)^4}\int d^4 p^\prime_1
 \frac{H^\prime_P H^\pp_P}{N_1^\prime N_1^\pp N_2} S^{PP}_{V\mu},
 \label{eq:BPP}
 \en
where 
 \be
S^{PP}_{V\mu} &=&{\rm Tr}[\gamma_5(\not
\!p^\pp_1+m_1^\pp)\gamma_\mu
              (\not \!p^\prime_1+m_1^\prime)\gamma_5(-\not
              \!p_2+m_2)],
 \label{eq:SPPV}
 \en
$N_1^\pp=p_1^{\pp2}-m_1^{\pp2}+i\epsilon$ and the subscript of
$S_V$ stands for the transition vector current. As noted in the
Introduction we consider the $q^+=0$ frame~\cite{Jaus99}. As in
the ${\cal A}^P_\mu$ case, the $p_1^{\prime-}$ integration picks
up the residue $p_2=\hat p_2$ and leads to
 \be
 N_1^{\prime(\pp)}
      &\to&\hat N_1^{\prime(\pp)}=x_1(M^{\prime(\pp)2}-M_0^{\prime(\pp)2}),
\non\\
 H^{\prime(\pp)}_M
      &\to& h^{\prime(\pp)}_M,
\non\\
 W^\pp_M
      &\to& w^\pp_M,
\non\\
\int \frac{d^4p_1^\prime}{N^\prime_1 N^\pp_1 N_2}H^\prime_P
H^\pp_M S^{PM}
      &\to& -i \pi \int \frac{d x_2 d^2p^\prime_\bot}
                             {x_2\hat N^\prime_1
                             \hat N^\pp_1} h^\prime_P h^\pp_M \hat S^{PM},
 \label{eq:contourB}
 \en
where
 \be
 M^{\pp2}_0
          =\frac{p^{\pp2}_\bot+m_1^{\pp2}}
                {x_1}+\frac{p^{\pp2}_{\bot}+m_2^2}{x_2},
 \en
with $p^\pp_\bot=p^\prime_\bot-x_2\,q_\bot$. In general, after the
integration in ${\cal B}^{PM}$, $\hat p^\prime_1$ can be expressed
in terms of three external vectors, $P^\prime$, $q$ and
$\tilde\omega$. Furthermore, the inclusion of the zero mode
contribution cancels away the $\tilde\omega$ dependence and in
practice for $\hat p_1^\prime $ and $\hat N_2$ in $\hat S^{PM}$
under the integration, we have~\cite{Jaus99}
 \be
\hat p^\prime_{1\mu}
       &\doteq& P_\mu A_1^{(1)}+q_\mu A_2^{(1)},
 \non\\
\hat p^\prime_{1\mu} \hat p^\prime_{1\nu}
       &\doteq& g_{\mu\nu} A_1^{(2)}+P_\mu P_\nu A_2^{(2)}+(P_\mu
                q_\nu+ q_\mu P_\nu) A^{(2)}_3+q_\mu q_\nu A^{(2)}_4,
 \non\\
\hat p^\prime_{1\mu} \hat p^\prime_{1\nu} \hat p^\prime_{1\alpha}
       &\doteq& (g_{\mu\nu} P_\alpha+g_{\mu\alpha} P_\nu+g_{\nu\alpha} P_\mu) A_1^{(3)}
               +(g_{\mu\nu} q_\alpha+g_{\mu\alpha} q_\nu+g_{\nu\alpha} q_\mu) A_2^{(3)}
 \non \\
       &&       +P_\mu P_\nu P_\alpha A_3^{(3)}
                +(P_\mu P_\nu q_\alpha+ P_\mu q_\nu P_\alpha+q_\mu P_\nu P_\alpha) A^{(3)}_4
 \non \\
       &&       +(q_\mu q_\nu P_\alpha+ q_\mu P_\nu q_\alpha+P_\mu q_\nu q_\alpha)
                 A^{(3)}_5
                +q_\mu q_\nu q_\alpha  A^{(3)}_6,
  \non\\
\hat N_2
       &\to& Z_2,
       \quad
       x_1 \hat N_2\to 0,
 \non\\
 \hat p_{1\mu}^\prime \hat N_2
        &\to& q_\mu\left[A^{(1)}_2 Z_2+\frac{q\cdot P}{q^2} A^{(2)}_1\right],
 \non\\
\hat p_{1\mu}^\prime \hat p_{1\nu}^\prime \hat N_2
       &\to &g_{\mu \nu} A^{(2)}_1 Z_2+q_\mu q_\nu
             \left[A^{(2)}_4 Z_2+2\frac{q\cdot P}{q^2} A^{(1)}_2 A^{(2)}_1\right],
\non\\
 \hat p_{1\mu}^\prime \hat p_{1\nu}^\prime p_{1\alpha}^\prime\hat N_2
       &\to &(g_{\mu\nu} q_\alpha+g_{\mu\alpha} q_\nu+g_{\nu\alpha} q_\mu)
       \left[A_2^{(3)}Z_2+\frac{q\cdot P}{3\,q^2} (A^{(2)}_1)^2\right]
\non\\
        &&+q_\mu q_\nu q_\alpha
             \left\{A^{(3)}_6 Z_2+3\frac{q\cdot P}{q^2}
                       \left[A^{(1)}_2 A^{(3)}_2-\frac{1}{3\,q^2}
                       (A^{(2)}_1)^2
                       \right]
             \right\},
 \label{eq:p1B}
 \en
where $A^{(i)}_j$, $Z_2$ are functions of $x_{1,2}$,
$p^{\prime2}_\bot$, $p^\prime_\bot\cdot q_\bot$ and $q^2$, and
their explicit expressions are given in \cite{Jaus99}. We do not
show the spurious contributions in the above equation since they
vanish either after applying the above rules or after integration.
The last rule on $\hat p_{1\mu}^\prime \hat p_{1\nu}^\prime
p_{1\alpha}^\prime\hat N_2$ in the above equation, which is needed
in the $P\to T$ calculation, is extended in this work. One needs
to consider the product of four $\hat p_1^\prime$'s. For
completeness, the formulas for the product of four $\hat
p_1^\prime$'s and the expressions for $A^{(i)}_j$, $Z_2$ can be
found in Appendix B.

From Eqs.~(\ref{eq:BPP})--(\ref{eq:p1B}) one can obtain the form
factors $f_\pm(q^2)$ for $q^2=-q_\bot^2\leq0$ [see Eq.
(\ref{eq:fpm})]. We will return to the issue of the momentum
dependence of form factors in the next sub-section.  The explicit
expressions for $f_\pm$ can be evaluated readily by using the
explicit representations of $\hat N^{\prime(\pp)}_1$,
$h^{\prime(\pp)}$ given in Eqs.~(\ref{eq:contourB}) and
(\ref{eq:h}). At $q^2=0$, the form factor $f_+(0)$ is reduced to
the familiar form \cite{Jaus90,ODonnell}
 \be
 f_+(0)=\,{1\over 16\pi^3}\int
 dxd^2p'_\bot\,\varphi^{\pp*}(x,p'_\bot)\varphi'(x,p'_\bot)\,{{\cal
 A'A^\pp}+p^{\prime 2}_\bot\over \sqrt{{\cal A'}^2+p^{\prime 2}_\bot}\sqrt{{\cal
 A^\pp}^2+p^{\prime 2}_\bot}},
 \en
where
 \be \label{eq:A}
 {\cal A'}=m'_1x+m_2(1-x), \qquad {\cal A^\pp}=m^\pp_1 x+m_2(1-x),
 \en
with $x\equiv x_2$.

For the $P\to S$ transition amplitude, we have
 \be
 {\cal B}^{PS}_\mu=-i^3\frac{N_c}{(2\pi)^4}\int d^4 p^\prime_1
 \frac{H^\prime_P H^\pp_S}{N_1^\prime N_1^\pp N_2} S^{PS}_{A\mu},
 \label{eq:BPS}
 \en
with
 \be
 S^{PS}_{A\mu} &=&{\rm Tr}[(-i)(\not \!p^\pp_1+m_1^\pp)
                             \gamma_\mu\gamma_5
                             (\not \!p^\prime_1+m_1^\prime)\gamma_5(-\not \!p_2+m_2)]
 \non\\
           &=&-i\, S^{PP}_{V\mu}(m_1^\pp\to -m_1^\pp).
 \label{eq:SPSA}
 \en
 Thus, the $P\to S$ transition form factors are related to $f_\pm$ by
  \be 
  u_\pm=-f_\pm(m_1^\pp\to -m_1^\pp,h^\pp_P\to h^\pp_S).
 \en
To be specific, we give the explicit forms of $u_\pm(q^2)$
obtained in the covariant light-front model
 \be
 u_+(q^2)&=&\frac{N_c}{16\pi^3}\int dx_2 d^2p^\prime_\bot
            \frac{h^\prime_P h^\pp_S}{x_2 \hat N_1^\prime \hat N^\pp_1}
            \Big[-x_1 (M_0^{\prime2}+M_0^{\pp2})-x_2 q^2
 \non\\
         &&\qquad+x_2(m_1^\prime+m_1^\pp)^2 +x_1(m_1^\prime-m_2)^2+x_1(m_1^\pp+m_2)^2\Big],
 \non\\
 u_-(q^2)&=&\frac{N_c}{16\pi^3}\int dx_2 d^2p^\prime_\bot
            \frac{2h^\prime_P h^\pp_S}{x_2 \hat N_1^\prime \hat N^\pp_1}
            \Bigg\{ x_1 x_2 M^{\prime2}+p_\bot^{\prime2}+m_1^\prime m_2
                  +(m_1^\pp+m_2)(x_2 m_1^\prime+x_1 m_2)
\non\\
         &&\qquad -2\frac{q\cdot P}{q^2}\left(p^{\prime2}_\bot+2\frac{(p^\prime_\bot\cdot q_\bot)^2}{q^2}\right)
                  -2\frac{(p^\prime_\bot\cdot q_\bot)^2}{q^2}
                  +\frac{p^\prime_\bot\cdot q_\bot}{q^2}
                  \Big[M^{\pp2}-x_2(q^2+q\cdot P)
\non\\
         &&\qquad -(x_2-x_1) M^{\prime2}+2 x_1 M_0^{\prime
                  2}-2(m_1^\prime-m_2)(m_1^\prime-m_1^\pp)\Big]
           \Bigg\}.
 \label{eq:upm}
 \en
It is ready to evaluate these form factors by using the explicit
expressions of $\hat N$ and $h$. Numerical study of these form
factors will be given in the next sub-section.

We next turn to the $P\to V,A$ transition form factors. For the
$P\to V$ transition, we have
 \be
 {\cal B}^{PV}_\mu=-i^3\frac{N_c}{(2\pi)^4}\int d^4 p^\prime_1
 \frac{H^\prime_P (i H^\pp_V)}{N_1^\prime N_1^\pp N_2} S^{PV}_{\mu\nu}\,\vp^{\pp*\nu},
 \label{eq:BPV}
 \en
where 
 \be
S^{PV}_{\mu\nu} &=&(S^{PV}_{V}-S^{PV}_A)_{\mu\nu}
 \non\\
                &=&{\rm Tr}\left[\left(\gamma_\nu-\frac{1}{W^\pp_V}(p_1^\pp-p_2)_\nu\right)
                                 (\not \!p^\pp_1+m_1^\pp)
                                 (\gamma_\mu-\gamma_\mu\gamma_5)
                                 (\not \!p^\prime_1+m_1^\prime)\gamma_5(-\not
                                 \!p_2+m_2)\right].
   \label{eq:SPV}
 \en
As will be seen later, this expression of $S$ is also useful for
the $P\to T$ calculation, and hence its explicit representation is
included in Appendix B. By the aid of Eqs.~(\ref{eq:contourB}) and
(\ref{eq:p1B}), it is straightforward to obtain the $P\to V$ form
factors, $g(q^2),\,f(q^2),\,a_\pm(q^2)$~\cite{Jaus99}. For
reader's convenience, the explicit forms of these form factors are
summarized in Appendix B. Note that the vector form factor
$V(q^2=0)=-(M^\prime+M^\pp) g(q^2=0)$ is consistent with that in
\cite{ODonnell,Cheng97} obtained by a Taylor expansion of the
$h^\pp_V/\hat N_1^\pp$ term in $g(q^2)$ [see Eq.~(\ref{eq:PtoV})]
with respect to $p^{\pp2}_\bot$. To show this, we write
 \be \label{eq:Taylor}
 \frac{h^\pp_V}{\hat N_1^\pp}
 =\frac{h^\pp_V}{\hat  N_1^\pp}\Big |_{p^{\pp2}_\bot\to p^{\prime  2}_\bot}
 -2\, x_2 p^\prime_\bot\cdot q_\bot\Big(\frac{d}{dp^{\pp2}_\bot}\frac{h^\pp_V}
 {\hat N_1^\pp}\Big)_{p^{\pp2}_\bot\to p^{\prime
 2}_\bot}+{\cal O}(x^2_2q^2),
 \en
and see that the second term on the right-hand side is needed when
considering the $q_\bot\to0$ limit of the $p^\prime\cdot
q_\bot/q^2$ term in the integrand of $g(q^2)$, while ${\cal
O}(x_2^2 q^2)$ terms in the above equation vanish in the same
limit. We perform the angular integration in the $\vec
p^\prime_\bot$ plane before taking the $q^2\to0$ limit. After
these steps, we obtain the same expression of $V(q^2=0)$ as in
\cite{ODonnell,Cheng97}.

The extension to $P\to A$ transitions is straightforward and, as
we shall see shortly, the resulting form factors have very similar
expressions as that in the above case. For the $P\to\,
^3\!A,\,^1\!A$ transitions, we have
 \be
 {\cal B}^{P\,^3\!A}_\mu&=&-i^3\frac{N_c}{(2\pi)^4}\int d^4 p^\prime_1
 \frac{H^\prime_P (-i H^\pp_{^3\!A})}{N_1^\prime N_1^\pp N_2} S^{P\,^3\!A}_{\mu\nu}\,\vp^{\pp*\nu},
 \non\\
 {\cal B}^{P\,^1\!A}_\mu&=&-i^3\frac{N_c}{(2\pi)^4}\int d^4 p^\prime_1
 \frac{H^\prime_P (-i H^\pp_{^1\!A})}{N_1^\prime N_1^\pp N_2} S^{P\,^1\!A}_{\mu\nu}\,\vp^{\pp*\nu},
 \label{eq:BPA}
 \en
where
 \be
S^{P\,^3\!A}_{\mu\nu}
&=&(S^{P\,^3\!A}_{V}-S^{P\,^3\!A}_A)_{\mu\nu}
 \non\\
                &=&{\rm Tr}\left[\left(\gamma_\nu-\frac{1}{W^\pp_{^3\!A}}(p_1^\pp-p_2)_\nu\right)
                                 \gamma_5
                                 (\not \!p^\pp_1+m_1^\pp)
                                 (\gamma_\mu-\gamma_\mu\gamma_5)
                                 (\not \!p^\prime_1+m_1^\prime)\gamma_5(-\not
                                 \!p_2+m_2)\right]
 \non\\
                &=&{\rm Tr}\left[\left(\gamma_\nu-\frac{1}{W^\pp_{^3\!A}}(p_1^\pp-p_2)_\nu\right)
                                 (\not \!p^\pp_1-m_1^\pp)
                                 (\gamma_\mu\gamma_5-\gamma_\mu)
                                 (\not \!p^\prime_1+m_1^\prime)\gamma_5(-\not
                                 \!p_2+m_2)\right],
 \non\\
 S^{P\,^1\!A}_{\mu\nu} &=&(S^{P\,^1\!A}_{V}-S^{P\,^1\!A}_A)_{\mu\nu}
 \non\\
                &=&{\rm Tr}\left[\left(-\frac{1}{W^\pp_{^1\!A}}(p_1^\pp-p_2)_\nu\right)
                                 \gamma_5
                                 (\not \!p^\pp_1+m_1^\pp)
                                 (\gamma_\mu-\gamma_\mu\gamma_5)
                                 (\not \!p^\prime_1+m_1^\prime)\gamma_5(-\not
                                 \!p_2+m_2)\right]
  \non\\
                &=&{\rm Tr}\left[\left(-\frac{1}{W^\pp_{^1\!A}}(p_1^\pp-p_2)_\nu\right)
                                 (\not \!p^\pp_1-m_1^\pp)
                                 (\gamma_\mu\gamma_5-\gamma_\mu)
                                 (\not \!p^\prime_1+m_1^\prime)\gamma_5(-\not
                                 \!p_2+m_2)\right].
 \en
We therefore have $S^{P\,^3\!A,P\,^1\!A}_{V(A)}=S^{PV}_{A(V)}$
with the replacement $m_1^\pp\to -m_1^\pp,\,W^\pp_V\to
W^\pp_{^3\!A,^1\!A}$. Note that only the $1/W^\pp$ terms are kept
in $S^{P\,^1\!A}$. Consequently,
 \be \label{eq:PtoA}
 \ell^{^3\!A,^1\!A}(q^2)&=&f(q^2) \,\,\,{\rm with}\,\,\,
                         (m_1^\pp\to -m_1^\pp,\,h^\pp_V\to h^\pp_{^3\!A,^1\!A},\,w^\pp_V\to w^\pp_{^3\!A,^1\!A}),
 \non\\
 q^{^3\!A,^1\!A}(q^2)&=&g(q^2) \,\,\,{\rm with}\,\,\,
                         (m_1^\pp\to -m_1^\pp,\,h^\pp_V\to h^\pp_{^3\!A,^1\!A},\,w^\pp_V\to  w^\pp_{^3\!A,^1\!A}),
 \non\\
 c_+^{^3\!A,^1\!A}(q^2)&=&a_+(q^2) \,\,\,{\rm with}\,\,\,
                          (m_1^\pp\to -m_1^\pp,\,h^\pp_V\to h^\pp_{^3\!A,^1\!A},\,w^\pp_V\to  w^\pp_{^3\!A,^1\!A}),
 \non\\
 c_-^{^3\!A,^1\!A}(q^2)&=&a_-(q^2) \,\,\,{\rm with}\,\,\,
                           (m_1^\pp\to -m_1^\pp,\,h^\pp_V\to h^\pp_{^3\!A,^1\!A},\,w^\pp_V\to  w^\pp_{^3\!A,^1\!A}),
 \en
where only the $1/W^\pp$ terms in $P\to\, ^1\!A$ form factors are
kept. It should be cautious that the replacement of $m_1^\pp\to
-m_1^\pp$ should not be applied to $m_1^\pp$ in $w^\pp$ and
$h^\pp$. These form factors can be expressed in the $P^{3/2}_1$
and $P^{1/2}_1$ basis by using Eq.~(\ref{eq:Phalf}).

Finally we turn to the $P\to T$ transition given by
 \be
 {\cal B}^{PT}_\mu=-i^3\frac{N_c}{(2\pi)^4}\int d^4 p^\prime_1
 \frac{H^\prime_P (i H^\pp_T)}{N_1^\prime N_1^\pp N_2} S^{PT}_{\mu\nu\lambda}\,\vp^{\pp*\nu\lambda},
 \label{eq:BPT}
 \en
where
 \be
S^{PT}_{\mu\nu\lambda} =S^{PV}_{\mu\nu}(-q+p^\prime_1)_\lambda.
 \label{eq:SPT}
 \en
The contribution from the $S^{PV}_{\mu\nu}(-q)_\lambda$ part is
trivial, since $q_\lambda$ can be taken out from the integration,
which is already done in the $P\to V$ case. Contributions from the
$\hat S^{PV}_{\mu\nu}\hat p^\prime_{1\lambda}$ part can be worked
out by using Eq.~(\ref{eq:p1B}). In particular, the calculation of
$k(q^2)$ and $b_-(q^2)$ needs to use the $\hat p^\prime_1 \hat
p^\prime_1 \hat p^\prime_1 \hat N_2$ formula. Putting all these
together leads to
 \be \label{eq:PtoT}
 h(q^2)&=&-g(q^2)\Big|_{h^\pp_V\to h^\pp_T}+\frac{N_c}{16\pi^3}\int dx_2 d^2 p^\prime_\bot
            \frac{2 h^\prime_P h^\pp_T}{x_2 \hat N^\prime_1 \hat N^\pp_1}
            \Bigg[(m_1^\prime-m_1^\pp)(A^{(2)}_3+A^{(2)}_4)
 \non\\
       &&          +(m_1^\pp+m_1^\prime-2 m_2)(A^{(2)}_2+A^{(2)}_3)
                   -m_1^\prime (A^{(1)}_1+A^{(1)}_2)
                   +\frac{2}{w^\pp_V}(2 A^{(3)}_1+2 A^{(3)}_2-A^{(2)}_1)
            \Bigg],
 \non\\
 k(q^2)&=&-f(q^2)\Big|_{h^\pp_V\to h^\pp_T}+\frac{N_c}{16\pi^3}\int dx_2 d^2 p^\prime_\bot
            \frac{ h^\prime_P h^\pp_T}{x_2 \hat N^\prime_1 \hat N^\pp_1}
            \Bigg\{2(A^{(1)}_1+A^{(1)}_2)[m_2(q^2-\hat N_1^\prime-\hat N^\pp_1-m_1^{\prime2}-m_1^{\pp2})
 \non\\
           &&      -m_1^\prime (M^{\pp2}-\hat N_1^\pp-m_1^{\pp2}-m_2^2)
                   -m^\pp_1(M^{\prime2}-\hat N^\prime_1-m_1^{\prime2}-m_2^2)-2 m_1^\prime m_1^\pp
                   m_2]
 \non\\
          &&       +2(m_1^\prime+m_1^\pp)\Bigg(A^{(1)}_2 Z_2+\frac{q\cdot P}{q^2} A^{(2)}_1\Bigg)
                   +16(m_2-m_1^\prime)(A^{(3)}_1+A^{(3)}_2)+4(2 m_1^\prime-m_1^\pp-m_2) A^{(2)}_1
 \non\\
          &&       +\frac{4}{w^\pp_V}\Bigg([M^{\prime2}+M^{\pp2}-q^2
                                           +2(m_1^\prime-m_2)(m_1^\pp+m_2)](2 A^{(3)}_1+2
                                           A^{(3)}_2-A^{(2)}_1)
 \non\\
          &&                                -4\Bigg[A^{(3)}_2 Z_2+\frac{q\cdot P}{3q^2}\Big(A^{(2)}_1\Big)^2
                                              \Bigg]
                                            +2A^{(2)}_1 Z_2
                                     \Bigg)
            \Bigg\},
 \non\\
 b_+(q^2)&=&-a_+(q^2)\Big|_{h^\pp_V\to h^\pp_T}+\frac{N_c}{16\pi^3}\int dx_2 d^2 p^\prime_\bot
            \frac{ h^\prime_P h^\pp_T}{x_2 \hat N^\prime_1 \hat N^\pp_1}
            \Bigg\{8(m_2-m_1^\prime)(A^{(3)}_3+2A^{(3)}_4+A^{(3)}_5)
 \non\\
          &&       -2m_1^\prime (A^{(1)}_1+A^{(1)}_2)
                   +4(2 m_1^\prime-m_1^\pp-m_2)
                   (A^{(2)}_2+A^{(2)}_3)
                   +2(m_1^\prime+m_1^\pp)(A^{(2)}_2+2A^{(2)}_3+A^{(2)}_4)
 \non\\
          &&       +\frac{2}{w^\pp_V}\Bigg[2[M^{\prime2}+M^{\pp2}-q^2
                                           +2(m_1^\prime-m_2)(m_1^\pp+m_2)]
                                           (A^{(3)}_3+2 A^{(3)}_4+A^{(3)}_5-A^{(2)}_2-A^{(2)}_3)
 \non\\
          &&
                                          +[q^2-\hat N_1^\prime-\hat N_1^\pp-(m_1^\prime+m_1^\pp)^2]
                                           (A^{(2)}_2+2A^{(2)}_3+A^{(2)}_4-A^{(1)}_1-A^{(1)}_2)
                                     \Bigg]
            \Bigg\},
  \non\\
 b_-(q^2)&=&-a_-(q^2)\Big|_{h^\pp_V\to h^\pp_T}+\frac{N_c}{16\pi^3}\int dx_2 d^2 p^\prime_\bot
            \frac{ h^\prime_P h^\pp_T}{x_2 \hat N^\prime_1 \hat N^\pp_1}
            \Bigg\{8(m_2-m_1^\prime)(A^{(3)}_4+2A^{(3)}_5+A^{(3)}_6)
 \non\\
          &&       -6m_1^\prime (A^{(1)}_1+A^{(1)}_2)
                   +4(2 m_1^\prime-m_1^\pp-m_2)
                   (A^{(2)}_3+A^{(2)}_4)
 \non\\
          &&       +2(3m_1^\prime+m_1^\pp-2m_2)(A^{(2)}_2+2A^{(2)}_3+A^{(2)}_4)
 \non\\
          &&       +\frac{2}{w^\pp_V}\Bigg[2[M^{\prime2}+M^{\pp2}-q^2
                                           +2(m_1^\prime-m_2)(m_1^\pp+m_2)]
                                           (A^{(3)}_4+2 A^{(3)}_5+A^{(3)}_6-A^{(2)}_3-A^{(2)}_4)
 \non\\
          &&
                                           +2Z_2(3A^{(2)}_4-2A^{(3)}_6-A^{(1)}_2)
                                           +2\frac{q\cdot P}{q^2}
                                             \Big(6 A^{(1)}_2 A^{(2)}_1-6A^{(1)}_2A^{(3)}_2
                                             +\frac{2}{q^2}\big(A^{(2)}_1\big)^2-A^{(2)}_1\Big)
 \non\\
          &&
                                          +[q^2-2M^{\prime2}+\hat N_1^\prime-\hat N_1^\pp
                                                -(m_1^\prime+m_1^\pp)^2+2(m_1^\prime-m_2)^2]
 \non\\
          &&                                \times(A^{(2)}_2+2A^{(2)}_3+A^{(2)}_4-A^{(1)}_1-A^{(1)}_2)
                                     \Bigg]
            \Bigg\}.
 \en

To summarize, equipped with the explicit expressions of the form
factors $f_+(q^2),f_-(q^2)$ [Eq. (\ref{eq:fpm})] for $P\to P$
transitions, $u_+(q^2),u_-(q^2)$ [Eq. (\ref{eq:upm})] for $P\to S$
transitions, $g(q^2),f(q^2),a_+(q^2),a_-(q^2)$ [Eq.
(\ref{eq:PtoV})] for $P\to V$ transitions,
$\ell(q^2),q(q^2),c_+(q^2),c_-(q^2)$ [Eq. (\ref{eq:PtoA})] for
$P\to A$ transition and $h(q^2),k(q^2),b_+(q^2),b_-(q^2)$ [Eq.
(\ref{eq:PtoT})] for $P\to T$ transitions, we are ready to perform
numerical studies of them. The $P\to S,A,T$ transition form
factors are the main new results in this work.

\subsection{Comments on zero-mode effects}
In the present paper we have followed and extended the work of
Jaus \cite{Jaus99} to the $p$-wave meson case. As stressed by
Jaus, there are two classes of form factors and decay constants.
There is one class of form factors like $F_1(q^2)$ for transitions
between pseudoscalar mesons, $V(q^2)$ and $A_2(q^2)$ for
transitions between pseudoscalar and vector mesons, and the
pseudoscalar decay constant $f_P$ that are free of zero mode
contributions. Another class of form factors like $A_1(q^2)$ (or
$f(q^2)$) and the vector decay constant $f_V$ are associated with
zero modes. The full vector vertex operator for $^3S_1$-state
meson has the expression (see Table I)
 \be \label{eq:vertexV}
 iH_V[\gamma_\mu-{1\over W}(p_1-p_2)_\mu].
 \en
To begin with, we first consider the ``simple" vector meson vertex
without the $1/W$ part in the above expression. Jaus employed a
simple multipole ansatz for the meson vertex function
 \be \label{eq:toyvertex}
 H_V(p_1^2,p_2^2)={g\over (p_1^2-\Lambda^2+i\epsilon)^n}
 \en
as the starting point of his simple covariant toy model. Then the
zero mode contributions can be systematically calculated in this
toy model. Note that the vertex function (\ref{eq:toyvertex}) is
not symmetric in the four momenta of the constituent quarks and
hence can hardly be considered a realistic approximation of the
meson vertex. To remedy this difficulty, Bakker, Choi and Ji (BCJ)
\cite{BCJ02} proposed to replace the point gauge-boson vertex
$\gamma_\mu(1-\gamma_5)$ by
 \be \label{eq:BCJvertex}
 \gamma_\mu(1-\gamma_5)\to {\Lambda_1^2\over
 p_1^2-\Lambda_1^2+i\epsilon}\,\gamma_\mu(1-\gamma_5) {\Lambda_2^2\over
 p_2^2-\Lambda_2^2+i\epsilon}.
 \en
It is easily seen that the two methods due to Jaus and BCJ should
give the same result for form factors, but may lead to different
results for decay constants. Indeed, Eq. (3.9) (without the $1/W$
part) of Jaus \cite{Jaus03} for the form factor $f(q^2)$ agrees
with Eqs. (37) and (38) of BCJ \cite{BCJ03}. Moreover, it is
interesting to notice that Eq. (3.16) of Jaus \cite{Jaus03} for
the decay constant $f_V$ (by considering the longitudinal
polarization case as in BCJ) also agrees with Eq. (41) of BCJ
\cite{BCJ02}, though the analytic expressions for the respective
vertex functions $H_V$ are different \cite{pc}. Since the
associated trace for $f_V$ is free of minus components of the
internal momenta and there are no zero modes in that case.
Therefore, to the level without the $1/W$ part in the vertex
operator (\ref{eq:vertexV}), there is no discrepancy between Jaus
and BCJ and both $f_V$ as well as $f(q^2)$ are free of zero mode
effects.

However, things are very different when the full vector meson
vertex (\ref{eq:vertexV}) is used. The $1/W$ part of the trace
contains minus components of the momenta and the zero mode problem
must be faced. While BCJ claimed that both $f_V$ and $f(q^2)$ are
immune to the zero mode even for the full vector meson vertex,
Jaus obtained non-trivial zero mode contributions. It appears to
us that the controversy about the role played by the zero mode
lies in the fact that Jaus and BCJ have different procedures for
identifying zero-mode effects. In the covariant light-front
approach of Jaus \cite{Jaus99}, the decomposition of the
current-induced matrix element into 4-vectors will require to
introduce a lightlike 4-vector $\tilde\omega$ which is not
covariant. Zero modes are required to eliminate the spurious
$\tilde\omega$ dependence. BCJ decompose the propagator into
on-shell and instantaneous (not on-shell) parts and show that only
the latter part can be the origin of a zero mode contribution.
More precisely, the contour integration over $p^-$ in Jaus is not
a regularized one, while in BCJ the contour can be regularly
closed due to the presence of the non-local boson vertex and the
zero modes display their effects at the level of $p^+$ (see Sec.
II.B.2 of \cite{BCJ02} for more detail about the BCJ approach for
zero modes).

The covariant toy model cannot be generalized beyond the simple
meson vertex given by (\ref{eq:toyvertex}), namely, there are some
possible residual $\tilde \omega$ contributions. In \cite{Jaus03}
Jaus has developed a method that permits the calculation of the
contribution of zero modes associated with the current-induced
matrix element. Through the study of the angular condition imposed
on helicity amplitudes, several consistency conditions can be
derived under some plausible assumptions and used to determine the
zero mode contributions. Within this approach, both $f_V$ and
$f(q^2)$ receive additional residual contributions (see Eqs.
(3.16) and (3.9) of \cite{Jaus03}, respectively) which can be
expressed in terms of $B_n^{(m)}$ and $C_n^{(m)}$ functions
defined in Appendix B.\footnote{The remaining spurious
$\tilde\omega$ contribution to the form factor $a_-(q^2)$ cannot
be determined in the same manner.}~~ These functions depend on
$p'^-_1$ and behavior like $(p'^-_1)^i(p'^+_1)^j$. Jaus then gave
a counting rule for detecting zero modes \cite{Jaus03}: For the
$B$ functions $i\leq j$, there is no zero mode contribution and
the value of $B_n^{(m)}$ can be calculated unambiguously at the
spectator quark pole. For the $C$ functions $i\geq j+1$ and the
value of $C_n^{(m)}$ is the sum of a spectator quark pole term and
an unknown zero mode contribution. These $B$ and $C$ functions
vanish in the covariant toy model, as it should be. Beyond the toy
model, the $C$ functions contain unknown zero-mode contributions.
Jaus used some consistency conditions to fix some of the $C$
functions.

Several remarks are in order. (i) Our meson light-front vertex
functions (\ref{eq:vertex}) are symmetric in quark momenta.
However, $B$ and $C$ functions do not appear in $f_V$ [Eq.
(\ref{eq:fV})] and $f(q^2)$ [Eq. (\ref{eq:PtoV})] for two reasons.
First of all, we have $C^{(1)}_1\doteq 0$ from Eqs. (\ref{eq:p1B})
and (\ref{eq:Aij}). Second, we contract ${\cal A}^V_\mu$ [see Eq.
(\ref{eq:AV})] and $S_{\mu\nu}^{PV}$ [Eq. (\ref{eq:SPV})] with the
transverse polarization vector $\vp^\mu(\pm)$. We have checked
explicitly that for the vertex functions given in (\ref{eq:h}),
the coefficients $B^{(i)}_j$ under integration (see Appendix B)
are numerically almost vanishing and the form factor $f(q^2)$ is
affected at most at one percent level. For our purposes, we can
therefore neglect all the residual contributions to the form
factors. (ii) The derivation of the decay constant $f_P$ and the
form factors $f_+(q^2),g(q^2),a_+(q^2)$,
$u_+(q^2),q(q^2),c_+(q^2)$, and $h(q^2),b_+(q^2)$ does not depend
on $\hat N_2$ and those relations connected to $\hat N_2$ (see Eq.
(\ref{eq:p1B}) and recall that $\hat N_2=Z_2-C_1^{(1)}$). These
form factors are free of zero mode effects and can be obtained
using the conventional light-front approach. (iii) Zero mode
effects vanish in the heavy quark limit (see Sec. IV). For
example, the HQS relation $f_P=f_V$ indicates that $f_V$ is immune
to the zero mode contribution.

\subsection{Form-factor momentum dependence and numerical results}

\begin{figure}[t!]
\centerline{
            {\epsfxsize3 in \epsffile{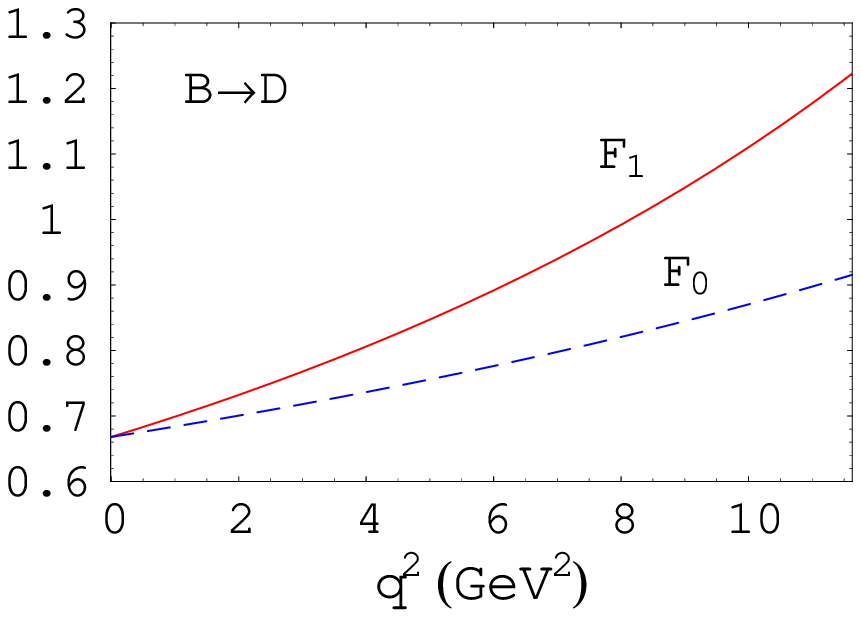}}
            {\epsfxsize3 in \epsffile{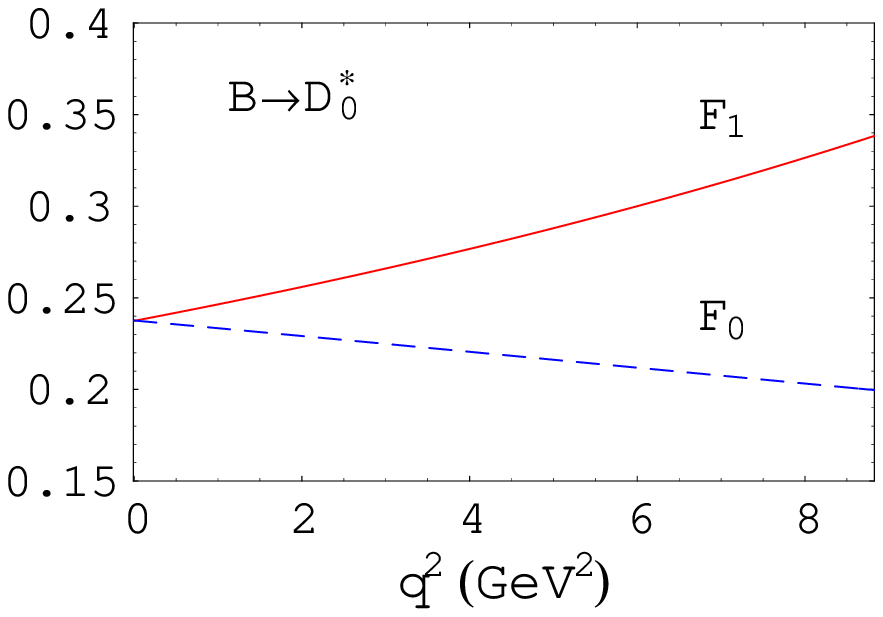}}}
\caption{Form factors $F_1(q^2)$ and $F_0(q^2)$ for $B\to D$ and
$B\to D_0^*$ transitions.} \label{fig:BD} 
\end{figure}

\begin{table}[b]
\caption{Form factors of
$D\to\pi,\rho,a_0(1450),a_1(1260),b_1(1235),a_2(1320)$ transitions
obtained in the covariant light-front model are fitted to the
3-parameter form Eq. (\ref{eq:FFpara}) except for the form factor
$V_2$ denoted by $^{*}$ for which the fit formula Eq.
(\ref{eq:FFpara1}) is used. All the form factors are dimensionless
except for $h,b_+,b_-$ with dimensions GeV$^{-2}$. For the
parameter $\beta_T$ appearing in the tensor-meson wave function,
we assume that it is the same as the $\beta$ parameter of $p$-wave
meson with the same quark content. }
 \label{tab:LFDtopi}
\begin{ruledtabular}
\begin{tabular}{| c c c c c || c c c c c |}
~~~$F$~~~~~
    & $F(0)$~~~~~
    & $F(q^2_{\rm max})$~~~~
    &$a$~~~~~
    & $b$~~~~~~
& ~~~ $F$~~~~~
    & $F(0)$~~~~~
    & $F(q^2_{\rm max})$~~~~~
    & $a$~~~~~
    & $b$~~~~~~
 \\
    \hline
$F^{D\pi}_1$
    & $0.67$
    & $2.71$
    & 1.19
    & 0.36
& $F^{D\pi}_0$
    & 0.67
    & 1.16
    & 0.50
    & $0.01$ \\
$V^{D\rho}$
    & $0.86$
    & $1.36$
    & 1.24
    & 0.48
&$A^{D\rho}_0$
    & 0.64
    & 0.93
    & 1.07
    & 0.54
    \\
$A^{D\rho}_1$
    & 0.58
    & 0.71
    & 0.51
    & 0.03
&$A^{D\rho}_2$
    & $0.48$
    & $0.68$
    & 0.95
    & 0.30
    \\
$F^{Da_0}_1$
    & $0.52$
    & $0.54$
    & 1.07
    & 0.26
&$F^{Da_0}_0$
    & 0.52
    & 0.52
    & $-0.08$
    & 0.03
    \\
 $A^{Da_1}$
    & $0.20$
    & $0.22$
    & 0.98
    & 0.20
& $V^{Da_1}_0$
    & $0.31$
    & $0.34$
    &  0.85
    &  0.49
    \\
 $V^{Da_1}_1$
    & $1.54$
    & $1.53$
    & $-0.05$
    & 0.05
&$V^{Da_1}_2$
    & $0.06$
    & $0.06$
    & $0.12$
    & $0.10$
    \\
$A^{Db_1}$
    & $0.11$
    & $0.13$
    & 1.08
    & 0.54
&$V_0^{Db_1}$
    & $0.49$
    & $0.54$
    & $0.89$
    & $0.28$
    \\
$V^{Db_1}_1$
    & $1.37$
    & $1.45$
    & $0.46$
    & 0.05
& $V_2^{Db_1}$
    & $-0.10^*$
    & $-0.11^*$
    & $0.21^*$
    & $0.67^*$
    \\
$h$
    & 0.188
    & 0.208
    & 1.21
    & 1.09
& $k$
    & 0.340
    & 0.338
    & $-0.07$
    & 0.12
    \\
$b_+$
    & $-0.084$
    & $-0.091$
    & $0.97$
    & 0.58
& $b_-$
    & 0.120
    & 0.133
    & $1.15$
    & 0.66 \\
\end{tabular}
\end{ruledtabular}
\end{table}

Because of the condition $q^+=0$ we have imposed during the course
of calculation, form factors are known only for spacelike momentum
transfer $q^2=-q^2_\bot\leq 0$, whereas only the timelike form
factors are relevant for the physical decay processes. It has been
proposed in \cite{Jaus96} to recast the form factors as explicit
functions of $q^2$ in the spacelike region and then analytically
continue them to the timelike region. Another approach is to
construct a double spectral representation for form factors at
$q^2<0$ and then analytically continue it to $q^2>0$
region~\cite{Melikhov96}. It has been shown recently that, within
a specific model, form factors obtained directly from the timelike
region (with $q^+>0$) are identical to the ones obtained by the
analytic continuation from the spacelike region~\cite{BCJ03}.

In principle, form factors at $q^2>0$ can be evaluated directly in
the frame where the momentum transfer is purely longitudinal,
i.e., $q_\bot=0$, so that $q^2=q^+q^-$ covers the entire range of
momentum transfer \cite{Cheng97}. The price one has to pay is
that, besides the conventional valence-quark contribution, one
must also consider the non-valence configuration (or the so-called
$Z$-graph) arising from quark-pair creation from the vacuum.
However, a reliable way of estimating the $Z$-graph contribution
is still lacking unless one works in a specific model, for
example, the one advocated in \cite{BCJ03}. Fortunately, this
additional non-valence contribution vanishes in the frame where
the momentum transfer is purely transverse i.e., $q^+=0$.

\begin{figure}[t!]
\centerline{
            {\epsfxsize3 in \epsffile{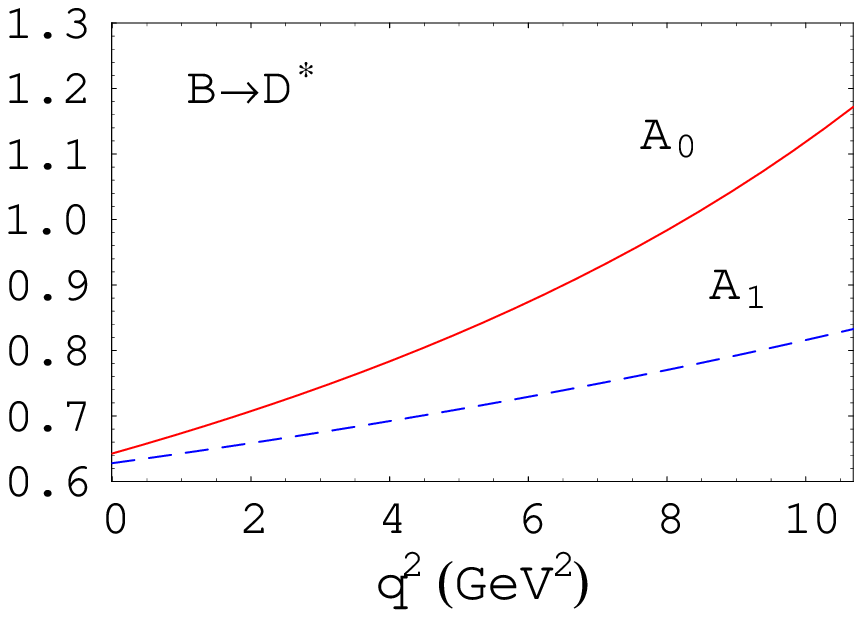}}
            {\epsfxsize3 in \epsffile{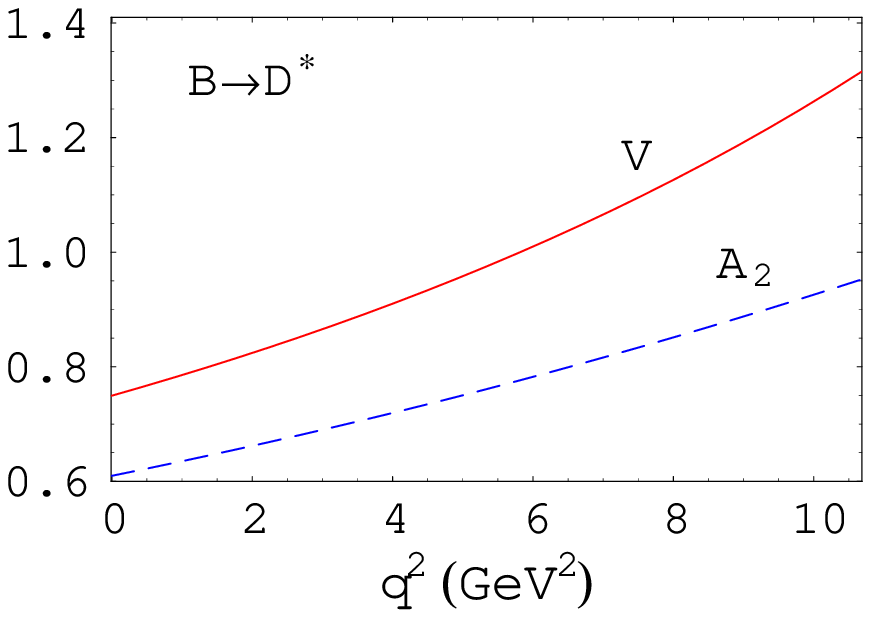}}}
%
\caption{Form factors $V(q^2)$, $A_0(q^2)$, $A_1(q^2)$ and
$A_2(q^2)$ for $B\to D^*$ transitions.} \label{fig:BDst} 
\end{figure}

\begin{table}[b]
\caption{Same as Table \ref{tab:LFDtopi} except for $D\to
K,K^*,K^*_0(1430),K_{^1\!P_1},K_{^3\!P_1},K_2^*(1430)$
transitions.} \label{tab:LFDtoK}
\begin{ruledtabular}
\begin{tabular}{| c c c c c || c c c c c |}
~~~$F$~~~~~
    & $F(0)$~~~~~
    & $F(q^2_{\rm max})$~~~~
    &$a$~~~~~
    & $b$~~~~~~
& ~~~ $F$~~~~~
    & $F(0)$~~~~~
    & $F(q^2_{\rm max})$~~~~~
    & $a$~~~~~
    & $b$~~~~~~
 \\
    \hline
$F^{DK}_1$
    & $0.78$
    & $1.57$
    & 1.05
    & 0.23
&$F^{DK}_0$
    & 0.78
    & 0.99
    & 0.38
    & $0.00$ \\
$V^{DK^*}$
    & $0.94$
    & $1.33$
    & 1.17
    & 0.42
&$A^{DK^*}_0$
    & 0.69
    & 0.92
    & 1.04
    & 0.44
    \\
$A^{DK^*}_1$
    & 0.65
    & 0.75
    & 0.50
    & 0.02
&$A^{DK^*}_2$
    & $0.57$
    & $0.75$
    & 0.94
    & 0.27
    \\
$F^{DK^*_0}_1$
    & $0.48$
    & $0.51$
    & 1.01
    & 0.24
&$F^{DK^*_0}_0$
    & 0.48
    & 0.50
    & $-0.11$
    & 0.02 \\
$A^{DK_{^1\!P_1}}$
    & $0.10$
    & $0.11$
    & $1.03$
    & 0.48
&$V^{DK_{^1\!P_1}}_0$
    & $0.44$
    & $0.47$
    & $0.80$
    & $0.27$
    \\
$V^{DK_{^1\!P_1}}_1$
    & $1.53$
    & $1.58$
    & $0.39$
    & 0.05
&$V^{DK_{^1\!P_1}}_2$
    & $-0.09^*$
    & $-0.09^*$
    & $-0.16^*$
    & $0.51^*$
    \\
$A^{DK_{^3\!P_1}}$
    & $0.98$
    & $1.05$
    & 0.92
    & 0.17
&$V^{DK_{^3\!P_1}}_0$
    & $0.34$
    & $0.38$
    &  1.44
    &  0.15
    \\
$V^{DK_{^3\!P_1}}_1$
    & $2.02$
    & $2.02$
    & $-0.01$
    & 0.03
&$V^{DK_{^3\!P_1}}_2$
    & $0.03$
    & $0.03$
    & $-0.18$
    & $0.10$
    \\
$h$
    & 0.192
    & 0.205
    & 1.17
    & 0.99
& $k$
    & 0.368
    & 0.367
    & $-0.04$
    & 0.11
    \\
$b_+$
    & $-0.096$
    & $-0.102$
    & 1.05
    & 0.58
& $b_-$
    & 0.137
    & 0.147
    & 1.17
    & 0.69 \\
\end{tabular}
\end{ruledtabular}
\end{table}

\begin{figure}[t!]
\centerline{
            {\epsfxsize3 in \epsffile{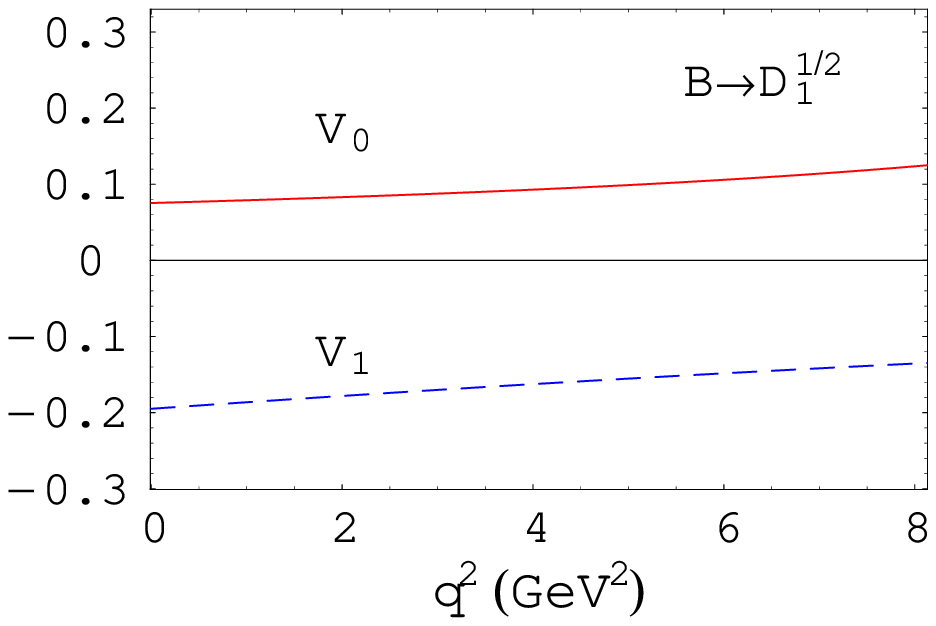}}
            {\epsfxsize3 in \epsffile{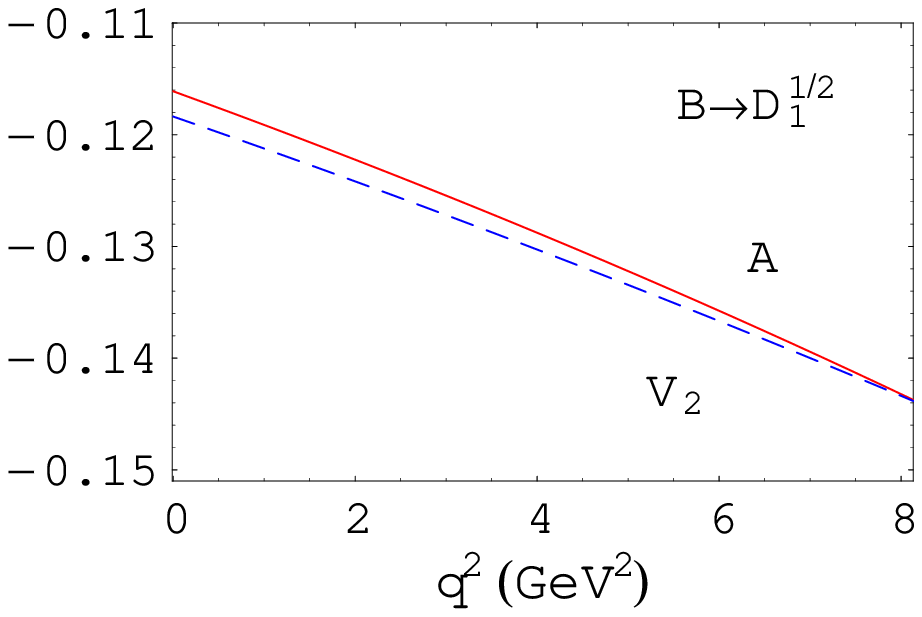}}}
\smallskip
\centerline{
            {\epsfxsize3 in \epsffile{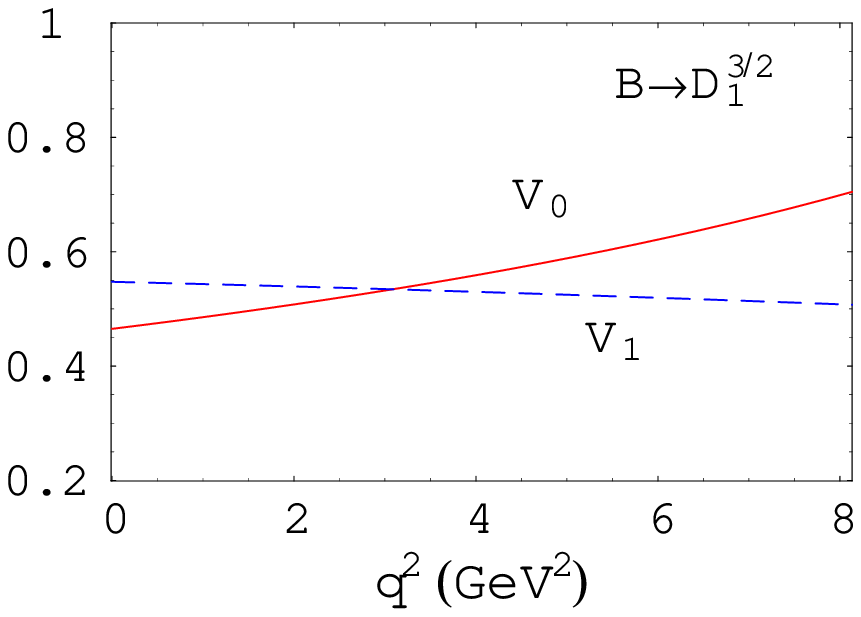}}
            {\epsfxsize3 in \epsffile{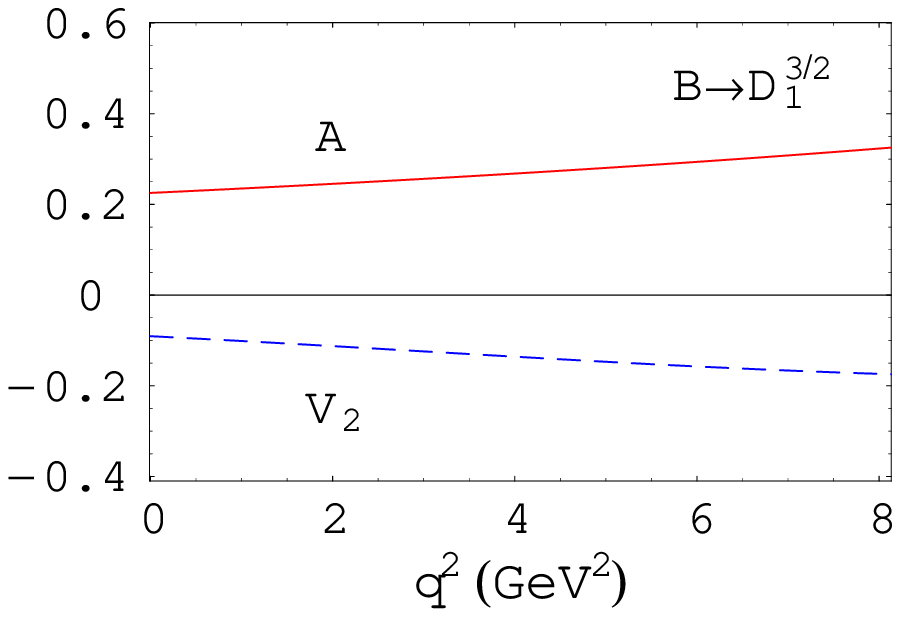}}}
\caption{Form factors $A(q^2)$, $V_0(q^2)$, $V_1(q^2)$ and
$V_2(q^2)$ for $B\to D_1^{1/2,3/2}$ transitions.} \label{fig:BD1}
\end{figure}

\begin{table}[b]
\caption{Same as Table \ref{tab:LFDtopi} except for
$B\to\pi,\rho,a_0(1450),a_1(1260),b_1(1235),a_2(1320)$
transitions.} \label{tab:LFBtopi}
\begin{ruledtabular}
\begin{tabular}{| c c c c c || c c c c c |}
~~~$F$~~~~~
    & $F(0)$~~~~~
    & $F(q^2_{\rm max})$~~~~
    &$a$~~~~~
    & $b$~~~~~~
& ~~~ $F$~~~~~
    & $F(0)$~~~~~
    & $F(q^2_{\rm max})$~~~~~
    & $a$~~~~~
    & $b$~~~~~~
 \\
    \hline
$F^{B\pi}_1$
    & $0.25$
    & $1.16$
    & 1.73
    & 0.95
& $F^{B\pi}_0$
    & 0.25
    & 0.86
    & 0.84
    & $0.10$
    \\
$V^{B\rho}$
    & $0.27$
    & $0.79$
    & 1.84
    & 1.28
&$A^{B\rho}_0$
    & 0.28
    & 0.76
    & 1.73
    & 1.20
    \\
$A^{B\rho}_1$
    & 0.22
    & 0.53
    & 0.95
    & 0.21
&$A^{B\rho}_2$
    & $0.20$
    & $0.57$
    & 1.65
    & 1.05
 \\
$F^{Ba_0}_1$
    & $0.26$
    & $0.68$
    & 1.57
    & 0.70
&$F^{Ba_0}_0$
    & 0.26
    & 0.35
    & $0.55$
    & 0.03 \\
$A^{Ba_1}$
    & $0.25$
    & $0.76$
    & 1.51
    & 0.64
&$V^{Ba_1}_0$
    & $0.13$
    & $0.32$
    &  1.71
    &  1.23
    \\
$V^{Ba_1}_1$
    & $0.37$
    & $0.42$
    & $0.29$
    & 0.14
&$V^{Ba_1}_2$
    & $0.18$
    & $0.36$
    & $1.14$
    & $0.49$
    \\
$A^{Bb_1}$
    & $0.10$
    & $0.23$
    & 1.92
    & 1.62
&$V^{Bb_1}_0$
    & $0.39$
    & $0.98$
    & $1.41$
    & 0.66
    \\
$V^{Bb_1}_1$
    & $0.18$
    & $0.36$
    & $1.03$
    & 0.32
&$V^{Bb_1}_2$
    & $-0.03^*$
    & $-0.15^*$
    & $2.13^*$
    & $2.39^*$
    \\
$h$
    & 0.008
    & 0.015
    & 2.20
    & 2.30
& $k$
    & $0.031$
    & $0.010$
    & $-2.47$
    & 2.47 \\
$b_+$
    & $-0.005$
    & $-0.011$
    & $1.95$
    & 1.80
& $b_-$
    & 0.0016
    & 0.0011
    & $-0.23$
    & 1.18 \\
\end{tabular}
\end{ruledtabular}
\end{table}

\begin{figure}[t!]
\centerline{
            {\epsfxsize3.15 in \epsffile{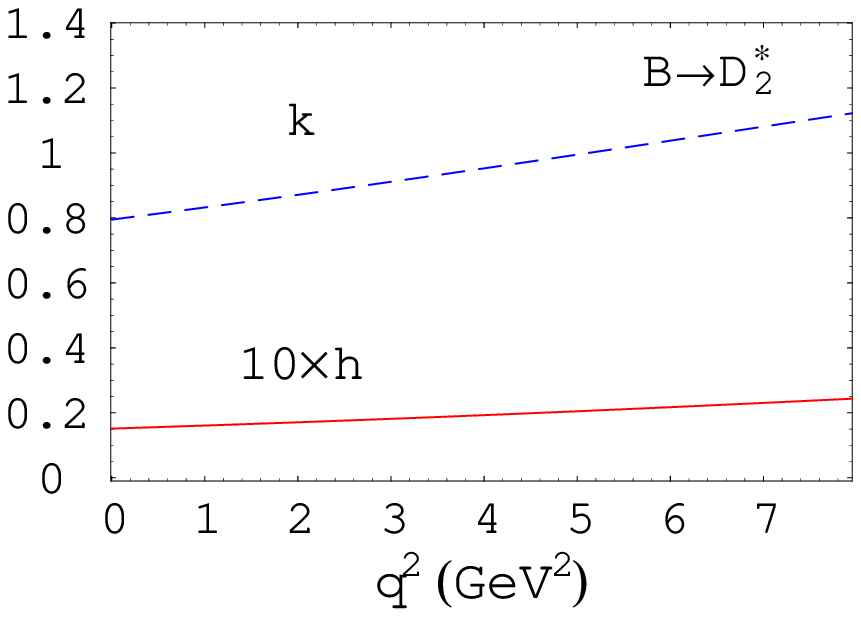}}
            {\epsfxsize3.15 in \epsffile{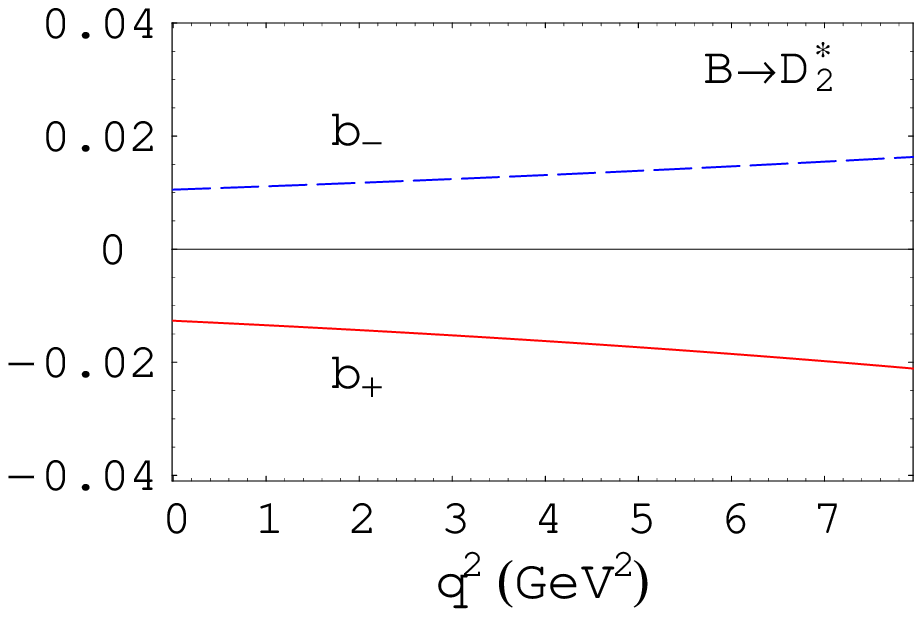}}}
%
\caption{Form factors $k(q^2)$, $h(q^2)$, $b_+(q^2)$ and
$b_-(q^2)$ for $B\to D_2^*$ transitions. Except for the
dimensionless $k(q^2)$, all other form factors are in units of
GeV$^{-2}$.} \label{fig:BD2} 
\end{figure}

\begin{table}[t]
\caption{Same as Table \ref{tab:LFDtopi} except for $B\to
K,K^*,K^*_0(1430),K_{^1\!P_1},K_{^3\!P_1},K_2^*(1430)$
transitions.} \label{tab:LFBtoK}
\begin{ruledtabular}
\begin{tabular}{| c c c c c || c c c c c |}
~~~$F$~~~~~
    & $F(0)$~~~~~
    & $F(q^2_{\rm max})$~~~~
    &$a$~~~~~
    & $b$~~~~~~
& ~~~ $F$~~~~~
    & $F(0)$~~~~~
    & $F(q^2_{\rm max})$~~~~~
    & $a$~~~~~
    & $b$~~~~~~
 \\
    \hline
$F^{BK}_1$
    & $0.35$
    & $2.17$
    & 1.58
    & 0.68
& $F^{BK}_0$
    & 0.35
    & 0.80
    & 0.71
    & $0.04$
    \\
$V^{BK^*}$
    & $0.31$
    & $0.96$
    & 1.79
    & 1.18
&$A^{BK^*}_0$
    & 0.31
    & 0.87
    & 1.68
    & 1.08
    \\
$A^{BK^*}_1$
    & 0.26
    & 0.58
    & 0.93
    & 0.19
&$A^{BK^*}_2$
    & $0.24$
    & $0.70$
    & 1.63
    & 0.98
    \\
$F^{BK^*_0}_1$
    & $0.26$
    & $0.70$
    & 1.52
    & 0.64
&$F^{BK^*_0}_0$
    & 0.26
    & 0.33
    & 0.44
    & 0.05
    \\
$A^{BK_{^3\!P_1}}$
    & $0.26$
    & $0.69$
    & 1.47
    & 0.59
&$V^{BK_{^3\!P_1}}_0$
    & $0.14$
    & $0.31$
    &  1.62
    &  1.14
    \\
$V^{BK_{^3\!P_1}}_1$
    & $0.39$
    & $0.42$
    & $0.21$
    & 0.16
&$V^{BK_{^3\!P_1}}_2$
    & $0.17$
    & $0.30$
    & $1.02$
    & $0.45$
    \\
$A^{BK_{^1\!P_1}}$
    & $0.11$
    & $0.25$
    & 1.88
    & 1.53
&$V^{BK_{^1\!P_1}}_0$
    & $0.41$
    & $0.99$
    & $1.40$
    & $0.64$
    \\
$V^{BK_{^1\!P_1}}_1$
    & $0.19$
    & $0.35$
    & $0.96$
    & 0.30
&$V^{BK_{^1\!P_1}}_2$
    & $-0.05^*$
    & $-0.16^*$
    & $1.78^*$
    & $2.12^*$
    \\
$h$
    & 0.008
    & 0.018
    & 2.17
    & 2.22
& $k$
    & 0.015
    & 0.004
    & $-3.70$
    & 1.78 \\
$b_+$
    & $-0.006$
    & $-0.013$
    & 1.96
    & 1.79
& $b_-$
    & 0.002
    & 0.002
    & 0.38
    & 0.92 \\
\end{tabular}
\end{ruledtabular}
\end{table}

To proceed we find that except for the form factor $V_2$ to be
discussed below, the momentum dependence of form factors in the
spacelike region can be well parameterized and reproduced in the
three-parameter form:
 \be \label{eq:FFpara}
 F(q^2)=\,{F(0)\over 1-a(q^2/m_{B(D)}^2)+b(q^2/m_{B(D)}^2)^2}\,,
 \en
 for $B(D)\to M$ transitions.
The parameters $a$, $b$ and $F(0)$ are first determined in the
spacelike region. We then employ this parametrization to determine
the physical form factors at $q^2\geq 0$. In practice, the
parameters $a,b$ and $F(0)$ are obtained by performing a
3-parameter fit to the form factors in the range $-20\,{\rm
GeV}^2\leq q^2\leq 0$ for $B$ decays and $-10\,{\rm GeV}^2\leq
q^2\leq 0$ for $D$ decays. These parameters are generally
insensitive to the $q^2$ range to be fitted except for the form
factor $V_2(q^2)$ in $B(D)\to\, ^1P_1,P_1^{3/2}$ transitions. The
corresponding parameters $a$ and $b$ are rather sensitive to the
chosen range for $q^2$. This sensitivity is attributed to the fact
that the form factor $V_2(q^2)$ approaches to zero at very large
$-|q^2|$ where the three-parameter parametrization
(\ref{eq:FFpara}) becomes questionable. To overcome this
difficulty, we will fit this form factor to the form
 \be \label{eq:FFpara1}
 F(q^2)=\,{F(0)\over (1-q^2/m_{B(D)}^2)[1-a(q^2/m_{B(D)}^2)+b(q^2/m_{B(D)}^2)^2]}
 \en
and achieve a substantial improvement. For example, we have
$a=2.18$ and $b=6.08$ when $V_2^{BK_{^1P_1}}$ is fitted to Eq.
(\ref{eq:FFpara}) and they become $a=1.78$ and $b=2.12$ (see Table
\ref{tab:LFBtoK}) when the fit formula Eq. (\ref{eq:FFpara1}) is
employed.

In Tables~\ref{tab:LFDtopi}$-$\ref{tab:LFBtoD}
we show the form factors and their $q^2$ dependence for the
transitions
$B(D)\to\pi,\rho,a_0(1450),a_1(1260),b_1(1235),a_2(1320)$,
$B(D)\to K,K^*,K^*_0(1430),K_{^1\!P_1},K_{^3\!P_1},K_2^*(1430)$
and $B\to D,D^*,D_0^*(2308),D_1^{1/2},D_1^{3/2},D_2^*(2460)$. The
$b\to c$ transition form factors are plotted in
Figs.~\ref{fig:BD}$-$\ref{fig:BD2}.
Because the quark contents of the $f_0,f_1,f_2$ mesons lying in
the mass region of $1.3-1.7$ GeV are not well established, we will
not consider them in this work. In calculations, we have taken the
meson masses from \cite{PDG}. The masses of $D_0^*$ and $D_1$ have
been measured recently by Belle to be $2308\pm17\pm15\pm28$ MeV
and $2427\pm26\pm20\pm15$ MeV, respectively \cite{BelleD}. Since
$D_1(2427)$ and $D'_1(2420)$ are almost degenerate, we shall take
$m_{D^{1/2}_1}\approx m_{D^{3/2}_1}\approx 2427$ MeV.

\begin{table}[t]
\caption{Same as Table \ref{tab:LFDtopi} except for $B\to
D,\,D^*,D^*_0,D_1^{1/2},D_1^{3/2},D_2^*$ transitions. For the
purpose of comparing with heavy quark symmetry, the form factors
$u_\pm,c_\pm,\ell,q$ are also shown.} \label{tab:LFBtoD}
\begin{ruledtabular}
\begin{tabular}{| c c c c c || c c c c c |}
~~~$F$~~~~~
    & $F(0)$~~~~~
    & $F(q^2_{\rm max})$~~~~
    &$a$~~~~~
    & $b$~~~~~~
& ~~~ $F$~~~~~
    & $F(0)$~~~~~
    & $F(q^2_{\rm max})$~~~~~
    & $a$~~~~~
    & $b$~~~~~~
 \\
    \hline
$F^{BD}_1$
    & $0.67$
    & $1.22$
    & 1.25
    & 0.39
& $F^{BD}_0$
    & 0.67
    & 0.92
    & 0.65
    & $0.00$ \\
$V^{BD^*}$
    & $0.75$
    & $1.32$
    & 1.29
    & 0.45
&$A^{BD^*}_0$
    & 0.64
    & 1.17
    & 1.30
    & 0.31 \\
$A^{BD^*}_1$
    & 0.63
    & 0.83
    & 0.65
    & 0.02
&$A^{BD^*}_2$
    & $0.61$
    & $0.95$
    & 1.14
    & 0.52
    \\
$F^{BD^*_0}_1$
    & $0.24$
    & $0.34$
    & 1.03
    & 0.27
& $F^{BD^*_0}_0$
    & 0.24
    & 0.20
    & $-0.49$
    & 0.35 \\
$A^{BD^{1/2}_1}$
    & $-0.12$
    & $-0.14$
    & 0.71
    & 0.18
&$V^{BD^{1/2}_1}_0$
    & 0.08
    & 0.13
    & 1.28
    & $-0.29$
    \\
$V^{BD^{1/2}_1}_1$
    & $-0.19$
    & $-0.13$
    & $-1.25$
    & 0.97
& $V^{BD^{1/2}_1}_2$
    & $-0.12$
    & $-0.14$
    & 0.67
    & 0.20
   \\
$A^{BD^{3/2}_1}$
    & $0.23$
    & $0.33$
    & 1.17
    & 0.39
&$V^{BD^{3/2}_1}_0$
    & $0.47$
    & $0.70$
    &  1.17
    &  0.03
    \\
$V^{BD^{3/2}_1}_1$
    & $0.55$
    & $0.51$
    & $-0.19$
    & 0.27
&$V^{BD^{3/2}_1}_2$
    & $-0.09^*$
    & $-0.17^*$
    & $2.14^*$
    & $4.21^*$
    \\
    \hline
$u_+$
    & $-0.24$
    & $-0.34$
    & 1.03
    & 0.27
& $u_-$
    & 0.31
    & 0.42
    & 0.86
    & 0.20 \\
$\ell_{1/2}$
    & 0.56
    & $0.38$
    & $-1.25$
    & 0.97
& $q_{1/2}$
    & 0.041
    & 0.050
    & 0.71
    & 0.18 \\
$c_+^{1/2}$
    & $-0.042$
    & $-0.050$
    & 0.67
    & 0.20
& $c_-^{1/2}$
    & 0.045
    & 0.055
    & 0.71
    & 0.20 \\
$\ell_{3/2}$
    & $-1.56$
    & $-1.45$
    & $-0.19$
    & 0.27
& $q_{3/2}$
    & $-0.079$
    & $-0.114$
    & 1.17
    & 0.39 \\
$c_+^{3/2}$
    & $-0.032^*$
    & $-0.061^*$
    & $2.14^*$
    & $4.21^*$
& $c_-^{3/2}$
    & $-0.027$
    & $-0.026$
    &  0.03
    &  0.45 \\
$h$
    & 0.015
    & 0.024
    & 1.67
    & 1.20
& $k$
    & 0.79
    & 1.12
    & 1.29
    & 0.93
    \\
$b_+$
    & $-0.013$
    & $-0.021$
    & 1.68
    & 0.98
& $b_-$
    & 0.011
    & 0.016
    & 1.50
    & 0.91 \\
\end{tabular}
\end{ruledtabular}
\end{table}

Several remarks are in order:
 \begin{enumerate}
 \item
Many form factors contain terms like $(p'_\bot\cdot q_\bot)/q^2$
in their integrands. At first sight, it appears that linear
$p'_\bot$ terms will not make contributions after integrating over
$p'_\bot$. But this is not the case. As noted before, a Taylor
expansion of the $h^\pp_M/\hat N_1^\pp$ term with respect to
$p^{\pp2}_\bot$ will generate a term proportional to $p'_\bot\cdot
q_\bot$ [cf. Eq. (\ref{eq:Taylor})]. When combined with the
$(p'_\bot\cdot q_\bot)/q^2$ term in the integrand of transition
form factors, this leads to
 \be \label{eq:ident}
 \int d^2p'_\bot\frac{(p'_\bot\cdot q_\bot)^2}{q^2}=-{1\over 2}\int
 d^2p'_\bot\,p'^2_\bot
 \en
in the $q^+=0$ frame. In analytic studies, however, it is more
convenient to utilize the identity obtained in \cite{Jaus99}
 \be \label{eq:identity}
\int dx_2\, d^2p^\prime_\bot \frac{h^\prime_P h^\pp_M}{x_2 \hat
N_1^\prime \hat N^\pp_1}2B_1^{(2)}=\int dx_2\, d^2p^\prime_\bot
\frac{h^\prime_P h^\pp_M}{x_2 \hat N_1^\prime \hat
N^\pp_1}\left(x_1Z_2-2A_1^{(2)}\right)=0.
 \en
Using the expressions of $Z_2$ and $A_1^{(2)}$ given in Eq.
(\ref{eq:Aij}), it is easily seen that the $(p'_\bot\cdot
q_\bot)/q^2$ term under integration can be related to other
$q$-independent quantities. The above identity allows us to
integrate out the $p'_\bot\cdot q_\bot$ term without explicitly
performing the Taylor expansion of $h^\pp_M/\hat N_1^\pp$. Instead
of using Eq. (\ref{eq:ident}) or (\ref{eq:identity}) we have taken
into account such effects in numerical calculations by
substituting the relation $p^\pp_\bot=p'_\bot-x_2q_\bot$ into
$h^\pp_M/\hat N^\pp_1$.
 \item
Owing to the less energy release, form factors for
$D\to\pi,\rho,\cdots$ and $D\to K,K^*,K^{**}$ transitions are more
sensitive to the masses of charmed and light quarks. For this we
can utilize the form-factor ratios $r_V\equiv
V^{PV}(0)/A_1^{PV}(0)$ and $r_2\equiv A_2^{PV}(0)/A_1^{PV}(0)$
measured in $D\to V\ell\bar\nu$ decays to constrain the quark
masses. The most recent and most precise measurement of $D^+\to
\ov K^{*0}\ell^+\bar\nu$ by FOCUS yields \cite{FOCUS}
 \be
 r_V(D\to K^*)=1.504\pm0.057\pm0.039\,,
 \qquad r_2(D\to K^*)=0.875\pm0.049\pm0.064\,.
 \en
The best quark masses $m_u,m_s$ and $m_c$ obtained in this manner
are listed in Eq. (\ref{eq:quarkmass}). Using this set of quark
masses and fixing $\beta_\phi=0.3073$~GeV from $f_\phi=237$~MeV we
have also computed the similar form factor ratios for
$D_s^+\to\phi$ transition and found
 \be
 r_V(D_s\to\phi)=1.569\,, \qquad r_2(D_s\to\phi)=0.865\,,
 \en
in good agreement with the very recent FOCUS measurement of the
$D_s^+\to\phi\mu^+\nu$ form factor ratios \cite{FOCUSDs}
 \be
 r_V(D_s\to\phi)=1.549\pm0.250\pm0.145\,, \qquad
 r_2(D_s\to\phi)=0.713\pm0.202\pm0.266\,.
 \en
 \item
In the absence of any information for the parameter $\beta_T$
appearing in the wave function of tensor mesons, we have taken
$\beta_T$ to be the same as the $\beta$ parameter of the $p$-wave
meson with the same quark content, for example,
$\beta(D_2^*)=\beta(D_0^*)=0.331$. Note that among the four $P\to
T$ transition form factors, the one $k(q^2)$ is particularly
sensitive to $\beta_T$. It is not clear to us if the complicated
analytic expression for $k(q^2)$ in Eq. (\ref{eq:PtoT}) is not
complete. To overcome this difficulty, we apply the heavy quark
symmetry relations in Eq. (\ref{eq:HQS2}) below to obtain $k(q^2)$
for $B\to D_2^*$ transition
 \be
 k(q^2)=\,m_Bm_{D_2^*}\left(1+{m_B^2+m_{D_2^*}^2-q^2\over
 2m_Bm_{D_2^*}}\right)\left[ h(q^2)-{1\over 2}b_+(q^2)+{1\over 2}b_-(q^2)\right].
 \en
This can be tested in $B^-\to D_2^{*0}\pi^-$ decays to be
discussed below in Sec. III.E.

 \item
For heavy-to-heavy transitions such as $B\to D,D^*,D^{**}$, the
sign of various form factors can be checked by heavy quark
symmetry. In the heavy quark limit, heavy quark symmetry requires
that the form factors $u_-,\ell_{1/2},q_{1/2},c_-^{1/2},h,k$ and
$b_-$ be positive, while $u_+,\ell_{3/2}, q_{3/2},c_+^{1/2},
c_+^{3/2},c_-^{3/2}$ and $b_+$ be negative [see Eqs.
(\ref{eq:HQS1})$-$(\ref{eq:HQS3})]. Our results are indeed in
accordance with HQS (see Table \ref{tab:LFBtoD}).
 \item
For $P\to A$ transitions, the form factor $V_0$ is always
positive, while the sign of other form factors $A,V_1,V_2$ depends
on the process under consideration, for example, they are all
positive in $B(D)\to a_1,K_{^3\!P_1}$ transitions and negative in
$B\to D_1$ transitions.
 \item
The form factors of $B$ to light axial-vector meson transitions
obey the relations $V_0^{Ba_1}<A_0^{B\rho}<V_0^{Bb_1}$ and
$V_0^{BK_{^3\!P_1}}<A_0^{BK^*}<V_0^{BK_{^1\!P_1}}$.
 \item
It is pointed out in \cite{Cheng97} that for $B\to D,D^*$
transitions, the form factors $F_1,A_0,A_2,V$ exhibit a dipole
behavior, while $F_0$ and $A_1$ show a monopole dependence.
According to the three-parameter parametrization
(\ref{eq:FFpara}), the dipole behavior corresponds to $b=(a/2)^2$,
while $b=0$ and $a\neq 0$ induces a monopole dependence. An
inspection of Tables \ref{tab:LFDtopi}$-$\ref{tab:LFBtoD}
indicates that form factors $F^{BD}_0$, $A_1^{BD^*}$, $F_0^{BK}$,
$F_0^{Ba_0}$ and $F_0^{DK(\pi)}$ have a monopole behavior, while
$F_1^{BD}$, $V^{BD^*}$, $A^{BD_1^{3/2}}$, $F_1^{B(D)K}$,
$A^{B(D)K_{^3\!P_1}}$, $V_0^{BK_{^1\!P_1}}$, $F_1^{B(D)a_0}$ and
$F_1^{DK^*_0}$ have a dipole dependence.
 \item
In the heavy quark limit,
$F_1^{BD_0^*}(q^2)=F_0^{BD_0^*}/[1-q^2/(m_B-m_{D_0^*})^2]$, while
$F_1^{BD}(q^2)=F_0^{BD}/[1-q^2/(m_B+m_D)^2]$ (see Eqs.
(\ref{eq:HQSxi}) and (\ref{eq:HQStau1half})). This explains why
$F_1$ and $F_0$ in the $B\to D_0^*$ transition deviate at large
$q^2$ faster than that in the $B\to D$ case (Fig. \ref{fig:BD}).
 \item
Unlike the form factor $F_0$ in $P\to a_0,K_0^*$ transitions which
is almost flat in its $q^2$ behavior, $F_0^{BD_0^*}$ is decreasing
with $q^2$ as it must approach to zero at the maximum $q^2$ when
$m_Q\to\infty$ [see Eq. (\ref{eq:HQStau1half})]. In general, form
factors for $P\to S$ transitions increase slowly with $q^2$
compared to that for $P\to P$ ones. For example, $F^{Ba_0}(0)\sim
F^{B\pi}(0)$ at $q^2=0$, while at zero recoil $F^{Ba_0}(q^2_{\rm
max})\ll F^{B\pi}(q^2_{\rm max})$. Note that the form factors of
$B\to a_0$ or $B\to K_0^*$ are similar to that of $B\to \pi$ or
$B\to K$ at $q^2=0$, while $F_{1,0}^{BD_0^*}(0)\ll
F_{1,0}^{BD}(0)$. This can be understood from the fact that $P\to
S$ form factors are the same as $P\to P$ ones except for the
replacement of $m_1^\pp\to -m_1^\pp$ and $h_P^\pp\to -h_S^\pp$
[see Eq. (\ref{eq:upm})].  Consequently, the ${\cal A}^\pp$ term
in Eq. (\ref{eq:A}) is subject to more suppression in
heavy-to-heavy transitions than in heavy-to-light ones. We shall
see in Sec. III.E that the suppression of the $B\to D^*_0$ form
factor relative to the $B\to D$ one is supported by experiment.
 \item
To determine the physical form factors for $B(D)\to
K_1(1270),K_1(1400)$, $B\to
D_1(2427),D_1(2420),D_{s1}(2460),D_{s1}(2536)$ transitions, one
needs to know the mixing angles of $^1P_1-^3\!P_1$ [see Eq.
(\ref{eq:K1mixing})] and $D_1^{1/2}-D^{3/2}_1$. As noted in
passing, the mixing angle for $K_1$ systems is about $-58^\circ$
as implied from the study of $D\to K_1(1270)\pi.~K_1(1400)\pi$
decays \cite{Cheng:2003bn}. A mixing angle
$\theta_{D_1}=(5.7\pm2.4)^\circ$ is obtained by Belle through a
detailed $B\to D^*\pi\pi$ analysis \cite{BelleD}, while
$\theta_{D_{s1}}\approx 7^\circ$ is determined from the quark
potential model \cite{Cheng:2003id} as the present upper limits on
the widths of $D_{s1}(2460)$ and $D'_{s1}(2536)$ do not provide
any  constraints on the $D_{s1}^{1/2}\!-\!D_{s1}^{3/2}$ mixing
angle.

\end{enumerate}

\subsection{Comparison with other model calculations}

\begin{table}[t]
\caption{Form factors of $D\to \pi,\rho,K,K^*$ transitions at
$q^2=0$ in various models.} \label{tab:Dtopi}
\begin{ruledtabular}
\begin{tabular}{| c | c c c c c || c c c c c | }
 Model & $F_{1,0}^{D\pi}(0)$ & $A_0^{D\rho}(0)$ & $A_1^{D\rho}(0)$
 & $A_2^{D\rho}(0)$ & $V^{D\rho}(0)$ & $F_{1,0}^{DK}(0)$ & $A_0^{DK^*}(0)$ & $A_1^{DK^*}(0)$
 & $A_2^{DK^*}(0)$ & $V^{DK^*}(0)$
 \\ \hline
 This work & 0.67 & 0.64 & 0.58 & 0.48 & 0.86 & 0.78 & 0.69 & 0.65
 & 0.57 & 0.94 \\
 MS \cite{Melikhov} & 0.69 & 0.66 & 0.59 & 0.49 & 0.90 & 0.78 &
 0.76 & 0.66 & 0.49 & 1.03 \\
 QSR \cite{Ball91} &0.5 &0.6 &0.5 &0.4 &1.0 &0.6 &0.4 &0.5 &0.6 &1.1 \\
 BSW \cite{BSW} &0.69 &0.67 &0.78 &0.92 &1.23
                &0.76 &0.73 &0.88 &1.15 &1.23   \\
\end{tabular}
\end{ruledtabular}
\end{table}

\begin{table}[b]
\caption{Form factors of $B\to \pi,\rho,K,K^*$ transitions at
$q^2=0$ in various models.} \label{tab:Btopi}
\begin{ruledtabular}
\begin{tabular}{| c | c c c c c || c c c c c | }
 Model & $F_{1,0}^{B\pi}(0)$ & $A_0^{B\rho}(0)$ & $A_1^{B\rho}(0)$
 & $A_2^{B\rho}(0)$ & $V^{B\rho}(0)$ & $F_{1,0}^{BK}(0)$ & $A_0^{BK^*}(0)$ & $A_1^{BK^*}(0)$
 & $A_2^{BK^*}(0)$ & $V^{BK^*}(0)$
 \\ \hline
 This work & 0.25 & 0.28 & 0.22 & 0.20 & 0.27 & 0.35 & 0.31 & 0.26
 & 0.24 & 0.31 \\
 MS \cite{Melikhov} & 0.29 & 0.29 & 0.26 & 0.24 & 0.31 & 0.36 &
 0.45 & 0.36 & 0.32 & 0.44 \\
 LCSR \cite{LCSR} & 0.26 & 0.37 & 0.26 & 0.22 & 0.34 & 0.34 & 0.47
 & 0.34 & 0.28 & 0.46 \\
 BSW \cite{BSW} &0.33 &0.28 &0.28 &0.28 &0.33
                &0.38 &0.32 &0.33 &0.33 &0.37 \\
\end{tabular}
\end{ruledtabular}
\end{table}

\begin{table}[t]
\caption{Form factors of $B\to D,D^*$ transitions at $q^2=0$ in
various models.} \label{tab:BtoD}
\begin{ruledtabular}
\begin{tabular}{| c | c c c c c | }
 Model & $F_{1,0}^{BD}(0)$ & $A_0^{BD^*}(0)$ & $A_1^{BD^*}(0)$
 & $A_2^{BD^*}(0)$ & $V^{BD^*}(0)$
 \\ \hline
 This work & 0.67 & 0.64 & 0.63 & 0.62 & 0.75 \\
 MS \cite{Melikhov} & 0.67 & 0.69 & 0.66 & 0.62 & 0.76 \\
 BSW \cite{BSW} & 0.69 &0.62 &0.65 &0.69 &0.71\\
\end{tabular}
\end{ruledtabular}
\end{table}

It is useful to compare our results based on the covariant
light-front model with other theoretical calculations. Except for
the Isgur-Scora-Grinstein-Wise (ISGW) quark model \cite{ISGW}, all
the existing studies on mesonic form factors focus mainly on the
ground-state $s$-wave to $s$-wave transitions. For $P\to P,V$ form
factors we choose the BSW model \cite{BSW}, the Melikhov-Stech
(MS) model \cite{Melikhov}, QCD sum rule (QSR) \cite{Ball91} and
light-cone sum rules (LCSR) \cite{LCSR} for comparison. Shown in
Tables \ref{tab:Dtopi}$-$\ref{tab:BtoD} are $(D,B)\to
\pi,\rho,K,K^*,D,D^*$ transition form factors calculated in
various models. We see that the covariant light-front model
predictions are most close to that of the MS model except for
$B\to K^*$ transitions.

\vspace{0.5cm} \noindent {\it ISGW model} :

\begin{table}[b]
\caption{Form factors of $B\to D^{**}$ transitions calculated in
the ISGW2 model.} \label{tab:ISGW}
\begin{center}
\begin{tabular}{| c c c c c || c c c c c | }
\hline ~~~$F$~~~~~
    & $F(0)$~~~~~
    & $F(q^2_{\rm max})$~~~~
    &$a$~~~~~
    & $b$~~~~~~
    & ~~~ $F$~~~~~
    & $F(0)$~~~~~
    & $F(q^2_{\rm max})$~~~~~
    & $a$~~~~~
    & $b$~~~~~~
 \\ \hline
  $F_1^{BD_0^*}$ & 0.18 & 0.24 & 0.28 & 0.25 & $F_0^{BD_0^*}$ & 0.18 & $-0.008$ & -- & -- \\
 $A^{BD_1^{1/2}}$ & $-0.16$ & $-0.21$ & 0.87 & 0.24 &
  $V_0^{BD_1^{1/2}}$ & $0.18$ & $0.23$ & 0.89 & 0.25 \\
 $V_1^{BD_1^{1/2}}$ & $-0.19$ & $0.006$ & -- & -- & $V_2^{BD_1^{1/2}}$ &
  $-0.18$ & $-0.24$ & 0.87 & 0.24\\
 $A^{BD_1^{3/2}}$ & $0.16$ & $0.19$ & 0.46 & 0.065 &
  $V_0^{BD_1^{3/2}}$ & $0.43$ & $0.51$ & 0.54 & 0.074 \\
 $V_1^{BD_1^{3/2}}$ & $0.40$ & $0.32$ & $-0.60$ & 1.15 & $V_2^{BD_1^{3/2}}$ &
  $-0.12$ & $-0.19$ & 1.45 & 0.83 \\
 \hline
 $u_+$ & $-0.18$ & $-0.24$ & 0.88 & 0.25 & $u_-$ & 0.46 & 0.62 & 0.87 & 0.25 \\
 $\ell_{1/2}$ & 0.54 & $-0.016$ & -- & -- & $q_{1/2}$ & 0.057 & 0.074 & 0.87 & 0.24 \\
 $c_+^{1/2}$ & $-0.064$ & $-0.083$ & 0.87 & 0.24 & $c_-^{1/2}$ & 0.068 & 0.088 & 0.87
 & 0.24 \\
 $\ell_{3/2}$ & $-1.15$ & $-0.90$ & $-0.60$ & 1.15 & $q_{3/2}$ & $-0.057$
 & $-0.066$ & 0.46 & 0.065 \\
 $c_+^{3/2}$ & $-0.043$ & $-0.066$ & 1.45 & 0.83 & $c_-^{3/2}$ & $-0.018$
 & $-0.013$ &  0.23 & 5.38 \\
 $h$ & 0.011 & 0.014 & 0.86 & 0.23 & $k$ & 0.60 & 0.68 & 0.40 & 0.68 \\
 $b_+$ & $-0.010$ & $-0.013$ & 0.86 & 0.23 & $b_-$ & 0.010 & 0.013 & 0.86 & 0.23 \\
 \hline
\end{tabular}
\end{center}
\end{table}

 Before our work, the ISGW quark model \cite{ISGW} is the only model
that can provide a systematical estimate of the transition of a
ground-state $s$-wave meson to a low-lying $p$-wave meson.
However, this model is based on the non-relativistic constituent
quark picture. In general, the form factors evaluated in the
original version of the ISGW model are reliable only at
$q^2=q^2_m$, the maximum momentum transfer. The reason is that the
form-factor $q^2$ dependence in the ISGW model is proportional to
exp[$-(q^2_m-q^2)$] and hence the form factor decreases
exponentially as a function of $(q^2_m-q^2)$. This has been
improved in the ISGW2 model \cite{ISGW2} in which the form factor
has a more realistic behavior at large $(q^2_m-q^2)$ which is
expressed in terms of a certain polynomial term. In addition to
the form-factor momentum dependence, the ISGW2 model incorporates
a number of improvements, such as the constraints imposed by heavy
quark symmetry, hyperfine distortions of wave functions, $\cdots$,
etc. \cite{ISGW2}.

The ISGW2 model predictions for $B\to D^{**}$ transition form
factors are shown in Table \ref{tab:ISGW}. Note that form factors
$F_0^{BD_0^*}(q^2)$, $V_1^{BD_1^{1/2}}(q^2)$ (or
$\ell_{1/2}(q^2)$) cannot be parameterized in the form of
(\ref{eq:FFpara}) or (\ref{eq:FFpara1}) since they vanish at
certain $q^2$, e.g. around $q^2\approx 8\,{\rm GeV}^2$ for
$V_1^{BD_1^{1/2}}(q^2)$. We see from the comparison of
Table~\ref{tab:ISGW} with Table~\ref{tab:LFBtoD} that (i) the form
factors at small $q^2$ obtained in the covariant light-front and
ISGW2 models agree within 40$\%$, and (ii) as $q^2$ increases,
$F_1^{BD_0^*}(q^2)$, $A^{BD_1^{3/2}}(q^2)$,
$V_0^{BD_1^{3/2}}(q^2)$, $h(q^2)$, $|b_+(q^2)|$ and $b_-(q^2)$
increase more rapidly in the LF model than those in the ISGW2
model, whereas $F_0^{BD_0^*}(q^2)$ and $|V_1^{BD_1^{1/2}}(q^2)|$
decrease more sharply in the latter model so that they even flip a
sign near the zero recoil point.

The fact that both LF and ISGW2 models have similar $B\to D^{**}$
form factors at small $q^2$ implies that relativistic effects
could be mild in $B\to D^{**}$ transitions. Nevertheless,
relativistic effects may manifest in heavy-to-light transitions,
especially at the maximum recoil. An example is provided shortly
below.

\vspace{0.5cm}\noindent{\it Others} :

\begin{table}[b]
\caption{$B\to a_1(1260)$ transition form factors at $q^2=0$ in
various models. The results of CQM and QSR have been rescaled
according to the form-factor definition in Eq.
(\ref{eq:ffpdimless}).} \label{tab:Ba1}
\begin{center}
\begin{tabular}{| c | c c c c | }
\hline
 ~~~~~~Model ~~~~~~ & ~~~$A^{Ba_1}(0)~~~$ & ~~~$V_0^{Ba_1}(0)$~~~ & ~~~$V_1^{Ba_1}(0)$~~~ &
 ~~~$V_2^{Ba_1}(0)$~~~
 \\ \hline
  This work & $0.25$ & $0.13$ & $0.37$ & $0.18$ \\
  ISGW2 \cite{ISGW2} & $0.21$ & $1.01$ & $0.54$ & $-0.05$ \\
 CQM \cite{Deandrea} & 0.09 & 1.20 & 1.32 & 0.34 \\
 QSR \cite{Aliev} & $-0.41\pm0.06$ & $-0.23\pm0.05$ & $-0.68\pm0.08$ & $-0.33\pm0.03$
 \\
 \hline
\end{tabular}
\end{center}
\end{table}

Based on the light-cone sum rules, Chernyak \cite{Chernyak} has
estimated the $B\to a_0(1450)$ transition form factor and obtained
$F_{1,0}^{Ba_0}(0)=0.46$, while our result is 0.26 and is similar
to the $B\to\pi$ form factor at $q^2=0$. For $B\to a_1(1260)$ form
factors, there are two existing calculations: one in a quark-meson
model (CQM) \cite{Deandrea} and the other based on the QCD sum
rule (QSR) \cite{Aliev}. The results are quite different, for
example, $V_0^{Ba_1}(0)$ obtained in the quark-meson model,
1.20\,, is larger than the sum-rule prediction, $-0.23\pm 0.05$\,,
by a factor of five apart from a sign difference. Predictions in
various models are summarized in Table \ref{tab:Ba1}. If
$a_1(1260)$ behaves as the scalar partner of the $\rho$ meson, it
is expected that $V_0^{Ba_1}$ is similar to $A_0^{B\rho}$.
Therefore, it appears to us that a magnitude of order unity for
$V_0^{Ba_1}(0)$ as predicted by the ISGW2 model and CQM is very
unlikely. Notice that the sign of the form factors predicted by
QSR is opposite to ours. In hadronic $B\to a_1P$ decays, the
relevant form factors are $V_0^{Ba_1}$ and $F_1^{BP}$ under the
factorization approximation. Presumably, the measurement of $\ov
B^0\to a_1^+\pi^-$ will enable us to test $V_0^{Ba_1}$.

\subsection{Comparison with experiment}
There are several experimentally measured decay modes, namely,
$B\to \ov D D_s^{**}$ and $B^-\to D^{**}\pi^-$ decays, which allow
to test our model calculations of decay constants and form factors
for $p$-wave charmed mesons $D^{**}$ and $D_s^{**}$.

For $\ov B\to D\ov D_s^{**}$ decays, they proceed only via
external $W$-emission and hence can be used to determine the decay
constant of $D_s^{**}$. More precisely, their factorizable
amplitudes are simply given by
 \be \label{eq:ampBtoDDs}
 A(\ov B\to D\ov D_s^{**})={G_F\over\sqrt{2}}V_{cb}V_{cs}^*\,a_1\la
 \ov D_s^{**}|(\bar sc)|0\ra\la D|(\bar c b)|\ov B\ra,
 \en
where $(\bar q_1q_2)\equiv \bar q_1\gamma_\mu(1-\gamma_5)q_2$ and
$a_1$ is a parameter of order unity. The recent Belle measurements
read \cite{BelleDs-1}
 \be \label{eq:dataBtoDDs}
 \B[B\to \ov DD_{s0}^*(2317)]\B[D_{s0}^*(2317)\to D_s\pi^0] &=&
 (8.5^{+2.1}_{-1.9}\pm2.6)\times 10^{-4}, \non \\
 \B[B\to \ov DD_{s1}(2460)]\B[D_{s1}(2460)\to D_s^*\pi^0] &=&
 (17.8^{+4.5}_{-3.9}\pm5.3)\times 10^{-4}.
 \en
The $D_{s0}^*(2317)$ width is dominated by its hadronic decay to
$D_s\pi^0$ as the upper limit on the ratio $\Gamma(D_{s0}^*\to
D_s^*\gamma)/\Gamma(D_{s0}^*\to D_s\pi^0)$ was set to be 0.059
recently by CLEO \cite{CLEO}. Therefore,
$0.94\lsim\B[D_{s0}^*(2317)\to D_s\pi^0]\lsim 1.0$. It follows
from Eqs. (\ref{eq:dataBtoDDs}) and ({\ref{eq:ampBtoDDs}) that
(see \cite{Cheng:2003id} for detail)
 \be \label{eq:fDs0}
 f_{D_{s0}^*}\approx 47-73~{\rm MeV},
 \en
for $a_1=1.07$. To estimate the branching ratios of $D_s^*\pi^0$
in the $D_{s1}(2460)$ decay, we need some experimental and
theoretical inputs. There are two different measurements of the
radiative mode by Belle: a value of $0.38\pm0.11\pm0.04$ for the
ratio $D_s\gamma/D_s^*\pi^0$ is determined from $B\to \ov DD_{s1}$
decays \cite{BelleDs-1}, while the result of $0.55\pm0.13\pm0.08$
is obtained from the charm fragmentation of $e^+e^-\to c\,\bar c$
\cite{BelleDs-2}. These two measurements are consistent with each
other, though the central values are somewhat different. We shall
take the averaged value of $0.44\pm0.09$ for
$D_s\gamma/D_s^*\pi^0$. The ratio $D_s\pi^+\pi^-/D_s^*\pi^0$ is
measured to be $0.14\pm0.04\pm0.02$ by Belle \cite{BelleDs-2}. As
for $D_s^*\gamma/D_s^*\pi^0$, it is found to be less than 0.22,
0.31 and 0.16, respectively, by BaBar \cite{BaBarDs1}, Belle
\cite{BelleDs-2} and CLEO \cite{CLEO}. Theoretically, the $M1$
transition $D_{s1}\to D_{s0}^*\gamma$ turns out to be quite small
\cite{GodfreyDs}. Assuming that the $D_{s1}(2460)$ decay is
saturated by $D_s^*\pi^0,~D_s\gamma,~D_s^*\gamma$ and $D_s\pi\pi$,
we are led to
 \be
 0.53\lsim \B(D_{s1}(2460)\to D_s^*\pi^0)\lsim 0.68\,.
 \en
This in turn implies $\B[B\to \ov DD_{s1}(2460)]=(1.6\sim
4.6)\times 10^{-3}$. As a result, the decay constant of
$D_{s1}(2460)$ is found to be
 \be
 f_{D_{s1}(2460)}\approx 110-190\,{\rm MeV}.
 \en
Our predictions $f_{D_{s0}^*}=71$ MeV and $f_{D_{s1}}=117$ MeV
with the latter being obtained from the relation
   \be
 f_{D_{s1}}=
 f_{D_{s1}^{1/2}}\cos\theta_s+f_{D_{s1}^{3/2}}\sin\theta_s
 \en
with $\theta_s\approx 7^\circ$ inferred from the potential model
\cite{Cheng:2003id}, are in agreement with experiment.

\begin{table}[b]
\caption{The predicted branching ratios for $B^-\to D^{**0}\pi^-$
decays in the covariant light-front (CLF) and ISGW2 models. Since
the decay constants of $p$-wave charmed mesons are not provided in
the latter model, we employ the CLF decay constants and the ISGW2
form factors for the ISGW2 results quoted below. Experimental
results are taken from  Belle \cite{BelleD}, BaBar \cite{BaBarD}
and PDG \cite{PDG}. The axial-vector meson mixing angle is taken
to be $\theta=12^\circ$ \cite{Cheng:2003id} and the parameters
$a_{1,2}$ are given by $a_1=1.07$ and $a_2=0.27$.}
\label{tab:BtoDpi}
\begin{ruledtabular}
\begin{tabular}{| l  c c l |}
~~~Decay~~~ & ~~~This work~~~ & ~~~ISGW2~~~ & ~~~Expt~~~ \\
\hline $B^-\to D^*_0(2308)^0\pi^-$ & $7.3\times 10^{-4}$ &
 $4.8\times 10^{-4}$ & $(9.2\pm2.9)\times 10^{-4}$ \cite{BelleD} \\
 $B^-\to D_1(2427)^0\pi^-$ & $ 4.6\times 10^{-4}$ & $9.4\times 10^{-4}$ &
 $(7.5\pm1.7)\times 10^{-4}$ \cite{BelleD} \\
 $B^-\to D'_1(2420)^0\pi^-$ & $1.1\times 10^{-3}$ & $8.2\times 10^{-4}$
 & $(9.3\pm1.4)\times 10^{-4}$  \cite{BelleD,BaBarD}  \\
 & & & $(1.5\pm0.6)\times 10^{-3}$ \cite{PDG} \\
 $B^-\to D^*_2(2460)^0\pi^-$ & $1.0\times 10^{-3}$ & $5.7\times 10^{-4}$ &
 $(7.4\pm0.8)\times 10^{-4}$ \cite{BelleD,BaBarD}  \\
\end{tabular}
\end{ruledtabular}
\end{table}

Ideally, the neutral $B$ decays $\ov B^0\to D^{**+}\pi^-$ that
receive only color-allowed contributions can be used to extract
$B\to D^{**}$ transition form factors. Unfortunately, such decays
have not yet been measured. Nevertheless, the decays $B^-\to
D^{**0}\pi^-$ that receive both contributions from color-allowed
and color-suppressed diagrams provide a nice ground for testing
the $B\to D^{**}$ form factors and the $D^{**}$ decay constants.
Following \cite{Cheng:2003id}, we show the predicted branching
ratios in Table \ref{tab:BtoDpi}. The experimental results are
taken from Belle \cite{BelleD}, BaBar \cite{BaBarD} and PDG
\cite{PDG}. For $B^-\to D_2^{*0}\pi^-$ we combine the Belle and
BaBar measurements
 \be
 \B(B^-\to D_2^{*0}\pi^-)\B(D_2^{*0}\to D^+\pi^-) &=& (3.1\pm0.4)\times
 10^{-4},
 \non \\
 \B(B^-\to D_2^{*0}\pi^-)\B(D_2^{*0}\to D^{*+}\pi^-) &=& (1.8\pm0.4)\times
 10^{-4},
 \en
to arrive at
 \be
 \B(B^-\to D_2^{*0}\pi^-)\B(D_2^{*0}\to D^+\pi^-,D^{*+}\pi^-)=(4.9\pm
 0.6)\times 10^{-4}.
 \en
Using $\B(D_2^{*0}\to D^+\pi^-,D^{*+}\pi^-)=2/3$ following from
the assumption that the $D_2^{*0}$ width is saturated by $D\pi$
and $D^*\pi$, we are led to $\B(B^-\to D_2^{*0}\pi^-)=(7.4\pm
0.8)\times 10^{-4}$.  We see from the Table \ref{tab:BtoDpi} that
the agreement between theory and experiment is generally good. In
particular, the suppression of the $D_0^{*0}\pi^-$ production
relative to the $D^0\pi^-$ one (the branching ratio for the latter
being $(5.3\pm0.5)\times 10^{-3}$ \cite{PDG}) clearly indicates a
smaller $B\to D_0^*$ form factor relative to $B\to D$ one. For
comparison, we also show the ISGW2 predictions in the same Table.
Since the decay constants of $p$-wave charmed mesons are not
provided in the ISGW2 model, we employ the decay constants in this
work and the ISGW2 form factors to obtain the ISGW2 results quoted
in Table \ref{tab:BtoDpi}. Predictions in the other models are
summarized in \cite{Cheng:2003id}.

Since the tensor meson cannot be produced from the $V-A$ current,
the decay $B^-\to D_2^{*0}\pi^-$ can be used to determine the form
factor combination $\eta(q^2)\equiv
k(q^2)+b_+(q^2)(m_B^2-m_{D_2^*}^2)+b_-(q^2)q^2$ at $q^2=m_\pi^2$.
The measured rate implies that $\eta(m_\pi^2)=0.43\pm0.02$\,, to
be compared with the predictions of 0.52 and 0.38 in the covariant
LF and ISGW2 models, respectively.

It is worth mentioning that the ratio
 \be
 R={\B(B^-\to D_2^*(2460)^0\pi^-)\over \B(B^-\to D'_1(2420)^0\pi^-)}
 \en
is measured to be $0.80\pm0.07\pm0.16$ by BaBar \cite{BaBarD},
$0.77\pm0.15$ by Belle \cite{BelleD} and $1.8\pm0.8$ by
CLEO~\cite{Galik}. The early prediction by Neubert \cite{Neubert}
yields a value of 0.35. The predictions of $R=0.91$ in the
covariant light-front model and 0.67 in the ISGW2 model are in
accordance with the data.

\section{Heavy Quark Limit}

In the heavy quark limit, heavy quark symmetry (HQS) \cite{IW89}
provides model-independent constraints on the decay constants and
form factors. For example, pseudoscalar and vector mesons would
have the same decay constants and all the heavy-to-heavy mesonic
decay form factors are reduced to some universal Isgur-Wise
functions. Therefore, it is important to study the heavy quark
limit behavior of these physical quantities to check the
consistency of calculations. Since the analysis of heavy hadron
structures and their dynamics in the infinite quark mass limit has
been tremendously simplified by heavy quark symmetry and heavy
quark effective theory (HQET) developed from QCD in terms of
$1/m_Q$ expansion \cite{Georgi90}, it would be much simpler to
study the decay constants and form factors directly within the
framework of a covariant light-front model of heavy mesons fully
based on HQS and HQET. Indeed, we have constructed such a model in
\cite{CCHZ} which can be viewed as the heavy quark limit of the
covariant light-front approach discussed in Sec. II. We shall show
explicitly that the decay constants and form factors obtained in
the covariant light-front model and then extended to the heavy
quark limit do agree with those derived directly in the
light-front model based on HQET.

Before proceeding, it is worth making a few remarks: (i) Just as
in the conventional light-front model, it is assumed in
\cite{CCHZ} that the valance quarks of the meson are on their mass
shell in the covariant light-front model based on HQET. However,
this is not in contradiction to the covariant light-front approach
discussed in Sec. II. As stressed before, the antiquark is on its
mass shell after $p^-$ integration in the covariant light-front
calculation. Moreover, the off shellness of the heavy quark
vanishes in the strict heavy quark limit. Therefore, the
calculation based on the light-front model in \cite{CCHZ} is
covariant. (ii) Since the heavy quark-pair creation is forbidden
in the $m_Q\to\infty$ limit, the $Z$-graph is no longer a problem
in the reference frame where $q^+\geq 0$. This allows us to
compute the Isgur-Wise functions directly in the timelike region.

In this work, we will adopt two different approaches to elaborate
on the heavy quark limit behavior of physical quantities: one from
top to bottom and the other from bottom to top. In the
top-to-bottom approach, we will derive the decay constants and
form factors in the covariant light-front model within HQET and
obtain model-independent HQS relations. In the bottom-to-top
approach, we study the heavy quark limit behavior of the decay
constants and transition form factors of heavy mesons obtained in
Secs. II and III and show that they do match the covariant model
results based on HQET.

\subsection {Heavy quark symmetry relations}

In the infinite quark mass limit, the decay constants of heavy
mesons must satisfy the HQS relations given by Eq.
(\ref{eq:HQSf}), while all the heavy-to-heavy mesonic decay form
factors  are reduced to three universal Isgur-Wise (IW) functions,
$\xi$ for $s$-wave to $s$-wave and $\tau_{1/2}$ as well as
$\tau_{3/2}$ for $s$-wave to $p$-wave transitions. Specifically,
$B\to D,D^*$ form factors are related to the IW function
$\xi(\omega)$ by \cite{IW89}
 \be \label{eq:FFxi}
\xi(\omega) &=& {1\over
2\sqrt{m_Bm_D}}\,\left[(m_B+m_D)f_+(q^2)+(m_B-m_D)f_-(q^2)\right]
\non \\
&=& -{1\over\sqrt{m_Bm_{D^*}}}\,{f(q^2)\over
1+\omega}=-2\sqrt{m_Bm_{D^*}}\,g(q^2) \non \\
&=& \sqrt{m_Bm_{D^*}}\,\left(a_+(q^2)-a_-(q^2)\right),
 \en
and obey two additional HQS relations
 \be \label{eq:FFxi1}
 a_+(q^2)+a_-(q^2)=0, \qquad
 (m_B-m_D)f_+(q^2)+(m_B+m_D)f_-(q^2)=0,
 \en
where $\omega=(m_B^2+m_{D^{(*)}}^2-q^2)/(2m_Bm_{D^{(*)}})$. The
$B\to D_0^*$ and $B\to D_1^{1/2}$ form factors in the heavy quark
limit are related to $\tau_{1/2}(\omega)$ via \cite{IW91}
 \be \label{eq:HQS1}
 \tau_{1/2}(\omega) &=& {1\over 4\sqrt{m_Bm_{D^*_0}}}\left[
 (m_B-m_{D^*_0})u_+(q^2)+(m_B+m_{D^*_0})u_-(q^2)\right] \non \\
 &=& {1\over 2\sqrt{m_Bm_{D^{1/2}_1}}}{\ell_{1/2}(q^2)\over \omega-1}=
 \sqrt{m_Bm_{D^{1/2}_1}}\,q_{1/2}(q^2) \non \\
 &=&
 -{\sqrt{m_Bm_{D^{1/2}_1}}\over 2}\left(c^{1/2}_+(q^2)-c^{1/2}_-(q^2)\right),
 \en
while $B\to D_1^{3/2}$ and $B\to D_2^*$ transition form factors
are related to $\tau_{3/2}(\omega)$ by
 \be \label{eq:HQS2}
 \tau_{3/2}(\omega) &=& -\sqrt{2\over
 m_Bm_{D^{3/2}_1}}\,{\ell_{3/2}(q^2)\over \omega^2-1}=-{1\over 3}\sqrt{2m_B^3
 \over m_{D^{3/2}_1}}\left(c^{3/2}_+(q^2)+c^{3/2}_-(q^2)\right)
 \non \\
 &=& \sqrt{2m_B^3\over m_{D^{3/2}_1}}\,{c^{3/2}_+(q^2)-c^{3/2}_-(q^2)\over
 \omega-2} = 2\,\sqrt{m_B^3m_{D^*_2}\over 3}\,h(q^2) \non \\
 &=& \sqrt{m_B\over 3m_{D^*_2}}{k(q^2)\over 1+\omega}
 =-{2\sqrt{2}\over 1+\omega}{\sqrt{m_Bm_{D^{3/2}_1}}}~q_{3/2}(q^2)
 \non \\
 &=& - \sqrt{m_B^3m_{D^*_2}\over
 3}\left(b_+(q^2)-b_-(q^2)\right),
 \en
and subject to the HQS relations
 \be \label{eq:HQS3}
 && b_+(q^2)+b_-(q^2)=0, \qquad\quad
 c^{1/2}_+(q^2)+c^{1/2}_-(q^2)=0, \non \\
 && (m_B+m_{D^*_0})u_+(q^2)+(m_B-m_{D^*_0})u_-(q^2)=0.
 \en

In terms of the dimensionless form factors defined in
(\ref{eq:ffsdimless}) and (\ref{eq:ffpdimless}), Eqs.
(\ref{eq:FFxi}) and (\ref{eq:FFxi1}) can be recast to
 \be \label{eq:HQSxi}
\xi(\omega) &=& {2\sqrt{m_Bm_D}\over
m_B+m_D}\,F_1^{BD}(q^2)=\,{2\sqrt{m_Bm_D}\over
m_B+m_D}\,{F_0^{BD}(q^2)\over \left[1-{q^2\over
(m_B+m_D)^2}\right]}
\non \\
&=& {2\sqrt{m_Bm_{D^*}}\over m_B+m_{D^*}}\,V^{BD^*}(q^2)=\,
{2\sqrt{m_Bm_{D^*}}\over m_B+m_{D^*}}\,A_0^{BD^*}(q^2)   \non \\
&=& {2\sqrt{m_Bm_{D^*}}\over m_B+m_{D^*}}\,A_2^{BD^*}(q^2)=\,
{2\sqrt{m_Bm_{D^*}}\over
m_B+m_{D^*}}\,{A_1^{BD^*}(q^2)\over\left[1-{q^2\over
(m_B+m_{D^*})^2} \right]}\,.
 \en
Likewise, Eqs. (\ref{eq:HQS1}) and (\ref{eq:HQS3}) can be
rewritten as
 \be \label{eq:HQStau1half}
 \tau_{1/2}(\omega) &=& {\sqrt{m_Bm_{D_0^*}}\over
m_B-m_{D_0^*}}\,F_1^{BD_0^*}(q^2)=\,{\sqrt{m_Bm_{D_0^*}}\over
m_B-m_{D_0^*}}\,{F_0^{BD_0^*}(q^2)\over \left[1-{q^2\over
(m_B-m_{D_0^*})^2}\right]}
\non \\
&=& -{\sqrt{m_Bm_{D^{1/2}_1}}\over
m_B-m_{D^{1/2}_1}}\,A^{BD^{1/2}_1}(q^2)=\,
{\sqrt{m_Bm_{D^{1/2}_1}}\over m_B-m_{D^{1/2}_1}}\,V_0^{BD^{1/2}_1}(q^2)   \non \\
&=& -{\sqrt{m_Bm_{D^{1/2}_1}}\over
m_B-m_{D^{1/2}_1}}\,V_2^{BD^{1/2}_1}(q^2)=\,
-{\sqrt{m_Bm_{D^{1/2}_1}}\over
m_B-m_{D^{1/2}_1}}\,{V_1^{BD^{1/2}_1}(q^2)\over\left[1-{q^2\over
(m_B-m_{D^{1/2}_1})^2} \right]}\,.
 \en
In next subsections we will derive the above HQS relations for
form factors and IW functions using the covariant light-front
model based on HQET.

We see from Table~\ref{tab:LFBtoD} that HQS relations
(\ref{eq:HQS3}) for form factors $b_\pm$ and $c^{1/2}_\pm$ are
respected even for finite heavy quark masses. From the numerical
results of $\tau_{1/2}(\omega=1)=0.31$ and
$\tau_{3/2}(\omega=1)=0.61$ to be presented below in Sec. VI.F,
one can check the HQS relations (\ref{eq:HQStau1half}) and
(\ref{eq:HQS2}) at the zero-recoil point. It turns out that, among
the fourteen $B\to D^{**}$ form factors, the covariant light-front
model predictions for
$A^{BD_1^{1/2(3/2)}},V_0^{BD_1^{1/2}},V_2^{BD_1^{1/2}},h,b_+,b_-$
are in good agreement with those in the heavy quark limit, while
the agreement is fair for $c_+^{3/2}$ and $c_-^{3/2}$. However,
the predictions for $F_{1,0}^{BD_0^*},V_1^{BD_1^{1/2(3/2)}}$ and
$k$ at zero recoil show a large deviation from the HQS
expectation. Indeed, Eqs. (\ref{eq:HQStau1half}) and
(\ref{eq:HQS2}) indicate that except for $F_1^{BD_0^*}$, these
form factors should approach to zero when $q^2$ reaches its
maximum value, a feature not borne out in the covariant
light-front calculations for finite quark masses. This may signal
that $\Lambda_{\rm QCD}/m_Q$ corrections are particularly
important in this case. Phenomenologically, it is thus dangerous
to determine all the form factors directly from the IW functions
and HQS relations since $1/m_Q$ corrections may play an essential
role for some of them and the choice of the $\beta$ parameters for
$s$-wave and $p$-wave wave functions will affect the IW functions.

\subsection{Covariant light-front model within HQET}

To begin with, we rescale the bound state 
of a heavy meson by $|P^+_H,P_{H\bot},J,
J_z\rangle=\sqrt{M_H}\,|H(v,J,J_z)\rangle$. It is well known that
in the heavy quark limit, the heavy quark propagator can be
replaced by
 \be
 \frac{i}{\not \! p_Q-m_Q+i\epsilon}\to \frac{i(1+\not \! v)}{2 v\cdot
 k+i\epsilon},
 \label{eq:HQpropagator}
 \en
where $p_Q=m_Q v+k$ and $k$ is the residual momentum of the heavy
quark. One can then redo all the calculations in Sec. II by using
the above propagator for $q_1^\prime$ and $q_1^\pp$ and perform
the contour integral as before. Since the contour integral forces
the antiquark to be on its mass shell, it is equivalent to using
the so-called on-shell Feynman rules~\cite{CCHZ} in calculations.
The zero mode effect arises from the $p_1^{\prime+}=p_1^{\pp+}=0$
region and it can be interpreted as virtual pair creation
processes~\cite{deMelo98}. In the infinite quark mass limit, both
quarks are close to their mass shell and far from the
$p_1^{\prime+}=p_1^{\pp+}=0$ region. Consequently, the pair
creation is forbidden and the zero mode contribution vanishes in
the heavy quark limit. Hence, we do not need to stick to the
$q^+=0$ frame and are able to study form factors directly in the
timelike region.

To extract the on-shell Feynman rules, we use the calculation of
the pseudoscalar meson annihilation [c.f. Fig.~1(a)] as an
illustration. By virtue of Eq.~(\ref{eq:HQpropagator}), the matrix
element of Eq.~(\ref{eq:AP}) can be rewritten as
 \be
 \la 0|A_\mu|P(v)\ra =\frac{{\cal A}^P_\mu}{\sqrt{ M_H}}
                     =-i^2\frac{N_c}{(2\pi)^4} \int d^4 p_q
                     \frac{H_P^\prime}{2 v\cdot k\,N_2
                     \sqrt{M_H}}
                     {\rm Tr}[\gamma_\mu\gamma_5(1+\not \!v)\gamma_5 (-\not\!p_q+m_q)],
 \en
where we have used $p^\prime_1=p_Q$, $p_2=p_q$, $m_2=m_q$,
$N_2=p_q^2-m_q^2+i\epsilon$ and $k=-p_q+(M_H-m_Q)v$ from
4-momentum conservation. As in Sec. II, we need to perform the
contour integral by closing the upper complex
$p_1^{\prime-}$-plane, or equivalently, the lower complex
$p_q^-$-plane. The integration forces $p_q^2=m_q^2$ and
consequently,
 \be
 \la 0|A_\mu|P(v)\ra =-i^2\frac{N_c}{(2\pi)^4} \int d^4 p_q (-2\pi i)\delta(p_q^2-m_q^2)
 \frac{h_P^\prime}{2 v\cdot k \sqrt {M_H}}
 {\rm Tr}[\gamma_\mu\gamma_5(1+\not \!v)\gamma_5
                            (-\not\!p_q+m_q)].\non\\
 \en
Since $x_2$ is of  order $\Lambda_{\rm QCD}/m_Q$ in the heavy
quark limit, it is useful to define $X\equiv m_Q x_2$, which is of
order $\Lambda_{\rm QCD}$ even if $m_Q\to \infty$. For on-shell
$p_q$ we have $p_q^- = (p_{q\bot}^2 + m_q^2)/p_q^+$ and
 \begin{equation}
    v \cdot p_q = {1\over 2X} \Big( p_\bot^2 + m_q^2
        + X^2 \Big) \, .
 \end{equation}
It is then straightforward to obtain
 \be
 \frac{h^\prime_P}{2 v\cdot k\sqrt{ M_H}}\to\frac{1}{2\sqrt
 N_c}\frac{1}{\sqrt{v\cdot p_q+m_q}}
 \Phi^\prime(X, p^2_\bot),
 \en
with the aid of Eqs. (\ref{eq:internalQ}), (\ref{eq:vertex}) and
the replacements
 \be \label{eq:HQL}
 \widetilde M_0 &\to& \sqrt{2(v\cdot p_q+m_q)m_Q}~, \non \\
 \varphi(x,p_\bot^2) &\to& \sqrt{m_Q\over X}\,\Phi(X, p^2_\bot)\,.
 \en
An important feature of the covariant model is the requirement
that the light-front wave function must be a function of $v \cdot
p_q$ \cite{CCHZ}:
 \begin{equation} \label{cwf}
    \Phi (X, p^2_\bot) \longrightarrow \Phi (v \cdot p_q) \, .
 \end{equation}
As we will see later, the widely used Gaussian-type wave functions
have such a structure in the heavy quark limit, while the BSW wave
function does not have one. The normalization condition of
$\Phi(v\cdot p_q)$ can be recast in a covariant form:
 \begin{equation} \label{cn}
    \int {d^4 p_q \over (2\pi)^4} (2\pi) \delta (p_q^2 - m_q^2)
        | \Phi ( v\cdot p_q) |^2 = 1,
 \end{equation}
or
 \be
 \int_0^\infty {dX\over X}\int {d^2 p_\bot\over
 2(2\pi)^3}\,|\Phi(X,p_\bot^2)|^2=1.
 \en

Putting everything together we have
 \be
 \la 0|A_\mu|P(v)\ra &=&-i^2\frac{N_c}{(2\pi)^4} \int d^4 p_q (-2\pi
 i)\delta(p_q^2-m_q^2) \non\\
&\times& \frac{1}{2\sqrt
 N_c}\frac{1}{\sqrt{v\cdot p_q+m_q}}
 \Phi(v\cdot p_q){\rm Tr}[\gamma_\mu\gamma_5(1+\not \!v)\gamma_5
                            (-\not\!p_q+m_q)].
 \en
In practice, we can use the following on-shell Feynman rules to
obtain the above and other amplitudes. The diagrammatic rule is
given as follows \cite{CCHZ}:

(i) The heavy meson bound state in the heavy quark limit gives a
vertex (wave function) as follows:
\begin{equation}
\begin{picture}(65,30)(0,38)
\put(20,41){\line(1,0){20}} \put(20,39){\line(1,0){20}}
\put(40,40){\circle*{6} }
\end{picture}
    : ~~~~~~~ {1\over \sqrt N_c} \sqrt{1 \over v \cdot p_q + m_q}
        ~\Phi(v \cdot p_q) i\Gamma_H \, ,
\end{equation}
\begin{equation}
\begin{picture}(65,30)(0,38)
\put(20,41){\line(1,0){20}} \put(20,39){\line(1,0){20}}
\put(20,40){\circle*{6} }
\end{picture}
    : ~~~~~~~ {1\over \sqrt N_c} \sqrt{1 \over v \cdot p_q + m_q}
        ~\Phi^*(v \cdot p_q) i\overline \Gamma_H \, ,
\end{equation}
with $\overline \Gamma_H=\gamma^0\Gamma^\dagger_H\gamma^0$.  For
$p$-wave mesons, we denote the covariant wave function by
$\Phi_p(v\cdot p_q)$.

(ii) The internal line attached to the bound state gives an
on-mass-shell propagator,
\begin{eqnarray}
\begin{picture}(65,30)(0,38)
\put(0,40.5){\line(1,0){40}} \put(19,40){\vector(1,0){2}}
\put(0,39.5){\line(1,0){40}} \put(19,28){$k$}
\end{picture}
    &:& ~~~~~~ i\left({1 + \not \! v\over 2}\right) ~~ ({\rm for~heavy~quarks}), ~~~ \\
\begin{picture}(65,30)(0,38)
\put(0,40){\line(1,0){40}} \put(21,40){\vector(-1,0){2}}
\put(17.5,30){$p_q$}
\end{picture}
    &:& ~~~~~~ { i(-\not \! p_q + m_q)} ~~ (\rm for ~light~antiquarks),
\end{eqnarray}
where  $v^2=1$ and $p_q^2 = m_q^2$.

(iii) For the internal antiquark line attached to the bound state,
sum over helicity and integrate the internal momentum using
\begin{equation}
   N_c \int {d^4 p_q \over (2\pi)^4 } (-2\pi i) \delta
        (p_q^2 -m_q^2),
\end{equation}
where the delta function comes from the on-mass-shell condition
and $N_c$ comes from the color summation.

(iv) For all other lines and vertices that do not attach to the
bound states, the diagrammatic rules are the same as the Feynman
rules in the conventional field theory.

These are the basic rules for the subsequent evaluations in the
covariant model. The vertex $\Gamma_H$ for the incoming heavy
meson can be read from Table I or from Eqs. (\ref{eq:Pvertex}) and
(\ref{eq:vertexAppB}) by applying Eq.~(\ref{eq:HQL}). Hence, the
vertex functions in the heavy quark limit have the expressions:
 \be
 ^1S_0: \qquad\qquad & -i\gamma_5 \non \\
 ^3S_1: \qquad\qquad & \not\!\varepsilon \non \\
 ^3P_0: \qquad\qquad & -{1\over\sqrt{3}}(v\cdot p_q+m_q) \non \\
 ^1P_1: \qquad\qquad & \varepsilon\cdot p_q\gamma_5   \\
 ^3P_1: \qquad\qquad & -{1\over\sqrt{2}}\left[(v\cdot p_q+m_q)(\not\!\varepsilon
  -{\varepsilon\cdot p_q\over v\cdot p_q+m_q})\right]\gamma_5 \non \\
 ^3P_2: \qquad\qquad & -\vp_{\mu\nu}\gamma^\mu p_q^\nu\,. \non
 \en
In terms of the $P_1^{1/2}$ and $P^{3/2}_1$ states, the relevant
vertex functions read
 \be
 P_1^{1/2}: \qquad\qquad & {1\over\sqrt{3}}(v\cdot p_q+m_q)\not\!\varepsilon\gamma_5  \non \\
 P_1^{3/2}: \qquad\qquad & -{1\over\sqrt{6}}[(v\cdot p_q+m_q)\not\!\varepsilon
  -3\varepsilon\cdot p_q]\gamma_5 \,.
 \en

\subsection{Decay constants}
In the infinite quark mass limit, the decay constants are defined
by
 \be
   && \langle 0 | \overline{q} \gamma^\mu \gamma_5 h_v | P(v)
        \rangle = i F_P v^\mu~~,
        ~~~ \langle 0 | \overline{q}\gamma^\mu h_v |
        P^*(v,\varepsilon) \rangle = F_V \varepsilon^\mu \, , \non \\
   &&  \la 0|\bar q\gamma^\mu h_v|S(v,\vp)\ra =F_Sv^\mu,~~
 \la 0|\bar q\gamma^\mu\gamma_5 h_v|A^{1/2}(v,\vp)\ra
 =F_{A^{1/2}}\varepsilon^\mu, \non \\
&& \la 0|\bar q\gamma^\mu\gamma_5 h_v|A^{3/2}(v,\vp)\ra
 =F_{A^{3/2}}\varepsilon^\mu,
 \en
where the decay constant $F_H$ is related to the usual one $f_H$
by $F_H=\sqrt{M_H}\,f_H$. Note that the tensor meson cannot be
created from the $V-A$ current. HQS demands
that~\cite{IW89,HQfrules}
 \be \label{eq:HQSF}
 F_{V}=F_{P},\qquad F_{A^{1/2}}=F_{S},\qquad
 F_{A^{3/2}}=0.
 \en

Using the Feynman rules shown above, it is ready to evaluate the
one-body matrix elements for heavy scalar and axial-vector mesons:
\begin{eqnarray}
    \langle 0 | \overline{q} \gamma^\mu h_v |\,S (v) \rangle
        &=& -{1\over \sqrt{3}} {\rm Tr}\Big\{\gamma^\mu {1 + \not \! v
        \over 2} {\cal M}_1 \Big\} \, , \non \\
    \langle 0 | \overline{q} \gamma^\mu\gamma_5 h_v |A^{1/2}(v, \varepsilon) \rangle
        &=& {1\over \sqrt{3}}{\rm Tr}\Big\{\gamma^\mu \gamma_5{1 + \not \! v
        \over 2} \not \! \varepsilon\gamma_5 {\cal M}_1 \Big\} \, ,
\end{eqnarray}
where
\begin{equation}  \label{covint1}
    {\cal M}_1 = \sqrt{N_c} \int {d^4 p_q \over (2\pi)^4} (2\pi)
        \delta(p_q^2- m_q^2) {\Phi_p(v \cdot p_q) \over
        \sqrt{v \cdot p_q +m_q}} (m_q- \not \! p_q)(v\cdot
        p_q+m_q)\,.
\end{equation}
Letting ${\cal M}_1=a_1+b_1\not \! v$, we obtain
 \begin{eqnarray}
    a_1 &=& \sqrt{N_c} \int {d^4 p_q \over (2\pi)^4} (2\pi)
        \delta(p_q^2- m_q^2) \Phi_p( v \cdot p_q)\,m_q\,
        \sqrt{v \cdot p_q +m_q}\,,  \non \\
    b_1 &=& -\sqrt{N_c} \int {d^4 p_q \over (2\pi)^4}(2\pi)
        \delta(p_q^2- m_q^2) \Phi_p(v \cdot p_q)(v \cdot p_q)
        \sqrt{v \cdot p_q +m_q} ~ \, .
\end{eqnarray}
Thus,
\begin{eqnarray} \label{eq:FS}
    F_{S} &=& F_{A^{1/2}}=-{2\over\sqrt{3}}(a_1+b_1)\non \\
    &=&  2\sqrt{N_c\over 3}\int {d^4 p_q \over (2\pi)^4}
        (2\pi) \delta(p_q^2- m_q^2) \Phi_p(v \cdot p_q)
        \sqrt{ v \cdot p_q + m_q}\,(v\cdot p_q-m_q) \nonumber \\
        &=& 2\sqrt{N_c\over 3} \int {dXd^2 p_\bot \over 2(2\pi)^3 X}
        ~\Phi_p(X,p^2_\bot) \sqrt{p_\bot^2 +
        (m_q+X)^2 \over 2 X}\,{p_\bot^2 +(m_q-X)^2 \over 2 X}.
 \en
Likewise, for the axial-vector $P^{3/2}_1$ meson
 \be
   \langle 0 | \overline{q} \gamma^\mu\gamma_5 h_v |A^{3/2}(v, \varepsilon) \rangle
        &=& -{1\over \sqrt{6}}{\rm Tr}\Big\{\gamma^\mu \gamma_5{1 + \not \! v
        \over 2}[\not \! \varepsilon(\gamma_\alpha+v_\alpha) -3\varepsilon_\alpha]
        \gamma_5 {\cal M}_2^\alpha \Big\},
\end{eqnarray}
with
 \begin{equation}
    {\cal M}_2^\alpha = \sqrt{N_c} \int {d^4 p_q \over (2\pi)^4} (2\pi)
        \delta(p_q^2- m_q^2) {\Phi_p(v \cdot p_q) \over
        \sqrt{v \cdot p_q +m_q}} (m_q- \not \! p_q)p_q^\alpha.
\end{equation}
The general expression of ${\cal M}_2^\alpha$ is
 \be
 {\cal M}_2^\alpha=a_2v^\alpha+b_2\gamma^\alpha+c_2\not \! v
 v^\alpha+d_2 \not \! v\gamma^\alpha\,.
 \en
Since $\vp\cdot v=0$ and the contraction of $\gamma_\mu$ with the
spin 3/2 field vanishes, namely,
 \be
  (1+\not\! v)[\not\!\varepsilon(\gamma_\alpha+v_\alpha)
  -3\varepsilon_\alpha]\gamma^\alpha=0,
  \en
we are led to $F_{A^{3/2}}=0$ in the heavy quark limit.

For completeness, the decay constants of the $s$-wave heavy mesons
are included here \cite{CCHZ}
 \begin{eqnarray} \label{eq:FP}
    F_{P} &=& F_{V} \non \\
        &=& 2\sqrt{N_c} \int {dXd^2 p_\bot \over 2(2\pi)^3 X}
        ~\Phi(X,p^2_\bot) \sqrt{
        (X+m_q)^2+p_\bot^2 \over 2 X}\,.
\end{eqnarray}

We now show that the decay constants obtained in the covariant
light-front model in Sec. II do respect the heavy quark symmetry
relations given in  (\ref{eq:HQSf}) or (\ref{eq:HQSF}) in the
infinite quark mass limit and have expressions in agreement with
Eqs. (\ref{eq:FS}) and (\ref{eq:FP}). To illustrate this, we
consider the decay constants of pseudoscalar and vector mesons. In
the $m'_1=m_Q\to\infty$ limit, Eqs. (\ref{eq:fP0}) and
(\ref{eq:fV}) are reduced to
 \be
 \sqrt{m_Q}f_P &\to& 4\sqrt{N_c\over 2}\int {dXd^2p_\bot\over
 2(2\pi)^3\sqrt{X}}\,\Phi(X,p_\bot^2)\,{X+m_q\over \sqrt{(X+m_q)^2+p^2_\bot}}, \non \\
 \sqrt{m_Q}f_V &\to& 4\sqrt{N_c\over 2}\int {dXd^2p_\bot\over
 2(2\pi)^3\sqrt{X}}\,\Phi(X,p^2_\bot)\,{1\over \sqrt{(X+m_q)^2+p^2_\bot}}
 \left({m_q^2+p_\bot^2\over X}+m_q\right),
 \en
where $m_q=m_2$, $X=m_Qx$ and use of Eq. (\ref{eq:HQL}) has been
made. Since the wave function is even in $p_z$, a quantity defined
in Eq. (\ref{eq:internalQ}), it follows that
 \be  \label{eq:HQLid1}
 && \int dx\,d^2p_\bot {\varphi(x,p_\bot)\over \sqrt{[m_Qx+m_q(1-x)]^2+p_\bot^2}}\,p_z
 =0 \non \\
 && \longrightarrow \int {dX d^2p_\bot\over\sqrt{X}}
 {\Phi(X,p^2_\bot)\over \sqrt{(X+m_q)^2+p^2_\bot}} \left(X-{m_q^2+p_\bot^2\over
 X}\right)=0.
 \en
Therefore, $f_V=f_P$ in the heavy quark limit. Moreover,
$\sqrt{m_Q}f_P$ is identical to $F_P$ in Eq. (\ref{eq:FP}) after
applying the identity
 \be
 \int {dXd^2p_\bot\over
 \sqrt{X}}\,\Phi(X,p^2_\bot)\,{2(X+m_q)\over\sqrt{(X+m_q)^2+p_\bot^2}} = \int {dXd^2p_\bot\over
 \sqrt{X}}\,\Phi(X,p^2_\bot)\,{\sqrt{(X+m_q)^2+p_\bot^2}\over X},
 \en
following from Eq. (\ref{eq:HQLid1}).

Likewise, Eq. (\ref{eq:fA}) for the decay constants of
axial-vector mesons is reduced in the heavy quark limit to
 \be
 \sqrt{m_Q}f_{^1\!A} &\to& 2\sqrt{N_c\over 2}\int
 {dXd^2p_\bot\over 2(2\pi)^3\sqrt{X}}\, {\Phi_p(X,p^2_\bot)\over
 \sqrt{(X+m_q)^2+p^2_\bot}}~p_\bot^2,  \non \\
 \sqrt{m_Q}f_{^3\!A} &\to& \sqrt{N_c}\int
 {dXd^2p_\bot\over 2(2\pi)^3\sqrt{X}}\, {\Phi_p(X,p^2_\bot)\over
 \sqrt{(X+m_q)^2+p^2_\bot}}\left({ (m_q^2+p_\bot^2)^2\over
 X^2}-m_q^2\right),
 \en
where use of Eq. (\ref{eq:HQLid1}) has been applied for deriving
the expression of $f_{^1\!A}$. By virtue of the identities
 \be \label{eq:HQLid2}
 && \int {dX d^2p_\bot\over\sqrt{X}}
 {\Phi_p(X,p^2_\bot)\over \sqrt{(X+m_q)^2+p^2_\bot}}
 (p_z^2-p^2_\bot/2)=0, \non \\
  && \int {dX d^2p_\bot\over\sqrt{X}}
 {\Phi_p(X,p^2_\bot)\over \sqrt{(X+m_q)^2+p^2_\bot}} (X^2-m^2_q)=0,
 \en
and
 \be
 && {3\over\sqrt{2}}\int {dX d^2p_\bot\over\sqrt{X}}
 {\Phi_p(X,p^2_\bot)\over \sqrt{(X+m_q)^2+p^2_\bot}}~p_\bot^2 \non
 \\ &=& \int {dXd^2 p_\bot \over X}
        ~\Phi_p(X,p^2_\bot) \sqrt{p_\bot^2 +
        (m_q+X)^2 \over 2 X}\,{p_\bot^2 +(m_q-X)^2 \over 2 X},
 \en
following from the first equation of (\ref{eq:HQLid2}), one can
show that $f_{^3\!A}=-\sqrt{2}f_{^1\!A}$ and hence $f_{A^{3/2}}=0$
and $\sqrt{m_Q}f_{A^{1/2}}= F_{A^{1/2}}=F_S$ in the $m_Q\to\infty$
limit.

\subsection{Isgur-Wise functions}
It is well known that the $s$-wave to $s$-wave meson transition in
the heavy quark limit is governed by a single universal IW
function $\xi(\omega)$ \cite{IW89}. Likewise, there exist two
universal functions $\tau_{1/2}(\omega)$ and $\tau_{3/2}(\omega)$
describing ground-state $s$-wave to $p$-wave transitions
\cite{IW91}. Since the IW function $\xi$ has been discussed in
detail in \cite{CCHZ}, we will focus on the other two IW functions
$\tau_{1/2}$ and $\tau_{3/2}$.

Let us first consider the function $\tau_{1/2}$, which can be
extracted from the $B\to D_0^*$ or $B\to D_1^{1/2}$ transition
 \be
 \la D^*_0(v')|\bar h^c_{v'} \Gamma h_v^b|B(v)\ra &=& -i{1\over\sqrt{3}}{\rm Tr}\left\{
 \left({1+\not \!v'\over 2}\right)\Gamma\left({1+\not\!v \over
 2}\right)\gamma_5 {\cal M}_3\right\}, \non \\
 \la D^{1/2}_1(v',\vp)|\bar h^c_{v'} \Gamma h_v^b|B(v)\ra &=& i{1\over\sqrt{3}}{\rm Tr}\left\{
 \not \! \vp \gamma_5
 \left({1+\not \!v'\over 2}\right)\Gamma\left({1+\not\!v \over
 2}\right)\gamma_5 {\cal M}_3\right\},
 \en
where ${\cal M}_3$ is the transition matrix element for the light
antiquark:
 \begin{eqnarray}
        {\cal M}_3 = \int [d^4p_q](m_q - \not \! p_q)(v'\cdot p_q+m_q) \, ,
        \label{bmk1}
 \end{eqnarray}
and we have introduced the short-hand notation
 \be
 [d^4p_q]\equiv {d^4 p_q \over (2\pi)^4} (2\pi) \delta
        (p_q^2 - m_q^2) {\Phi_p^*( v' \cdot p_q ) \Phi ( v
        \cdot p_q )  \over \sqrt{(v'\cdot
        p_q +m_q)(v\cdot p_q + m_q)}}.
  \en
The structure of ${\cal M}_3$ dictated by Lorentz invariance has
the form \cite{Yan92}
\begin{equation}
    {\cal M}_3 = a_3 + b_3 \not \! v + c_3 \not \! v' + d_3 \not \!
        v \not \! v' \, . \label{bmk}
\end{equation}
This covariant decomposition allows us to easily determine the
coefficients $a_3, b_3, c_3, d_3$ with the results:
\begin{eqnarray}
    a_3 &=& \int [d^4p_q]~ m_q(v'\cdot p_q+m_q) \, , \non \\
    b_3 &=& - \int [d^4p_q]~ (v'\cdot p_q+m_q){1\over 2}
        \Bigg\{ {(v+v')\cdot p_q \over 1 + \omega} +
        {(v-v')\cdot p_q \over 1 - \omega} \Bigg\} \, ,  \\
    c_3 &=& - \int [d^4p_q]~ (v'\cdot p_q+m_q){1\over 2}
        \Bigg\{ {(v+v')\cdot p_q \over 1 + \omega} -
        {(v-v')\cdot p_q \over 1 - \omega} \Bigg\} \, , \non  \\
    d_3 &=& 0  \, , \non
\end{eqnarray}
where $\omega\equiv v\cdot v'$.

Then $B\to D_0^*$ and $B\to D_1^{1/2}$ transitions are simplified
to
 \be \label{1/2me.b}
 \la D^*_0(v')|\bar h^c_{v'} \Gamma h_v^b|B(v)\ra &=& -i\,2\tau_{1/2}(\omega){\rm Tr}\left\{
 \left({1+\not \!v'\over 2}\right)\Gamma\left({1+\not\!v \over
 2}\right)\gamma_5\right\}, \non \\
  \la D^{1/2}_1(v',\vp)|\bar h^c_{v'} \Gamma h_v^b|B(v)\ra &=& i\,2\tau_{1/2}(\omega){\rm Tr}\left\{
  \not \! \vp \gamma_5
 \left({1+\not \!v'\over 2}\right)\Gamma\left({1+\not\!v \over
 2}\right)\gamma_5\right\},
 \en
with
 \be
 \tau_{1/2}(\omega) &=& {1\over 2\sqrt{3}}(a_3-b_3+c_3-d_3) \non \\
  &=& {1\over 2\sqrt{3}}\int [d^4p_q]\,(v'\cdot p_q+m_q)
 \left(m_q+{ (v-v')\cdot p_q\over
 1-\omega}\right).
 \en
Since
 \be
 v\cdot p_q = {1\over 2X}(p_\bot^2 + m_q^2 + X^2), \qquad
 v'\cdot p_q = {1\over 2X'}(p_\bot^2 + m_q^2 + {X'}^2),
 \en
the IW function $\tau_{1/2}$ can be explicitly expressed as
 \be \label{eq:tau1half}
 \tau_{1/2}(\omega) &=& {1\over 2\sqrt{3}}\int {dX d^2 p_\bot\over
 2(2\pi)^3X^2}\,{1\over \sqrt{z}(1-z)}\,\Phi_p^*(zX,p_\bot^2)\Phi(X,p_\bot^2) \non \\
 &\times & \sqrt{p_\bot^2+(m_q+zX)^2\over
 p_\bot^2+(m_q+X)^2}\left[p_\bot ^2+(m_q+X)(m_q-zX)\right],
 \en
where $z\equiv X'/X$ and it can be related to $v \cdot v'$ by
\begin{equation}
    z \rightarrow z_{\pm} = \omega \pm \sqrt{\omega^2
        - 1}~,  ~~~~~ z_+ = {1 \over z_-} \, ,
\end{equation}
with the + $(-)$ sign corresponding to $v^3$ greater (less) than
$v'^3$. Note that $v^3$ greater (less) than $v'^3$ corresponds the
daughter meson recoiling in the negative (positive) $z$ direction
in the rest frame of the parent meson. In other words, after
setting $v_\bot = v'_\bot=0$, the daughter meson recoiling in the
positive and negative $z$ directions are the only two possible
choices of Lorentz frames. It is easily seen that
$\tau_{1/2}(\omega)$ remains the same under the replacement of
$z\to 1/z$. This indicates that the Isgur-Wise function thus
obtained is independent of the recoiling direction, namely, it is
truly Lorentz invariant.

Next consider the $B\to D_2^*$ or $B\to D_1^{3/2}$ transition to
extract the second universal function $\tau_{3/2}$
 \be \label{3/2me.a}
  \la D^*_2(v',\vp)|\bar h^c_{v'} \Gamma h_v^b|B(v)\ra &=& -i{\rm Tr}\left\{
  \varepsilon _{\alpha\beta}\gamma^\alpha
 \left({1+\not \!v'\over 2}\right)\Gamma\left({1+\not\!v \over
 2}\right)\gamma_5 {\cal M}^\beta_4\right\},  \\
   \la D^{3/2}_1(v',\vp)|\bar h^c_{v'} \Gamma h_v^b|B(v)\ra &=& -{i\over\sqrt{6}}{\rm Tr}\left\{
 [(-\gamma_\alpha+ v'_\alpha)\not \! \vp +3\vp_\alpha]\gamma_5
 \left({1+\not \!v'\over 2}\right)\Gamma\left({1+\not\!v \over
 2}\right)\gamma_5 {\cal M}^\alpha_4\right\}, \non
 \en
where
 \be
 {\cal M}^\alpha_4=\int [d^4p_q](m_q-\not \!p_q)p_q^\alpha.
 \en
Its most  general expression is
 \be \label{eq:M4}
 {\cal M}_{4\alpha}=a_4 v_\alpha+b_4 v'_\alpha+c_4\not \!v v_\alpha
 +d_4
\not \!v' v_\alpha+ e_4\not \!v v'_\alpha +f_4 \not \!v' v'_\alpha
+ g_4\gamma_\alpha+h_4\not \!v\gamma_\alpha+ h'_4\not \!
v'\gamma_\alpha.
 \en
Although only terms proportional to $a_4$, $c_4$ and $d_4$ will
contribute to $B\to D_2^*$ and $B\to D^{3/2}_1$ transitions after
contracting with the vertex of the spin 3/2 particles, all the
terms in ${\cal M}_{4\alpha}$ have to be retained in order to
project out the coefficients. With ${\cal M}_{4\alpha}$
contracting with $v^\alpha$, $v'^\alpha$ and $\gamma^\alpha$ we
find the following equations:
 \be
 a_4 +b_4 \omega &=& \int [d^4p_q]\,m_q v\cdot p_q, \non \\
 a_4 \omega +b_4 &=& \int [d^4p_q]\,m_q v'\cdot p_q, \non \\
 c_4+2d_4\omega+f_4+4g_4 &=& -\int [d^4p_q]\,m_q^2, \non \\
 c_4+d_4\omega+g_4 &=& -\int [d^4p_q]\,{v\cdot p_q\over 2}\left(
 {(v+v')\cdot p_q\over 1+\omega}+{(v-v')\cdot p_q\over 1-\omega}\right),  \\
 d_4+f_4\omega &=& -\int [d^4p_q]\,{v\cdot p_q\over 2}\left(
 {(v+v')\cdot p_q\over 1+\omega}-{(v-v')\cdot p_q\over 1-\omega}\right), \non \\
 c_4\omega+d_4 &=& -\int [d^4p_q]\,{v'\cdot p_q\over 2}\left(
 {(v+v')\cdot p_q\over 1+\omega}+{(v-v')\cdot p_q\over 1-\omega}\right), \non \\
 d_4\omega+f_4+g_4 &=& -\int [d^4p_q]\,{v'\cdot p_q\over 2}\left(
 {(v+v')\cdot p_q\over 1+\omega}-{(v-v')\cdot p_q\over 1-\omega}\right), \non
 \en
and $e_4=d_4$, $h_4=h_4'=0$. Solving the above equations yields
 \be
 a_4 &=& {m_q\over 2}\int [d^4p_q](\lambda_++\lambda_-)\,, \non \\
 b_4 &=& {m_q\over 2}\int [d^4p_q](\lambda_+-\lambda_-)\,, \non \\
 c_4 &=& -{1\over 4}\int [d^4p_q] \left((\lambda_++\lambda_-)^2-
 {g_4\over (1+\omega)(1-\omega)}\right)\,,  \\
 d_4 &=& -{1\over 4}\int [d^4p_q] \left((\lambda_+^2-\lambda_-^2)
 +{g_4 \omega\over (1+\omega)(1-\omega)})\right)\,, \non \\
 f_4 &=& -{1\over 4}\int [d^4p_q] \left( (\lambda_+-\lambda_-)^2-
 {g_4\over (1+\omega)(1-\omega)}\right)\,, \non
 \en
and
 \be
 g_4=-{1\over 2}\int [d^4p_q]\,\left(m_q^2-{1\over 2}(1+\omega)\lambda_+^2
 -{1\over 2}(1-\omega)\lambda_-^2\right),
 \en
with
 \be
  \lambda_+\equiv {(v+v')\cdot p_q\over 1+\omega},\qquad\quad
  \lambda_-\equiv {(v-v')\cdot p_q\over 1-\omega}.
  \en

Since only $v_\alpha,\not \!v v_\alpha$ and $\not \!v'v_\alpha$
terms in $M_{4\alpha}$ survive after contracting with the vertex
of $D_2^*$ and $D^{3/2}_1$ particles, the matrix elements
(\ref{3/2me.a}) are simplified to
 \be \label{3/2me.b}
  \la D^*_2(v',\vp)|\bar h^c_{v'} \Gamma h_v^b|B(v)\ra &=& -i\sqrt{3}\,\tau_{3/2}
(\omega)\varepsilon _{\alpha\beta}v^\beta{\rm Tr}\left\{
  \gamma^\alpha \left({1+\not \!v'\over 2}\right)\Gamma\left({1+\not\!v \over
 2}\right)\gamma_5 \right\}, \non \\
 \la D^{3/2}_1(v',\vp)|\bar h^c_{v'} \Gamma h_v^b|B(v)\ra &=& -{i\over\sqrt{2}}\tau_{3/2}
(\omega)  \\
 &\times & {\rm Tr}\left\{
 [\not \! \vp (1+\omega)+3\vp\cdot v ]\gamma_5
 \left({1+\not \!v'\over 2}\right)\Gamma\left({1+\not\!v \over
 2}\right)\gamma_5 \right\}, \non
 \en
with
 \be
 \tau_{3/2}(\omega) &=& {1\over \sqrt{3}}(a_4-c_4-d_4)  \\
 &=& {1\over 2\sqrt{3}}\int [d^4p_q]\left[
 (\lambda_++\lambda_-)(m_q+\lambda_+)+{\lambda_+^2(1+\omega)
 +\lambda_-^2(1-\omega)-m_q^2\over 2(1+\omega)}\right]. \non
 \en
A more explicit expression of $\tau_{3/2}$ reads
 \be
  \tau_{3/2}(\omega) &=& {1\over \sqrt{3}}\int {dX d^2 p_\bot \over
  2(2\pi)^3}
  \,{2\sqrt{z}\,\Phi_p^*(zX, p^2_\bot)\Phi(X, p^2_\bot)\over \sqrt{[p_\bot^2 + (m_q+X)^2]
        [p^2_\bot + (m_q +zX)^2 ]}} \non \\
        &\times& \Bigg\{
  {1\over 2(1-\omega)(1+\omega)^2}\Big[ (1-2\omega)(v'\cdot p_q)(2v+v')\cdot p_q  \\
  &+&  3(v\cdot p_q)^2-(1-\omega^2)m_q^2\Big]+{1\over 1-\omega^2}
  \,[v\cdot p_q-\omega(v'\cdot p_q)]m_q  \Bigg\}. \non
  \en
After some manipulation, we obtain a simple relation between
$\tau_{3/2}$ and $\tau_{1/2}$:
 \be \label{eq:tau3half}
 \tau_{3/2}(\omega) &=&
 {2\over 1+\omega}\tau_{1/2}(\omega) +{\sqrt{3}\over
 1+\omega}\int {dX d^2p^\prime_\bot\over 2(2\pi)^3}\Phi^*_p(zX, p'^2_\bot)\Phi(X,
        p'^2_\bot) \non \\
 &\times& {\sqrt{z}\,p'^2_\bot \over \sqrt{[p'^2_\bot + (m_q+X)^2]
        [p'^2_\bot + (m_q +zX)^2 ]} }\,.
 \en
Finally we include the usual IW function $\xi(\omega)$ for the
sake of completeness \cite{CCHZ}
 \begin{eqnarray} \label{eq:xi}
    \xi (\omega) &=& \int {dX d^2 p_\bot \over 2(2\pi)^3X}
     ~ {2 \sqrt{z}\over 1+z}~ \Phi^*(zX, p^2_\bot)\Phi(X,
        p^2_\bot) \nonumber \\
        & \times &  {p_\bot^2 + (m_q + X)(m_q
         + zX) \over \sqrt{[p_\bot^2 + (m_q+X)^2]
        [p^2_\bot + (m_q +zX)^2 ]}}\,.
\end{eqnarray}
and the relevant matrix elements are given by
 \begin{eqnarray} \label{eq:ppme}
        & & \langle D (v') | \overline{h}_{v'}^c \Gamma
                h_v^b | B (v) \rangle =\xi(\omega) {\rm Tr}\Big\{
        \gamma_5 \Big( {1+ \not{\! v}' \over 2} \Big)\Gamma
        \Big( {1+\not{\! v} \over 2} \Big)\gamma_5
        \Big\}  \, , \non \\
        & & \langle D^* (v',\varepsilon) | \overline{h}_{v'}^c \Gamma
        h_v^b | B (v) \rangle = \xi(\omega) {\rm Tr}\Big\{\!
        \not{\! \varepsilon}^* \Big( {1 + \not{\! v}' \over 2} \Big)
        \Gamma \Big( {1+ \not{\! v} \over 2} \Big) i \gamma_5
        \Big\}  \, .
\end{eqnarray}

\subsection{Form factors in the heavy quark limit}
From Eqs. (\ref{eq:ppme}), (\ref{1/2me.b}) and (\ref{3/2me.b}) we
obtain the matrix elements of $B\to D,D^*,D^{**}$ transitions in
the heavy quark limit
 \be \label{formHQS}
 \la  D(v')|V_\mu|B(v)\ra &=& \xi(\omega)(v+v')_\mu,
 \non \\
 \la D^*(v',\vp)|V_\mu|B(v)\ra &=& -\xi(\omega)
 \epsilon_{\mu\nu\alpha\beta} \vp^{*\nu}{v'}^\alpha v^\beta, \non \\
 \la D^*(v',\vp)|A_\mu|B(v)\ra &=& i\xi(\omega)
 \Big[ (1+\omega)\vp^*_\mu-(\vp^*\cdot v)v'_\mu\Big], \non \\
 \la  D_0^*(v')|A_\mu|B(v)\ra &=& i\,2\tau_{1/2}(\omega)(v-v')_\mu,
 \non \\
 \la D_1^{1/2}(v',\vp)|V_\mu|B(v)\ra &=& -i\,2\tau_{1/2}(\omega)
 \Big[ (1-\omega)\vp^*_\mu+(\vp^*\cdot v)v'_\mu\Big],  \non \\
 \la D_1^{1/2}(v',\vp)|A_\mu|B(v)\ra &=& -2\tau_{1/2}(\omega)
 \epsilon_{\mu\nu\alpha\beta} \vp^{*\nu}{v'}^\alpha v^\beta,  \\
 \la D_1^{3/2}(v',\vp)|V_\mu|B(v)\ra &=& i{1\over\sqrt{2}}\,\tau_{3/2}(\omega)
 \Big\{ (1-\omega^2)\vp^*_\mu-(\vp^*\cdot v)[3v_\mu+(2-\omega)v'_\mu]\Big\}, \non \\
 \la D_1^{3/2}(v',\vp)|A_\mu|B(v)\ra &=& {1\over\sqrt{2}}\,\tau_{3/2}(\omega)
 (1+\omega)\epsilon_{\mu\nu\alpha\beta} \vp^{*\nu}{v'}^\alpha v^\beta,  \non \\
 \la D^*_2(v',\vp)|V_\mu|B(v)\ra &=& \sqrt{3}\,\tau_{3/2}(\omega)
 \epsilon_{\mu\nu\alpha\beta} \vp^{*\nu\gamma}v_\gamma {v'}^\alpha v^\beta, \non \\
 \la D^*_2(v',\vp)|A_\mu|B(v)\ra &=& -i\sqrt{3}\,\tau_{3/2}(\omega)
 \Big\{ (1+\omega)\vp^*_{\mu\nu}v^\nu-\vp^*_{\alpha\beta}v^\alpha v^\beta
 v'_\mu\Big\}. \non
 \en
It is easily seen that the $B\to D^{**}$ matrix elements of weak
currents vanish at the zero recoil point $\omega=1$ owing to the
orthogonality of the wave functions of $B$ and $D^{**}$. Setting
$p_B=m_Bv$ and $p_{D}=m_{D}v'$,$\cdots$, etc. in
Eqs.~(\ref{eq:ffs}) and (\ref{eq:ffp}) and comparing with
(\ref{formHQS}) yields all the form-factor HQS relations given in
Sec. IV.A.

We are ready to check the heavy quark limit behavior of form
factors to see if they satisfy the HQS constraints. Consider the
form factor $F_1^{BD_0^*}(q^2)=-u_+(q^2)$ first. Let
$x_2=x,x_1=1-x,m_2=m_q,X=m_bx,X'=m_cx$, it follows from Eq.
(\ref{eq:upm}) that
 \be
 u_+(q^2) &=& {N_c\over 16\pi^3}\int dx\, d^2p^\prime_\bot
 \frac{h^\prime_P h^\pp_S}{x^2 \hat N_1^\prime \hat N^\pp_1}\Big\{
 2(X+m_q)X'-2m_qX-2m_q^2(1-x)^2-x^2q^2   \non \\
 && +2m_q x(X-X')-2p_\bot^{\prime 2}+2xp'_\bot\cdot q_\bot \Big\},
 \en
where use of Eq. (\ref{eq:internalQ}) and
$p_\bot^\pp=p'_\bot-xq_\bot$ has been used. In the heavy quark
limit $x\sim {\cal O}(\Lambda_{\rm QCD}/m_Q)\to 0$, we have
 \be
 u_+(q^2) &\to& -{1\over 2\sqrt{3}} \int {dXd^2p'_\bot\over 2(2\pi)^3X}
 \,{\widetilde
 M_0^\pp\over \widetilde M'_0
 M^\pp_0}\,\varphi_p^*(x,p'^2_\bot)\varphi(x,p'^2_\bot)
  \left[p^{\prime 2}_\bot+(m_q+X)(m_q-X')\right].
 \en
Substituting the replacements
 \be \label{eq:replacement}
 && \widetilde M_0'\to \sqrt{ m_b [(X+m_q)^2+p'^2_\bot]/ X},
 \qquad \widetilde M_0^\pp \to \sqrt{ m_c [(X'+m_q)^2+p'^2_\bot]/
 X'},  \non \\
 && \varphi(x,p'^2_\bot) \to \sqrt{m_b\over X}\,\Phi(X,p'^2_\bot),
 \qquad \varphi_p(x,p'^2_\bot) \to \sqrt{m_c\over X'}\,\Phi_p(X',p'^2_\bot)
 \en
valid in the infinite quark mass limit and noting that $z=
X'/X=m_c/m_b$, we arrive at
 \be
 u_+(q^2) &\to&
 -{1\over 2\sqrt{3}}\int {dX d^2 p'_\bot\over
 2(2\pi)^3X^2}\,{1\over z}\,\Phi_p^*(zX,p'^2_\bot)\Phi(X,p'^2_\bot) \non \\
 &\times & \sqrt{p'^2_\bot+(m_q+zX)^2\over
 p'^2_\bot+(m_q+X)^2}\left[p'^2_\bot+(m_q+X)(m_q-zX)\right],
 \en
and hence
 \be
 {\sqrt{m_Bm_{D_0^*}}\over
 m_B-m_{D_0^*}}F_1^{BD_0^*}(q^2) \to \tau_{1/2}(\omega).
 \en
Likewise, it is easily shown that
 \be
 {\sqrt{m_Bm_D}\over
 m_B+m_D}F_1^{BD}(q^2)\to\xi(\omega).
 \en

In order to demonstrate that the $B\to D^*$ form factors are
related to the IW function $\xi(\omega)$, we need to apply the
identity (\ref{eq:identity}) which has the expression
 \be \label{eq:identity1}
\int dx\, d^2p^\prime_\bot \frac{2h^\prime_P h^\pp_V}{x \hat
N_1^\prime \hat N^\pp_1}\left((q\cdot P){p'_\bot\cdot q_\bot\over
q^2}- {1\over x}(p'^2_\bot+m_q^2-X^2)\right)=0
 \en
in the $m_Q\to\infty$ limit. This identity allows us to integrate
out the $(p'_\bot\cdot q_\bot)/q^2$ term. Then the form factor
$g(q^2)$ that reads [see Eq. (\ref{eq:PtoV})]
 \be
 g(q^2) &=& -{1\over m_B+m_{D^*}}\,{N_c\over 16\pi^3}\int dx\, d^2p^\prime_\bot
 \frac{2h^\prime_P h^\pp_V}{x^2 \hat N_1^\prime \hat N^\pp_1}\Bigg\{
 (X+X')(X+m_q)+x(q\cdot P){p'_\bot\cdot q_\bot\over
 q^2}  \non \\
 &+& 2{m_B+m_{D^*}\over w_V^\pp}\left(xp'^2_\bot+x{p'_\bot\cdot
 q_\bot\over q^2}\right)\Bigg\}
 \en
is reduced under the heavy quark limit to
 \be
 -{1\over m_B+m_{D^*}}\int {dX d^2p^\prime_\bot\over 2(2\pi)^3X}\Phi^*(zX, p'^2_\bot)\Phi(X,
        p'^2_\bot) {p'^2_\bot + (m_q + X)(m_q
         + zX) \over \sqrt{[p'^2_\bot + (m_q+X)^2]
        [p'^2_\bot + (m_q +zX)^2 ]}},
 \en
where use of Eqs. (\ref{eq:vertex}), (\ref{eq:identity}) and
(\ref{eq:replacement}) has been made. Comparing with Eq.
(\ref{eq:xi}) it is evident that the heavy quark limit of $g(q^2)$
has the same expression as $\xi(\omega)$ apart from a mass factor.
Therefore, we arrive at\footnote{If the Taylor expansion of
$h^\pp_V/\hat N_1^\pp$ is performed to take care of the
$p'_\bot\cdot q_\bot$ term in the integrand of $g(q^2)$, it turns
out that the heavy quark limit of the $B\to D^*$ form factors will
be related to the IW function
 \be
  \zeta(\omega) &=& \int {dX d^2p^\prime_\bot\over 2(2\pi)^3}\Phi^*(X', p'^2_\bot)\Phi(X,
        p'^2_\bot) { X'(X+m_q)+X'(X-X') p'^2_\bot \Theta_V\over \sqrt{[p'^2_\bot + (m_q+X)^2]
        [p'^2_\bot + (m_q +X')^2 ]}},
 \en
where
 \be
 \Theta_V={\hat N_1^\pp\over h_V^\pp}\Big(\frac{d}{dp^{\pp2}_\bot}\frac{h^\pp_V}
 {\hat N_1^\pp}\Big)_{p^{\pp2}_\bot\to p^{\prime
 2}_\bot}.
 \en
This function $\zeta(\omega)$ first obtained in \cite{Cheng97} was
found numerically identical to $\xi(\omega)$, as it should be.
However, one has to appeal to the identity (\ref{eq:identity1}) in
order to prove this equivalence analytically.}
 \be
 {2\sqrt{m_Bm_{D^*}}\over m_B+m_{D^*}}V^{BD^*}(q^2)=-2\sqrt{m_Bm_{D^*}}\,g(q^2) \to
 \xi(\omega).
 \en

We next turn to the form factors $q_{1/2}$ and $q_{3/2}$ and see
if they are related to the IW functions $\tau_{1/2}$ and
$\tau_{3/2}$, respectively. We first study the heavy quark limit
behavior of $q^{1A}$ and $q^{3A}$. It follows from Eq.
(\ref{eq:PtoA}) that
 \be
 \sqrt{m_Bm_{D_1^{1/2}}}\,q^{1A}(q^2) &=& \sqrt{m_Bm_{D_1^{1/2}}}
 {N_c\over 16\pi^3}\int dx\, d^2p^\prime_\bot
 \frac{2h^\prime_P h^\pp_{^1\!A}}{x \hat N_1^\prime \hat
 N^\pp_1}\left( p'^2_\bot+{(p'_\bot\cdot q_\bot)^2\over
 q^2}\right) \non \\
 &\to& -{1\over 2}\int {dX d^2p^\prime_\bot\over 2(2\pi)^3}\Phi^*_p(zX, p'^2_\bot)\Phi(X,
        p'^2_\bot) \non \\
 &\times & {\sqrt{z}\,p'^2_\bot \over \sqrt{[p'^2_\bot + (m_q+X)^2]
        [p'^2_\bot + (m_q +zX)^2 ]}},
 \en
and
 \be
\sqrt{m_Bm_{D_1^{3/2}}}\,q^{3A}(q^2) &\to& -{1\over 2\sqrt{2}}\int
{dX d^2p^\prime_\bot\over 2(2\pi)^3X^2}\,{\Phi^*_p(zX,
p'^2_\bot)\Phi(X,p'^2_\bot)\over \sqrt{z}(1-z)}   \non \\
&\times & \sqrt{p'^2_\bot+(m_q+zX)^2\over
p'^2_\bot+(m_q+X)^2}\left[p'^2_\bot+(m_q+X)(m_q-zX)\right] \non \\
 &-& {1\over 2\sqrt{2}}\int {dX d^2p^\prime_\bot\over 2(2\pi)^3}\Phi^*_p(zX, p'^2_\bot)\Phi(X,
        p'^2_\bot) \non \\
 &\times & {\sqrt{z}\,p'^2_\bot \over \sqrt{[p'^2_\bot + (m_q+X)^2]
        [p'^2_\bot + (m_q + zX)^2 ]}}.
 \en
Since
 \be
 q_{1/2}(q^2)={1\over\sqrt{3}}\,q^{1A}(q^2)-\sqrt{2\over
 3}\,q^{3A}(q^2), \qquad  q_{3/2}(q^2)=\sqrt{2\over 3}\,q^{1A}(q^2)+{1\over
 \sqrt{3}}\,q^{3A}(q^2),
 \en
following from Eq. (\ref{eq:Phalf}), we obtain
 \be
 \sqrt{m_Bm_{D_1^{1/2}}}\,q_{1/2}(q^2) \to \tau_{1/2}(\omega)
 \en
and
 \be
 -{2\sqrt{2}\over 1+\omega}\sqrt{m_Bm_{D_1^{3/2}}}\,q_{3/2}(q^2)
 &\to& \tau_{3/2}(\omega),
 \en
as promised before.

All other HQS relations in Eqs. (\ref{eq:FFxi}), (\ref{eq:HQS1})
and (\ref{eq:HQS2}) can be proved in the same manner except for
the $B\to D_2^*$ form factors $h,k,b_+,b_-$ for which we are not
able to show at present that they are related to
$\tau_{3/2}(\omega)$ in the heavy quark limit. Perhaps one needs
some identities in \cite{Jaus99} and those derived in Appendix B
to verify the HQS relations between $B\to D_2^*$ form factors and
$\tau_{3/2}(\omega)$. This remains to be investigated.

\subsection{Numerical results for IW functions and discussion}
Covariance requires that light-front wave functions be a function
of $v \cdot p_q$. Currently, there exist several phenomenological
light-front wave functions commonly utilized in the literature.
There are several popular phenomenological light-front wave
functions that have been employed to describe various hadronic
structures in the literature. Two of them, the Bauer-Stech-Wirbel
(BSW) wave function $\Phi_{\rm BSW}(x,p^2_\bot)$ \cite{BSW} and
the Gaussian-type wave function $\Phi_{G}(x,p^2_\bot)$
\cite{Gauss}, have been widely used in the study of heavy mesons.
In the heavy quark limit, we denote these wave functions as
follows:
\begin{eqnarray}
     \Phi_{\rm BSW} (X,p^2_\bot) &=& 4 \sqrt{2} \left({\pi \over \beta^2}
        \right) X \exp\Bigg\{-{p_\bot^2 + X^2 \over 2\beta^2 }
        \Bigg\}, \non \\
     \Phi_G(X,p^2_\bot) &=& 4 \Big({\pi\over \beta^2}\Big)^{3/4}
        \sqrt{{X^2 + m_q^2 + p_\bot^2 \over 2X}} \nonumber \\
    & \times & \exp\Bigg\{-{1\over 2\beta^2}
        \Big[ p_\bot^2 + \Big({X\over 2} - {m_q^2 +p_\bot^2\over 2X}\Big)^2
        \Big]\Bigg\} \, ,  \label{gslfw}
\end{eqnarray}
where $\Phi_G(X,p^2_\bot)$ is the heavy quark limit expression of
the Gaussian-type wave function given in Eq. (\ref{eq:Gauss}). For
$p$-wave heavy mesons, the wave functions are
 \be
 \Phi^{\rm BSW}_p (X,p^2_\bot)=\sqrt{2\over \beta^2}~\Phi^{\rm BSW} (X,p^2_\bot), \qquad
 \Phi^G_p (X,p^2_\bot)=\sqrt{2\over \beta^2}~\Phi^G(X,p^2_\bot).
 \en
As pointed out in \cite{CCHZ}, not all the phenomenological
light-front wave functions have the covariant property. We found
that the Gaussian wave function and the invariant-mass wave
function can be reexpressed as a pure function of $v \cdot p_q$.
The wave function $\Phi_G$ can be rewritten in terms of $v \cdot
p_q$:
\begin{eqnarray}
    \Phi_G(X,p^2_\bot) &=& 4 \Big({\pi\over \beta^2}\Big)^{3/4}
        \sqrt{{X^2 + m_q^2 + p_\bot^2 \over 2X}} \nonumber \\
    & \times &  \exp\Bigg\{-{1\over 2\beta^2} \Big[
        \Big({X\over 2} + {m_q^2 +p_\bot^2\over 2X}\Big)^2 - m_q^2 \Big]\Bigg\}
        \nonumber \\
    &=& 4 \Big({\pi\over \beta^2}\Big)^{3/4}
        \sqrt{ v \cdot p_q} \exp\Bigg\{-{1\over 2\beta^2}
        \Big[ (v \cdot p_q)^2 - m_q^2 \Big]\Bigg\} \, .
\end{eqnarray}
Therefore this wave function preserves the Lorentz covariance of
Eqs.~(\ref{eq:M4}), (\ref{bmk}) and (\ref{covint1}). This also can
be examined by a numerical check of the covariant condition
 \be
 \int {dX d^2 p_\bot \over 2 (2\pi)^3 X}\, \Phi (X,
        p^2_\bot) \, X = \int {dX d^2 p_\bot \over
        2 (2\pi)^3 X}\,  \Phi (X, p^2_\bot) \, {m_q^2
        + p_\bot^2 \over X} \, ,
 \en
which is satisfied if $\Phi(X,p_\bot^2)$ is a function of $v\cdot
p_q$. However, very surprisingly, the commonly used BSW wave
function cannot be recast as a pure function of $v \cdot p_q$.
Hence the BSW wave function breaks the Lorentz covariance. Indeed,
we have already found previously \cite{Cheng97} that there is some
inconsistent problem by using the BSW wave function to calculate
various transition form factors. Now we can understand why the BSW
wave function gives such results inconsistent with HQS found in
\cite{Cheng97,De97}. Hence, by demanding relativistic covariance,
we can rule out certain types of heavy meson light-front wave
functions.

\begin{figure}[t]
\vspace{-1cm}
 \includegraphics[angle=-90,width=11cm]{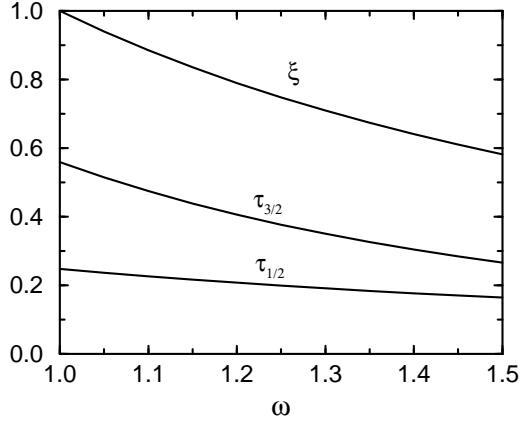}
\vspace{-1cm}
    \caption[]{\small The Isgur-Wise functions $\xi$, $\tau_{1/2}$ and $\tau_{3/2}$
    as a function of $\omega$.} \label{fig:IW}
\end{figure}

To perform numerical calculations of the decay constants and IW
functions in the heavy quark limit, we follow \cite{CCHZ} to use
the input $m_q=250$ MeV and $f_B=180$ MeV to fix the parameter
$\beta_\infty$ to be 0.49. For decay constants we then obtain
 \be
 F_{P}=F_{V}=413\,{\rm MeV}^{3/2},\qquad F_{S}=F_{A^{1/2}}=399\,{\rm
 MeV}^{3/2}.
 \en
The decay constant of the $P^{3/2}_1$ heavy meson vanishes in the
infinite quark mass limit. We see that the decay constants of
ground-state $s$-wave mesons and low-lying $p$-wave mesons are
similar in the heavy quark limit.

The IW functions (\ref{eq:xi}), (\ref{eq:tau1half}) and
(\ref{eq:tau3half}) can be fitted nicely to the form
 \be
 f(\omega)=f(1)[1-\rho^2(\omega-1)+{\sigma^2\over 2}(\omega-1)^2],
 \en
and it is found that (see Fig. \ref{fig:IW})
 \be
 \xi(\omega) &=& 1-1.22(\omega-1)+0.85(\omega-1)^2, \non \\
 \tau_{1/2}(\omega) &=& 0.31\left(1-1.18(\omega-1)+0.87(\omega-1)^2\right), \non \\
 \tau_{3/2}(\omega) &=&
 0.61\left(1-1.73(\omega-1)+1.46(\omega-1)^2\right),
 \en
where we have used the same $\beta_\infty$ parameter for both wave
functions $\Phi$ and $\Phi_p$. It should be stressed that unlike
$\tau_{1/2}(1)$ and $\tau_{3/2}(1)$, the normalization $\xi(1)=1$
at the zero recoil point is a model-independent consequence; that
is, it is independent of the structure of wave functions. In Table
\ref{tab:IW} we have compared this work for the IW functions
$\tau_{1/2}(\omega)$ and $\tau_{3/2}(\omega)$ with other model
calculations. It turns out that our results are similar to that
obtained in the ISGW model \cite{ISGW} (numerical results for the
latter being quoted from \cite{Morenas}). Our result $\rho^2=1.22$
for the slope parameter is consistent with the current world
average of $1.44\pm0.14$ extracted from exclusive semileptoic $B$
decays \cite{HFAG}.

\begin{table}[t]
\caption{The Isgur-Wise functions $\tau_{1/2}$ and $\tau_{3/2}$ at
zero recoil and their slope parameters. The numerical results for
\cite{ISGW,CCCN,Veseli,Godfrey} denoted by ``$*$" are quoted from
\cite{Morenas}.}
 \label{tab:IW}
\begin{center}
\begin{tabular}{| l l l l  l |}
\hline
 $\tau_{1/2}(1)$ & $\rho_{1/2}^2$ & $\tau_{3/2}(1)$
& $\rho_{3/2}^2$ & Ref. \\ \hline
 0.31 & 1.18 & 0.61 & 1.73 & This work \\
 0.06 & 0.73 & 0.52 & 1.45 & \cite{CCCN}$^*$  \\
 0.09 & 1.1 & 0.28 & 0.9 &   \cite{Deandrea98} \\
 0.13 & 0.57 & 0.43 & 1.39 & \cite{Veseli}$^*$ \\
 0.22 & 0.83 & 0.54 & 1.50 & \cite{Godfrey}$^*$ \\
 0.34 & 1.08 & 0.59 & 1.76 & \cite{ISGW}$^*$ \\
 $0.35\pm0.08$ & $2.5\pm1.0$ & -- & -- & \cite{Colangelo} \\
 $0.41\pm0.04$~~~ & $1.30\pm0.23$~~~ & $0.66\pm0.02$~~~ & $1.93\pm0.16$~~~ &
 \cite{Wambach} \\
 -- & -- & $0.74\pm0.15$ & $0.90\pm0.05$ & \cite{Huang}  \\
 \hline
\end{tabular}
\end{center}
\end{table}

It is interesting to notice that there is a sum rule derived by
Uraltsev \cite{Uraltsev}
 \be \label{eq:Uraltsev}
 \sum_n|\tau_{3/2}^{(n)}(1)|^2-\sum_n|\tau_{1/2}^{(n)}(1)|^2={1\over 4}\,,
 \en
where $n$ stands for radial excitations. This sum rule clearly
implies that $|\tau_{3/2}(1)|\gg |\tau_{1/2}(1)|$. Our results
indicate that this sum rule is slightly over-saturated even by
$n=0$ $p$-wave states. Another sum rule due to Bjorken
\cite{Bjorken} reads
 \be
 \rho^2={1\over
 4}+\sum_n|\tau_{1/2}^{(n)}(1)|^2+2\sum_n|\tau_{3/2}^{(n)}(1)|^2\,,
 \en
where $\rho^2$ is the slope of the IW function $\xi(\omega)$.
Combined with the Uraltsev sum rule (\ref{eq:Uraltsev}) leads to
 \be
 \rho^2={3\over 4}+3\sum_n|\tau_{1/2}^{(n)}(1)|^2.
 \en
Note that while the Bjorken sum rule receives perturbative
corrections \cite{Boyd}, the Uraltsev sum rule does not (for a
recent study, see \cite{Dorsten}).

\section{Conclusions}

In this work we have studied the decay constants and form factors
of the ground-state $s$-wave and low-lying $p$-wave mesons within
a covariant light-front approach. This formalism that preserves
the Lorentz covariance in the light-front framework has been
developed and applied successfully to describe various properties
of pseudoscalar and vector mesons. One of our main goals is to
extend this approach to the $p$-wave meson case. Our main results
are as follows:

\begin{itemize}

\item The main ingredients of the covariant light-front model,
namely, the vertex functions, are explicitly worked out for both
$s$-wave and $p$-wave mesons.

\item The decay constant of light scalar mesons is largely
suppressed relative to that of the pseudoscalar mesons and this
suppression becomes less effective for heavy scalar resonances.
The predicted decay constants $|f_{D_{s0}^*}|=71$ MeV and
$|f_{D_{s1}(2460)}|=117$ MeV are consistent with the corresponding
values of $47\sim 73$ MeV and $110\sim 190$ MeV inferred from the
measurement of $\ov D D_{s0}^*$ and $\ov DD_{s1}$ productions in
$B$ decays.

\item In the limit of SU(N)-flavor symmetry, the decay constants
of the scalar meson and the $^1P_1$ axial-vector meson are found
to be vanished, as it should be.

\item The analytic expressions for $P\to S,A$ transition form
factors can be obtained from that of $P\to P,V$ ones by some
simple replacements. We have also worked out the form factors in
$P\to T$ transitions.

\item The momentum dependence of the physical form factors is
determined by first fitting the form factors obtained in the
spacelike region to a 3-parameter function in $q^2$ and then
analytically continuing them to the timelike region. Some of the
$V_2(q^2)$ form factors in $P\to A$ transitions are fitted to a
different 3-parameter form so that the fit parameters are stable
within the chosen $q^2$ range.

\item Numerical results of the form factors for $B(D)\to\pi$,
$\rho$, $a_0(1450)$, $a_1(1260)$, $b_1(1235)$, $a_2(1320)$,
$B(D)\to K,~K^*,~K^{**}$ and $B\to D,~D^*,~D^{**}$ transitions are
presented in detail, where $K^{**}$ and $D^{**}$ denote
generically $p$-wave strange and charmed mesons, respectively.

\item Comparison of this work with the ISGW2 model based on the
nonrelativistic constituent quark picture is made for $B\to
D^{**}$ transition form factors. In general, the form factors at
small $q^2$ in both models agree within 40\%. However,
$F_0^{BD_0^*}(q^2)$ and $V_1^{BD_1^{1/2}}(q^2)$ have a very
different $q^2$ behavior in these two models as $q^2$ increases.
Relativistic effects are mild in $B\to D^{**}$ transitions but can
manifest in heavy-to-light transitions at maximum recoil. For
example, $V_0^{Ba_1}(0)$ is found to be 0.13 in the covariant LF
model, while it is as big as 1.01 in the ISGW2 model.

\item The decay amplitudes of $B^-\to D^{**0}\pi^-$ involve the
$B\to D^{**}$ form factors and $D^{**}$ decay constants. We have
compared the model calculations with experiment and found a good
agreement. In particular, the suppression of the $D_0^{*0}\pi^-$
production relative to $D^0\pi^-$ one clearly indicates a smaller
$B\to D_0^*$ form factor relative to the $B\to D$ one.

\item The heavy quark limit behavior of decay constants and form
factors is examined and it is found that the requirement of heavy
quark symmetry is satisfied.

\item Decay constants and form factors are also evaluated
independently in a covariant light-front formulism within the
framework of heavy quark effective theory. The resultant decay
constants and form factors agree with those obtained from the
covariant light-front model and then extended to the heavy quark
limit. The universal Isgur-Wise functions
$\xi(\omega),\tau_{1/2}(\omega)$ and $\tau_{3/2}(\omega)$ are
obtained and a relation between $\tau_{1/2}$ and $\tau_{3/2}$ is
found. In the infinite quark mass limit, all the form factors are
related to the Isgur-Wise functions. In addition to $\xi(1)=1$ at
zero recoil $\omega=1$, it is found that $\tau_{1/2}(1)=0.61$,
$\tau_{3/2}(1)=0.31$ and $\rho^2=1.22$ for the slope parameter of
$\xi(\omega)$. The Bjorken and Uraltsev sum rules for the
Isgur-Wise functions are fairly satisfied.

\end{itemize}

\vskip 2.5cm \acknowledgments We are grateful to Wolfgang Jaus for
very helpful clarification and to Chuang-Hung Chen and Chao-Qiang
Geng for valuable discussions. One of us (C.W.H.) wishes to thank
the Institute of Physics, Academia Sinica for its hospitality
during his summer visit where this work started. This research was
supported in part by the National Science Council of R.O.C. under
Grant Nos. NSC92-2112-M-001-016, NSC92-2811-M-001-054 and
NSC92-2112-M-017-001.

\newpage
\appendix
\section{Vertex functions in the conventional light-front approach}

In the conventional light-front approach, a meson bound state
consisting of a quark $q_1$ and an antiquark $\bar q_2$ with the
total momentum $P$ and spin $J$ can be written as (see, for
example \cite{Cheng97})
\begin{eqnarray}
        |M(P, ^{2S+1}L_J, J_z)\rangle
                =\int &&\{d^3p_1\}\{d^3p_2\} ~2(2\pi)^3 \delta^3(
                \tilde P -\tilde p_1-\tilde p_2)~\nonumber\\
        &&\times \sum_{\lambda_1,\lambda_2}
                \Psi^{JJ_z}_{LS}(\tilde p_1,\tilde p_2,\lambda_1,\lambda_2)~
                |q_1(p_1,\lambda_1) \bar q_2(p_2,\lambda_2)\rangle,
 \label{lfmbs}
\end{eqnarray}
where $p_1$ and $p_2$ are the on-mass-shell light-front momenta,
\begin{equation}
        \tilde p=(p^+, p_\bot)~, \quad p_\bot = (p^1, p^2)~,
                \quad p^- = {m^2+p_\bot^2\over p^+},
\end{equation}
and
\begin{eqnarray}
        &&\{d^3p\} \equiv {dp^+d^2p_\bot\over 2(2\pi)^3}, \nonumber \\
        &&|q(p_1,\lambda_1)\bar q(p_2,\lambda_2)\rangle
        = b^\dagger_{\lambda_1}(p_1)d^\dagger_{\lambda_2}(p_2)|0\rangle,\\
        &&\{b_{\lambda'}(p'),b_{\lambda}^\dagger(p)\} =
        \{d_{\lambda'}(p'),d_{\lambda}^\dagger(p)\} =
        2(2\pi)^3~\delta^3(\tilde p'-\tilde p)~\delta_{\lambda'\lambda}.
                \nonumber
\end{eqnarray}
In terms of the light-front relative momentum variables $(x,
p_\bot)$ defined by
\begin{eqnarray}
        && p^+_1=x_1 P^{+}, \quad p^+_2=x_2 P^{+}, \quad x_1+x_2=1, \nonumber \\
        && p_{1\bot}=x_1 P_\bot+p_\bot, \quad p_{2\bot}=x_2
                P_\bot-p_\bot,
\end{eqnarray}
the momentum-space wave-function $\Psi^{JJ_z}_{LS}$ for a
$^{2S+1}L_J$ meson can be expressed as
\begin{equation}
        \Psi^{ JJ_z}_{LS}(\tilde p_1,\tilde p_2,\lambda_1,\lambda_2)
                = \frac{1}{\sqrt N_c}\la L S; L_z S_z|L S;J J_z\ra
                  R^{SS_z}_{\lambda_1\lambda_2}(x,p_\bot)~ \varphi_{LL_z}(x, p_\bot),
\end{equation}
where $\varphi_{LL_z}(x,p_\bot)$ describes the momentum
distribution of the constituent quarks in the bound state with the
orbital angular momentum $L$, $\la L S; L_z S_z|L S;J J_z\ra$ is
the corresponding Clebsch-Gordan coefficient and
$R^{SS_z}_{\lambda_1\lambda_2}$ constructs a state of definite
spin ($S,S_z$) out of light-front helicity ($\lambda_1,\lambda_2$)
eigenstates.  Explicitly,
\begin{equation}
        R^{SS_z}_{\lambda_1 \lambda_2}(x,p_\bot)
              =\sum_{s_1,s_2} \langle \lambda_1|
                {\cal R}_M^\dagger(1-x,p_\bot, m_1)|s_1\rangle
                \langle \lambda_2|{\cal R}_M^\dagger(x,-p_\bot, m_2)
                |s_2\rangle
                \left\langle \frac{1}{2}\,\frac{1}{2};s_1
                s_2|\frac{1}{2}\frac{1}{2};SS_z\right\rangle,
\end{equation}
where $|s_i\rangle$ are the usual Pauli spinors, and ${\cal R}_M$
is the Melosh transformation
operator~\cite{Jaus90,deAraujo:1999cr}:
 \be
        \la s|{\cal R}_M (x,p_\bot,m_i)|\lambda\ra
        &=&\frac{\bar
        u_D(p_i,s) u(p_i,\lambda)}{2 m_i}=-\frac{\bar
        v_D(p_i,s) v(p_i,\lambda)}{2 m_i}
        \non\\
        &=&\frac{m_i+x_i M_0
                      +i\vec \sigma_{s\lambda}\cdot\vec p_\bot \times
                      \vec                n}
                {\sqrt{(m_i+x_i M_0)^2 + p^{2}_\bot}},
\en with $u_{(D)}$, a Dirac spinor in the light-front (instant)
form, $\vec n = (0,0,1)$, a unit vector in the $z$-direction, and
~[cf. Eq. (\ref{eq:internalQ})]
 \be
 M_0^2={m_1^2+p^2_\bot\over x_1}+{m_2^2+p^2_\bot\over x_2}.
 \en
Note that
 $u_D(p,s)=u(p,\lambda) \la \lambda|{\cal R}^\dagger_M|s\ra$
and, consequently, the state $|q(p,\lambda)\ra \la \lambda|{\cal
R}^\dagger_M|s\ra$ transforms like $|q(p,s)\ra$ under rotation,
i.e. its transformation does not depend on its momentum.

In practice it is more convenient to use the covariant form for
$R^{SS_z}_{\lambda_1\lambda_2}$ \cite{Jaus91}:
\begin{equation}
        R^{SS_z}_{\lambda_1\lambda_2}(x,p_\bot)
                =\frac{1}{\sqrt2~{\widetilde M_0}(M_0+m_1+m_2)}
        ~\bar u(p_1,\lambda_1)(\not\!\bar P+M_0)\Gamma\,v(p_2,\lambda_2),
        \label{covariant}
\end{equation}
with
\begin{eqnarray}
        &&\widetilde M_0\equiv\sqrt{M_0^2-(m_1-m_2)^2}, \non \\
        &&\bar P\equiv p_1+p_2, \non \\
        &&\hep^\mu(\pm 1) =
                \left[{2\over P^+} \vec \varepsilon_\bot (\pm 1) \cdot
                \vec P_\bot,\,0,\,\vec \varepsilon_\bot (\pm 1)\right],
                \quad \vec \varepsilon_\bot
                (\pm 1)=\mp(1,\pm i)/\sqrt{2}, \nonumber\\
        &&\hep^\mu(0)={1\over M_0}\left({-M_0^2+P_\bot^2\over
                P^+},P^+,P_\bot\right).   \label{polcom}
\end{eqnarray}
For the pseudoscalar and vector mesons, we have
\begin{eqnarray} \label{eq:Pvertex}
        &&\Gamma_P=\gamma_5 \qquad\quad ({\rm pseudoscalar}, S=0), \non \\
        &&\Gamma_V=-\not{\! \hep}(S_z) \qquad\quad~ ({\rm vector}, S=1).
\end{eqnarray}
It is instructive to derive the above expressions by using the
relations
 \be
 \bar u(p_1,\lambda_1)
 &=&\bar u(p_1,\lambda_1) \frac{u_D(p_1,s_1)\bar
 u_D(p_1,s_1)}{2 m_1},
 \non\\
 v(p_2,\lambda_2)
 &=&\frac{-v_D(p_2,s_2)\bar
 v_D(p_2,s_2)}{2 m_2}
 v(p_2,\lambda_2),
 \non\\
 \bar u_D(p_1,s_1)\frac{\not\!{\bar P}+M_0}{2 M_0}\gamma_5 v_D(p_2,s_2)
 &=&\bar u_D((e_1,\vec p),s_1)\frac{\gamma_0+1}{2}\gamma_5 v_D((e_2,-\vec
 p),s_2)
 \non\\
 &=&\sqrt{(e_1+m_1)(e_2+m_2)}\,i\chi_{s_1}^\dagger \sigma_2 \chi_{s_2}^*
 \non\\
  &=&\sqrt{2(e_1+m_1)(e_2+m_2)}\left\langle \frac{1}{2}\,\frac{1}{2};s_1
  s_2|\frac{1}{2}\,\frac{1}{2};00\right\rangle, \\
 \bar u_D(p_1,s_1)\frac{\not\!{\bar P}+M_0}{2 M_0}(-\not{\! \hep}(S_z)) v_D(p_2,s_2)
 &=&\bar u_D((e_1,\vec p),s_1)
               \frac{\gamma_0+1}{2} \vec \varepsilon(S_z)\cdot\vec\gamma\,
         v_D((e_2,-\vec p),s_2)
 \non \\
 &=&\sqrt{(e_1+m_1)(e_2+m_2)}\,
     i\chi_{s_1}^\dagger \vec \varepsilon(S_z)\cdot\vec\sigma \sigma_2 \chi_{s_2}^*
 \non\\
 &=&\sqrt{2(e_1+m_1)(e_2+m_2)}\left\langle \frac{1}{2}
 \,\frac{1}{2};s_1s_2|\frac{1}{2}\,\frac{1}{2};1S_z\right\rangle,
\non\\
 \sqrt{2(e_1+m_1)(e_2+m_2)}&=&\frac{\widetilde
 M_0(M_0+m_1+m_2)}{\sqrt2M_0}, \non
 \en
where~[cf. Eq.~(\ref{eq:internalQ})]
\begin{eqnarray}
 M_0=e_1+e_2,\quad
 e_i =\sqrt{m^{2}_i+p^{2}_\bot+p^{2}_z},\quad
 p_z=\frac{x_2 M_0}{2}-\frac{m_2^2+p^{2}_\bot}{2 x_2 M_0},
 \end{eqnarray}
$\chi_s$ is the usual Pauli spinor and we have used the usual
properties, especially, the covariant one, of Dirac spinors.
Applying equations of motion on spinors in Eq.~(\ref{covariant})
leads to
 \be
 \bar u(p_1)(\not\!\bar
 P+M_0)\gamma_5\,v(p_2)&=&(M_0+m_1+m_2) \bar u(p_1)\gamma_5\,v(p_2),
 \non\\
\bar u(p_1)(\not\!\bar
 P+M_0)\not{\! \hep}\,v(p_2)&=&\bar u(p_1)[(M_0+m_1+m_2) \not{\!
 \hep}-\hep\cdot (p_1-p_2)]\,v(p_2),
 \label{eq:eom}
 \en
and $R^{SS_z}_{\lambda_1\lambda_2}$ is reduced to a more familiar
form~\cite{Jaus91}.
It is, however, more convenient to use the form shown in
Eq.~(\ref{covariant}) when extending to the $p$-wave meson case.
Two remarks are in order. First, $p_1+p_2$ is not equal to the
meson's four-momentum in the conventional LF approach as both the
quark and antiquark are on-shell. On the contrary, the total
four-momentum is conserved at each vertex in the covariant LF
framework. Second, the longitudinal polarization 4-vector
$\hep^\mu(0)$ given above is not exactly the same as that of the
vector meson and we have $\hep\cdot\bar P=0$. We normalize the
meson state as
\begin{equation}
        \langle M(P',J',J'_z)|M(P,J,J_z)\rangle = 2(2\pi)^3 P^+
        \delta^3(\tilde P'- \tilde P)\delta_{J'J}\delta_{J'_z J_z}~,
\label{wavenor}
\end{equation}
so that
\begin{equation}
        \int {dx\,d^2p_\bot\over 2(2\pi)^3}~\varphi^{\prime*}_{L^\prime L^\prime_z}(x,p_\bot)
                                                   \varphi_{LL_z}(x,p_\bot)
        =\delta_{L^\prime,L}~\delta_{L^\prime_z,L_z}.
\label{momnor}
\end{equation}
Explicitly, we have
 \be
  \varphi_{00}=\varphi,\qquad
  \varphi_{1L_z}=p_{L_z} \varphi_p,
 \en
where $p_{L_z=\pm1}=\mp(p_{\bot x}\pm i p_{\bot y})/\sqrt2$,
$p_{L_z=0}=p_{z}$ are proportional to the spherical harmonics
$Y_{1L_z}$ in momentum space, and $\varphi$, $\varphi_p$ are the
distribution amplitudes of $s$-wave and $p$-wave mesons,
respectively. For a Gaussian-like wave function as shown in
Eq.~(\ref{eq:wavefn}) \cite{Cheng97}, one has
$\varphi_p=\sqrt{2/\beta^2}\varphi$.

For $p$-wave mesons, it is straightforward to obtain
 \be
 \la 1 S; L_z S_z|1 S;J J_z\ra\, p_{L_z} \,R^{SS_z}_{\lambda_1\lambda_2}(x,p_\bot)
                &=&\frac{1}{\sqrt2~{\widetilde M_0}(M_0+m_1+m_2)}
                \non\\
                &&\times
        \bar u(p_1,\lambda_1)(\not\!\bar P+M_0)\Gamma_{^{2S+1}\!P_J} v(p_2,\lambda_2),
         \label{covariantp}
\en
 with
 \be \label{eq:vertexAppB}
 \Gamma_{^3\!P_0}&=&\frac{1}{\sqrt3}\left(\frac{K\cdot
                                   \bar P}{M_0}-\not\!K\right),
 \non\\
 \Gamma_{^1\!P_1}&=&\hep\cdot K \gamma_5,
 \non\\
 \Gamma_{^3\!P_1}&=&\frac{1}{\sqrt2}\left((\not\!K-\frac{K\cdot
 \bar P}{M_0})\not\!\hep-\hep\cdot K\right)\gamma_5,
 \non\\
 \Gamma_{^3\!P_2}&=& \hat \vp_{\mu\nu}\gamma^\mu(- K^\nu),
 \en
where $K\equiv (p_2-p_1)/2$, $\hat \vp_{\mu\nu}(m)=\la 11;m^\prime
m^\pp|11;2 m\ra\, \hep_\mu(m^\prime)\hep_\nu(m^\pp)$. Note that
the polarization tensor of a tensor meson satisfies the relations:
$\hat\vp_{\mu\nu}=\hat\vp_{\nu\mu}$ and $\hat\vp_{\mu\nu} \bar
P^{\mu}=0=\hat\vp^{\mu}_{\mu}$ and that $\hep_\mu$,
$\hat\vp_{\mu\nu}$ are identical to $\vp_\mu$, $\vp_{\mu\nu}$,
respectively, for maximal transverse polarized states ($m=\pm J$).
The above expressions for $^3P_1$ and $^3P_0$ states are
consistent with \cite{Jaus91,Ji92} and \cite{DCJ03}, respectively.

The vertex functions shown in Table~1 and Eq.~(\ref{eq:h}) follow
from the above explicit expressions for $\Psi^{JJ_z}_{LS}$. For
example, by taking $\hat\vp_{\mu\nu}(-K^\nu)$ in place of
$\hep_\mu$ in Eq.~(\ref{eq:eom}) we obtain the $^3P_2$ vertex in
the form shown in Table~1. Note that there are an overall factor
and sign to be determined. The overall factor,
$(M^2-M_0^2)\sqrt{x_1 x_2}$ [cf. Eq.~(\ref{eq:h})], is fixed by
comparing the pseudoscalar decay constant $f_P$ obtained in both
covariant and conventional approaches [see Eqs. (\ref{eq:fP}) and
(\ref{eq:fP0})], while the overall sign can be fixed by the HQS
expectation for decay constants and form factors. For example, the
sign of the $P^{1/2}_1$ state relative to $^3P_0$ is fixed by the
HQS relation $f_{P^{1/2}_1}=f_S$. An additional factor of $i$ is
assigned in Table~1 as in the usual Feynman rules to ensure that
the corresponding operators are hermitian. For example, we have an
$i$ in front of $\gamma_\mu$ but not $\gamma_5$, just like the
usual QED and Yukawa vertices, respectively. Similarly,
polarization vectors are decoupled from the vertex Feynman rules
as usual.

\section{Some Useful Formulas}
In this Appendix we first collect some formulas in \cite{Jaus99}
relevant for the present work and then we proceed to summarize the
formula for the product of four $\hat p'_1$'s needed for the
calculation in Sec.~III.

The explicit representation of the traces in Eqs.~(\ref{eq:SPPV})
and (\ref{eq:SPV}) can be found in \cite{Jaus99}. For completeness
we collect them in below:
 \be
S^{PP}_{V\mu} &=&2 p^\prime_{1\mu} [M^{\prime 2}+M^{\pp2}-q^2-2
           N_2-(m_1^\prime-m_2)^2-(m^\pp_1-m_2)^2+(m_1^\prime-m_1^\pp)^2]
 \non\\
          &&+q_\mu[q^2-2 M^{\prime2}+N^\prime_1-N^\pp_1+2
          N_2+2(m_1^\prime-m_2)^2-(m_1^\prime-m_1^\pp)^2]
 \nonumber\\
          &&+P_\mu[q^2-N^\prime_1-N^\pp_1-(m_1^\prime-m_1^\pp)^2],
 \label{eq:SPPVappen}
 \en
and
 \be
S^{PV}_{\mu\nu} &=&(S^{PV}_{V}-S^{PV}_A)_{\mu\nu}
 \non\\
           &=&-2 i\epsilon_{\mu\nu\alpha\beta}
                 \Big\{p^{\prime\alpha}_1 P^\beta (m_1^\pp-m_1^\prime)
                   +p^{\prime\alpha}_1q^\beta(m_1^\pp+m_1^\prime-2
                   m_2)+q^\alpha P^\beta m_1^\prime
                 \Big\}
 \non\\
           &&+\frac{1}{W^\pp_V}(4 p^\prime_{1\nu}-3
           q_\nu-P_\nu)i\epsilon_{\mu\alpha\beta\rho}p^{\prime\alpha}_1 q^\beta P^\rho
 \non\\
           &&+2
           g_{\mu\nu}\Big\{m_2(q^2-N_1^\prime-N^\pp_1-m_1^{\prime2}-m_1^{\pp2})
                      -m_1^\prime (M^{\pp2}-N_1^\pp-N_2-m_1^{\pp2}-m_2^2)
 \non\\
           &&-m^\pp_1(M^{\prime2}-N^\prime_1-N_2-m_1^{\prime2}-m_2^2)-2 m_1^\prime m_1^\pp m_2\Big\}
 \non\\
           &&+8 p^\prime_{1\mu} p^\prime_{1\nu}(m_2-m_1^\prime)
             -2(P_\mu q_\nu+q_\mu P_\nu+2q_\mu q_\nu) m_1^\prime
             +2p^\prime_{1\mu} P_\nu (m_1^\prime-m_1^\pp)
 \non\\
           &&+2 p^\prime_{1\mu} q_\nu(3 m_1^\prime-m_1^\pp-2m_2)+2
           P_\mu p^\prime_{1\nu}(m_1^\prime+m_1^\pp)+2 q_\mu
           p^\prime_{1\nu}(3 m_1^\prime+m_1^\pp-2 m_2)
 \non\\
           &&+\frac{1}{2W^\pp_V}(4 p^\prime_\nu-3q_\nu-P_\nu)
              \Big\{2 p^\prime_{1\mu}[M^{\prime2}+M^{\pp2}-q^2
                -2 N_2+2(m_1^\prime-m_2)(m_1^\pp+m_2)]
 \non\\
           &&   +q_\mu[q^2-2 M^{\prime2}+N^\prime_1-N_1^\pp+2 N_2-(m_1+m_1^\pp)^2+2(m_1^\prime-m_2)^2]
 \non\\
           &&   +P_\mu[q^2-N_1^\prime-N_1^\pp-(m_1^\prime+m_1^\pp)^2]
              \Big\}.
   \label{eq:SPVappen}
 \en
Note that our convention for $\epsilon_{\mu\nu\alpha\beta}$,
namely, $\epsilon_{0123}=1$, is different from that
in~\cite{Jaus99}.

The analytic expressions for $P\to S,A$ transition form factors
can be obtained from that of $P\to P,V$ ones by some simple
replacements. Hence, we list the explicit expressions for $P\to P$
and $P\to V$ transition form factors in \cite{Jaus99}:
  \be
 f_+(q^2)&=&\frac{N_c}{16\pi^3}\int dx_2 d^2p^\prime_\bot
            \frac{h^\prime_P h^\pp_P}{x_2 \hat N_1^\prime \hat N^\pp_1}
            \Bigg[x_1 (M_0^{\prime2}+M_0^{\pp2})+x_2 q^2
 \non\\
         &&\qquad-x_2(m_1^\prime-m_1^\pp)^2 -x_1(m_1^\prime-m_2)^2-x_1(m_1^\pp-m_2)^2\Bigg],
 \non\\
 f_-(q^2)&=&\frac{N_c}{16\pi^3}\int dx_2 d^2p^\prime_\bot
            \frac{2h^\prime_P h^\pp_P}{x_2 \hat N_1^\prime \hat N^\pp_1}
            \Bigg\{- x_1 x_2 M^{\prime2}-p_\bot^{\prime2}-m_1^\prime m_2
                  +(m_1^\pp-m_2)(x_2 m_1^\prime+x_1 m_2)
\non\\
         &&\qquad +2\frac{q\cdot P}{q^2}\left(p^{\prime2}_\bot+2\frac{(p^\prime_\bot\cdot q_\bot)^2}{q^2}\right)
                  +2\frac{(p^\prime_\bot\cdot q_\bot)^2}{q^2}
                  -\frac{p^\prime_\bot\cdot q_\bot}{q^2}
                  \Big[M^{\pp2}-x_2(q^2+q\cdot P)
\non\\
         &&\qquad -(x_2-x_1) M^{\prime2}+2 x_1 M_0^{\prime
                  2}-2(m_1^\prime-m_2)(m_1^\prime+m_1^\pp)\Big]
           \Bigg\},
 \label{eq:fpm}
 \en
and
 \be \label{eq:PtoV}
 g(q^2)&=&-\frac{N_c}{16\pi^3}\int dx_2 d^2 p^\prime_\bot
           \frac{2 h^\prime_P h^\pp_V}{x_2 \hat N^\prime_1 \hat N^\pp_1}
           \Bigg\{x_2 m_1^\prime+x_1 m_2+(m_1^\prime-m_1^\pp)
           \frac{p^\prime_\bot\cdot q_\bot}{q^2}
           +\frac{2}{w^\pp_V}\left[p^{\prime2}_\bot+\frac{(p^\prime_\bot\cdot q_\bot)^2}{q^2}\right]
           \Bigg\},
 \non\\
  f(q^2)&=&\frac{N_c}{16\pi^3}\int dx_2 d^2 p^\prime_\bot
            \frac{ h^\prime_P h^\pp_V}{x_2 \hat N^\prime_1 \hat N^\pp_1}
            \Bigg\{2
            x_1(m_2-m_1^\prime)(M^{\prime2}_0+M^{\pp2}_0)-4 x_1
            m_1^\pp M^{\prime2}_0+2x_2 m_1^\prime q\cdot P
 \non\\
         &&+2 m_2 q^2-2 x_1 m_2
           (M^{\prime2}+M^{\pp2})+2(m_1^\prime-m_2)(m_1^\prime+m_1^\pp)^2
           +8(m_1^\prime-m_2)\left[p^{\prime2}_\bot+\frac{(p^\prime_\bot\cdot q_\bot)^2}{q^2}\right]
 \non\\
         &&
           +2(m_1^\prime+m_1^\pp)(q^2+q\cdot
           P)\frac{p^\prime_\bot\cdot q_\bot}{q^2}
           -4\frac{q^2 p^{\prime2}_\bot+(p^\prime_\bot\cdot q_\bot)^2}{q^2 w^\pp_V}
            \Bigg[2 x_1 (M^{\prime2}+M^{\prime2}_0)-q^2-q\cdot P
 \non\\
         &&-2(q^2+q\cdot P)\frac{p^\prime_\bot\cdot
            q_\bot}{q^2}-2(m_1^\prime-m_1^\pp)(m_1^\prime-m_2)
            \Bigg]\Bigg\},
 \non\\
a_+(q^2)&=&\frac{N_c}{16\pi^3}\int dx_2 d^2 p^\prime_\bot
            \frac{2 h^\prime_P h^\pp_V}{x_2 \hat N^\prime_1 \hat N^\pp_1}
            \Bigg\{(x_1-x_2)(x_2 m_1^\prime+x_1 m_2)-[2x_1
            m_2+m_1^\pp+(x_2-x_1)
            m_1^\prime]\frac{p^\prime_\bot\cdot q_\bot}{q^2}
 \non\\
         &&-2\frac{x_2 q^2+p_\bot^\prime\cdot q_\bot}{x_2 q^2
            w^\pp_V}\Big[p^\prime_\bot\cdot p^\pp_\bot+(x_1 m_2+x_2 m_1^\prime)(x_1 m_2-x_2 m_1^\pp)\Big]
            \Bigg\},
 \non\\
 a_-(q^2)&=&\frac{N_c}{16\pi^3}\int dx_2 d^2 p^\prime_\bot
            \frac{ h^\prime_P h^\pp_V}{x_2 \hat N^\prime_1 \hat N^\pp_1}
            \Bigg\{2(2x_1-3)(x_2 m_1^\prime+x_1
            m_2)-8(m_1^\prime-m_2)\left[\frac{p^{\prime2}_\bot}{q^2}+2\frac{(p^\prime_\bot\cdot
            q_\bot)^2}{q^4}\right]
 \non\\
         &&-[(14-12 x_1) m_1^\prime-2 m_1^\pp-(8-12 x_1) m_2]
           \frac{p^\prime_\bot\cdot q_\bot}{q^2}
 \non\\
         &&+\frac{4}{w^\pp_V}\Bigg([M^{\prime2}+M^{\pp2}-q^2+2(m_1^\prime-m_2)(m_1^\pp+m_2)]
                                   (A^{(2)}_3+A^{(2)}_4-A^{(1)}_2)
\non\\
         &&                         +Z_2(3 A^{(1)}_2-2A^{(2)}_4-1)
                                    +\frac{1}{2}[x_1(q^2+q\cdot P)
                                                 -2 M^{\prime2}-2 p^\prime_\bot\cdot q_\bot
\non\\
         &&                                      -2 m_1^\prime(m_1^\pp+m_2)-2
                                                 m_2(m_1^\prime-m_2)
                                                 ](A^{(1)}_1+A^{(1)}_2-1)
\non\\
        &&                          +q\cdot P\Bigg[\frac{p^{\prime2}_\bot}{q^2}
                                                  +\frac{(p^\prime_\bot\cdot q_\bot)^2}{q^4}\Bigg]
                                                  (4
                                                  A^{(1)}_2-3)
                             \Bigg)
            \Bigg\}.
 \en

We next give the results for $\hat p^\prime_1 \hat p^\prime_1 \hat
p^\prime_1 \hat p^\prime_1$ and $\hat p^\prime_1 \hat p^\prime_1
\hat p^\prime_1 \hat N_2$. In Eq.~(\ref{eq:contourB}), under the
typical integration
 \be
 \frac{N_c}{16 \pi^3}\int \frac{d x_2 d^2p^\prime_\bot}
                             {x_2\hat N^\prime_1
                             \hat N^\pp_1} h^\prime_P h^\pp_M \hat
                             S^{PM},
 \label{eq:int}
 \en
in a $P\to M$ transition matrix element, $\hat p^\prime_1 \hat
p^\prime_1 \hat p^\prime_1 \hat p^\prime_1$ in $\hat S^{PM}$ can
be expressed in terms of three external momenta, $P$, $q$ and
$\tilde\omega$. Up to the first order in $\tilde\omega$, we have
 \be
 \hat p^\prime_{1\mu} \hat p^\prime_{1\nu} \hat p^\prime_{1\alpha} \hat p^\prime_{1\beta}
 \doteq\sum_{i=1}^9 I_{i\mu\nu\alpha\beta} A^{(4)}_i
 +\sum_{j=1}^4 J_{j\mu\nu\alpha\beta} B^{(4)}_j
 +\sum_{k=1}^2 K_{k\mu\nu\alpha\beta} C^{(4)}_k+O(\tilde\omega^2),
  \en
where
 \be
 I_{1\mu\nu\alpha\beta}&=&(gg)_{\mu\nu\alpha\beta}=g_{\mu\nu} g_{\alpha\beta}+g_{\mu\alpha}
                          g_{\nu\beta}+g_{\mu\beta} g_{\nu\alpha},
 \non\\
 I_{2\mu\nu\alpha\beta}&=&(gPP)_{\mu\nu\alpha\beta}=g_{\mu\nu}
 P_\alpha P_\beta+g_{\mu\alpha} P_\nu P_\beta+g_{\mu\beta}P_\nu
 P_\alpha+g_{\alpha\beta}P_\mu P_\nu+g_{\nu\beta} P_\mu
 P_\alpha+g_{\nu\alpha} P_\mu P_\beta,
 \non\\
 I_{3\mu\nu\alpha\beta}&=&(gPq)_{\mu\nu\alpha\beta}=g_{\mu\nu}
 (P_\alpha q_\beta+q_\alpha P_\beta)+{\rm permutations},
 \non\\
 I_{4\mu\nu\alpha\beta}&=&(gqq)_{\mu\nu\alpha\beta}=g_{\mu\nu}
 q_\alpha q_\beta++g_{\mu\alpha} q_\nu q_\beta+g_{\mu\beta}q_\nu
 q_\alpha+g_{\alpha\beta}q_\mu q_\nu+g_{\nu\beta} q_\mu
 q_\alpha+g_{\nu\alpha} q_\mu q_\beta,,
 \non\\
 I_{5\mu\nu\alpha\beta}&=&(PPPP)_{\mu\nu\alpha\beta}=P_\mu P_\nu P_\alpha P_\beta,
 \non\\
 I_{6\mu\nu\alpha\beta}&=&(PPPq)_{\mu\nu\alpha\beta}=P_\mu P_\nu P_\alpha
 q_\beta+P_\mu P_\nu q_\alpha P_\beta
                         +P_\mu q_\nu P_\alpha P_\beta+q_\mu P_\nu P_\alpha P_\beta,
 \non\\
 I_{7\mu\nu\alpha\beta}&=&(PPqq)_{\mu\nu\alpha\beta}=P_\mu P_\nu
 q_\alpha q_\beta+{\rm permutations},
 \non\\
 I_{8\mu\nu\alpha\beta}&=&(Pqqq)_{\mu\nu\alpha\beta}=P_\mu q_\nu q_\alpha q_\beta+q_\mu P_\nu q_\alpha q_\beta
                         +q_\mu q_\nu P_\alpha q_\beta+q_\mu q_\nu q_\alpha P_\beta,
 \non\\
 I_{9\mu\nu\alpha\beta}&=&(qqq)_{\mu\nu\alpha\beta}=q_\mu q_\nu q_\alpha q_\beta,
 \\
 J_{1\mu\nu\alpha\beta}&=&(gP\tilde\omega)_{\mu\nu\alpha\beta}=\frac{1}{\tilde\omega\cdot P}
                           [g_{\mu\nu}(P_\alpha\tilde\omega_\beta+\tilde\omega_\alpha
                           P_\beta)+{\rm permutations}],
 \non\\
 J_{2\mu\nu\alpha\beta}&=&(PPP\tilde\omega)_{\mu\nu\alpha\beta}=
                         \frac{1}{\tilde\omega\cdot P}
                           ( P_\mu P_\nu P_\alpha \tilde\omega_\beta+P_\mu P_\nu \tilde\omega_\alpha P_\beta
                         +P_\mu \tilde\omega_\nu P_\alpha P_\beta+\tilde\omega_\mu P_\nu P_\alpha P_\beta),
 \non\\
 J_{3\mu\nu\alpha\beta}&=&(PPq\tilde\omega)_{\mu\nu\alpha\beta}
                        = \frac{1}{\tilde\omega\cdot P}
                        [(P_\mu P_\nu q_\alpha
                         +P_\mu q_\nu P_\alpha
                         +q_\mu P_\nu P_\alpha )\tilde\omega_\beta
                         +{\rm permutations}],
 \non\\
  J_{43\mu\nu\alpha\beta}&=&(Pqq\tilde\omega)_{\mu\nu\alpha\beta}
                        = \frac{1}{\tilde\omega\cdot P}
                        [(P_\mu q_\nu q_\alpha
                         +q_\mu P_\nu q_\alpha
                         +q_\mu q_\nu P_\alpha )\tilde\omega_\beta
                         +{\rm permutations}],
 \non\\
 K_{1\mu\nu\alpha\beta}&=&(gq\tilde\omega)_{\mu\nu\alpha\beta}=\frac{1}{\tilde\omega\cdot P}
                           [g_{\mu\nu}(q_\alpha\tilde\omega_\beta+\tilde\omega_\alpha
                           q_\beta)+{\rm permutations}],
 \non\\
 K_{2\mu\nu\alpha\beta}&=&(qqq\tilde\omega)_{\mu\nu\alpha\beta}=
                         \frac{1}{\tilde\omega\cdot P}
                           ( q_\mu q_\nu q_\alpha \tilde\omega_\beta+q_\mu q_\nu \tilde\omega_\alpha q_\beta
                         +q_\mu \tilde\omega_\nu q_\alpha q_\beta+\tilde\omega_\mu q_\nu q_\alpha
                         q_\beta). \non
 \en
By contracting $\hat p^\prime_{1\mu} \hat p^\prime_{1\nu} \hat
p^\prime_{1\alpha} \hat p^\prime_{1\beta}$ with
$\tilde\omega^\beta$, $q^\beta$ and $g^{\alpha\beta}$, and
comparing with the complete expressions of $\hat p^\prime_{1\mu}
\hat p^\prime_{1\nu} \hat p^\prime_{1\alpha}$ and $\hat
p^\prime_{1\mu} \hat p^\prime_{1\nu}$ shown in \cite{Jaus99}, we
obtain
 \be
 A^{(4)}_1&=&\frac{1}{3}\big(A^{(2)}_1\big)^2,
 \qquad
 A^{(4)}_2=A^{(1)}_1 A^{(3)}_1,
 \quad
 A^{(4)}_3=A^{(1)}_1 A^{(3)}_2,
 \non\\
 A^{(4)}_4&=&A^{(1)}_2 A^{(3)}_2 -\frac{1}{q^2}A^{(4)}_1,
 \qquad
 A^{(4)}_5=A^{(1)}_1 A^{(3)}_3,
 \quad
 A^{(4)}_6=A^{(1)}_1 A^{(3)}_4,
 \non\\
 A^{(4)}_7&=&A^{(1)}_1 A^{(3)}_5,
 \qquad
 A^{(4)}_8=A^{(1)}_1 A^{(3)}_6,
 \quad
 A^{(4)}_9=A^{(1)}_1 A^{(3)}_6 -\frac{3}{q^2}A^{(4)}_4,
 \\
 B^{(4)}_1&=&A^{(1)}_1 C^{(3)}_1-A^{(4)}_1,
 \quad
 B^{(4)}_2=A^{(1)}_1 B^{(3)}_1-A^{(4)}_2,
 \quad
 B^{(4)}_3=A^{(1)}_1 B^{(3)}_2-A^{(4)}_3,
 \non\\
 B^{(4)}_4&=&A^{(1)}_1 C^{(3)}_2-A^{(4)}_4,
 \quad
 C^{(4)}_1=A^{(3)}_2 C^{(1)}_1+\frac{q\cdot P}{q^2} A^{(4)}_1,
 \quad
 C^{(4)}_2=A^{(3)}_6 C^{(1)}_1+3\frac{q\cdot P}{q^2} A^{(4)}_4,
 \non
 \en
where~\cite{Jaus99}
 \be \label{eq:Aij}
 A^{(1)}_1&=&\frac{x_1}{2},
 \quad
 A^{(1)}_2=A^{(1)}_1-\frac{p^\prime_\bot\cdot q_\bot}{q^2},
 \quad
 C^{(1)}_1=-\hat N_2+Z_2,
 \non\\
 Z_2&=&\hat N_1^\prime+m_1^{\prime2}-m_2^2+(1-2x_1)M^{\prime2}
 +(q^2+q\cdot P)\frac{p^\prime_\bot\cdot q_\bot}{q^2}, \non\\
 A^{(2)}_1&=&-p^{\prime2}_\bot-\frac{(p^\prime_\bot\cdot q_\bot)^2}{q^2},
 \quad
 A^{(2)}_2=\big(A^{(1)}_1\big)^2,
 \quad
 A^{(2)}_3=A^{(1)}_1 A^{(1)}_2,
 \non\\
 A^{(2)}_4&=&\big(A^{(1)}_2\big)^2-\frac{1}{q^2}A^{(2)}_1,
 \quad
 A^{(3)}_1=A^{(1)}_1 A^{(2)}_1,
 \quad
 A^{(3)}_2=A^{(1)}_2 A^{(2)}_1,
 \\
 A^{(3)}_3&=&A^{(1)}_1 A^{(2)}_2,
 \quad
 A^{(3)}_4=A^{(1)}_2 A^{(2)}_2,
 \quad
 A^{(3)}_5=A^{(1)}_1 A^{(2)}_4,
  \non\\
 A^{(3)}_6&=&A^{(1)}_2 A^{(2)}_4-\frac{2}{q^2}A^{(1)}_2 A^{(2)}_1,
 \non\\
 B^{(2)}_1&=&A^{(1)}_1 Z_2-A^{(2)}_1,
  \quad
 B^{(3)}_1=A^{(1)}_1 (B^{(2)}_1-A^{(2)}_1),
  \quad
 B^{(3)}_2=A^{(1)}_2 B^{(2)}_1+\frac{q\cdot P}{q^2} A^{(2)}_1.
 \non
 \en
Following the prescription in \cite{Jaus99}, the spurious
contributions $C^{(4)}_{1,2}$ should be vanished by including the
zero mode contribution and we have
 \be
 A^{(3)}_2 \hat N_2\to A^{(3)}_2 Z_2+\frac{q\cdot P}{q^2} A^{(4)}_1,
 \quad
 A^{(3)}_6 \hat N_2\to A^{(3)}_6 Z_2+3\frac{q\cdot P}{q^2} A^{(4)}_4,
 \en
which lead to the $\hat p^\prime_{\mu} \hat p^\prime_{\nu} \hat
p^\prime_{\alpha} \hat N_2$ formula shown in Eq.~(\ref{eq:p1B}).
Note that in general $B^{(i)}_j$ are non-vanishing by themselves,
but they do vanish under integration in some choice of vertex
function~\cite{Jaus99}. There are some attempts to include these
effects for generic vertex functions~\cite{Jaus03}. The important
of these effects can be checked numerically. For example, we have
checked numerically that the integral of Eq.~(\ref{eq:int}) with
$\hat S^{PM}$ replaced by $B^{(j)}_{i}$ are vanishingly small. In
practice, one only needs $A^{(i)}_j$ terms for $\hat p^\prime_1
\dots \hat p^\prime_1$ formulas.


\begin{thebibliography}{99}
 \newcommand{\bi}{\bibitem}

 \bi{Ter} M.V. Terent'ev, Sov. J. Phys. {\bf 24}, 106 (1976); V.B.
Berestetsky and M.V. Terent'ev, {\sl ibid.} {\bf 24}, 547 (1976);
{\sl ibid.} {\bf 25}, 347 (1977).

 \bibitem{Chung} P.L. Chung, F. Coester, and W.N. Polyzou,
        \pl B {\bf 205}, 545 (1988).

 \bi{Jaus90} W. Jaus, \pr D {\bf 41}, 3394 (1990).

 \bibitem{Jaus91}
     W.~Jaus,
     Phys.\ Rev.\ D {\bf 44}, 2851 (1991).

 \bibitem{Ji92}
              C.~R.~Ji, P.~L.~Chung, and S.~R.~Cotanch,
              Phys.\ Rev.\ D {\bf 45}, 4214 (1992).

 \bibitem{Jaus96}
       W.~Jaus,
       Phys.\ Rev.\ D {\bf 53}, 1349 (1996) [Erratum-ibid.\ D {\bf 54},
       5904 (1996)].

 \bibitem{Cheng97} H. Y. Cheng, C. Y. Cheung, and C. W. Hwang, Phys.
        Rev. D {\bf 55}, 1559 (1997).


 \bi{BCJ02}
      B.~L.~Bakker, H.~M.~Choi, and C.~R.~Ji, Phys. Rev. D {\bf
      65}, 116001 (2002).

 \bi{CCHZ} H.Y. Cheng, C.Y. Cheung, C.W. Hwang, and W.M. Zhang,
           \pr D {\bf 57}, 5598 (1998).

 \bibitem{Jaus99}
     W.~Jaus,
     Phys.\ Rev.\ D {\bf 60}, 054026 (1999).

 \bi{BCJ03}
      B.~L.~Bakker, H.~M.~Choi, and C.~R.~Ji,
      Phys.\ Rev.\ D {\bf 67}, 113007 (2003).

 \bibitem{CM69}
      S.~J.~Chang and S.~K.~Ma,
      Phys.\ Rev.\  {\bf 180}, 1506 (1969).

 \bibitem{zeromode}
    S.~J.~Chang, R.~G.~Root, and T.~M.~Yan,
   Phys.\ Rev.\ D {\bf 7}, 1133 (1973);
    T.~M.~Yan,
    {\it ibid.}~{\bf 7}, 1780 (1973).

 \bibitem{Jaus03}
    W.~Jaus,
    Phys.\ Rev.\ D {\bf 67}, 094010 (2003).

 \bi{BaBar} BaBar Collaboration, B. Aubert {\it et al.,} \prl {\bf
 90}, 242001 (2003).

 \bi{CLEO} CLEO Collaboration, D. Besson {\it et al.,}
 \pr D {\bf 68}, 032002 (2003).

 \bi{BelleD} Belle Collaboration, K. Abe {\it et al.,}
 hep-ex/0307021.

 \bi{PDG} Particle Data Group,  K. Hagiwara {\it et al.,} \pr D {\bf
 66}, 010001 (2002).

 \bi{ISGW} N. Isgur, D. Scora, B. Grinstein, and M.B. Wise, \pr
          D {\bf 39}, 799 (1989).

 \bibitem{deMelo98}
     J.~P.~de Melo, J.~H.~Sales, T.~Frederico, and P.~U.~Sauer,
     Nucl.\ Phys.\ A {\bf 631}, 574C (1998).



  \bibitem{IW89} N. Isgur and M. B. Wise, Phys. Lett. B {\bf 232},
                113 (1989); {\bf 237}, 527 (1990).

 \bi{Bjorken} J.D. Bjorken, SLAC-PUB-5278 (1990); J.D. Bjorken, J.
              Dunietz, and J. Taron, \np B {\bf 371}, 111 (1992).

 \bi{Uraltsev} N. Uraltsev, \pl B {\bf 501}, 86 (2001).

 \bi{Suzuki} M. Suzuki, \pr D {\bf 47}, 1252 (1993).

 \bi{IW91} 
           N.~Isgur and M.~B.~Wise,
           Phys.\ Rev.\ D {\bf 43}, 819 (1991).

 \bibitem{HQfrules}
   A.~Le Yaouanc, L.~Oliver, O.~Pene, and J.~C.~Raynal,
   Phys.\ Lett.\ B {\bf 387}, 582 (1996);
   S.~Veseli and I.~Dunietz,
   Phys.\ Rev.\ D {\bf 54}, 6803 (1996).

 \bibitem{Gauss} P. L. Chung, F. Coester, and W. N. Polyzou, Phys. Lett.
       B {\bf 205}, 545 (1988).


 \bi{CDKM} J. Carbonell, B. Desplanques, V.A. Karmanon, and J.F.
 Mathiot, Phys. Rep. {\bf 300}, 215 (1998).

  \bibitem{Hwang02}
     C.~W.~Hwang,
     Eur.\ Phys.\ J.\ C {\bf 23}, 585 (2002).

 \bi{ISGW2} D. Scora and N. Isgur, \pr D {\bf 52}, 2783 (1995).

  \bi{Bloch} J.C.R. Bloch, Yu.L. Kalinovsky, C.D. Roberts, and S.M.
              Schmidt, \pr D {\bf 60}, 111502 (1999).

  \bi{BaBarDs} BaBar Collaboration, B. Aubert {\it et al.,}
               \pr D {\bf 67}, 092003 (2003).

  \bi{Bernard} C. Bernard {\it et al.,} Phys. Rev. D {\bf 65}, 014510 (2002).

 \bi{Close} F.E. Close and N.A. T\"ornqvist, J. Phys. G {\bf
 28}, R249 (2002).

 \bi{Maltman} K. Maltman, \pl B {\bf 462}, 14 (1999).

 \bi{Chernyak} V. Chernyak, \pl B {\bf 509}, 273 (2001).

 \bi{BelleDs-1} Belle Collaboration, P. Krokovny {\it et al.,}
 hep-ex/0308019.

 \bibitem{Cheng:2003id}
      H.~Y.~Cheng,
      Phys. Rev. D {\bf 68}, 094005 (2003).

  \bibitem{Cheng:2003bn}
      H.~Y.~Cheng,
      Phys.\ Rev.\ D {\bf 67}, 094007 (2003).


 \bibitem{BSW} M. Wirbel, S. Stech, and M. Bauer, Z. Phys. C {\bf 29},
        637 (1985); M. Bauer, B. Stech, and M. Wirbel, {\it ibid},
        {\bf 34}, 103 (1987); M. Bauer, B. Stech, and M. Wirbel, {\it ibid},
        {\bf 42}, 671 (1989).

 \bi{ODonnell} P.J. O'Donnell, Q.P.
        Xu, and H.K.K. Tung, Phys. Rev. D {\bf 52}, 3966 (1995).

 \bi{pc} W. Jaus, private communication.

 \bi{Melikhov96} D. Melikhov, \pr D {\bf 53}, 2460 (1996); \pl B {\bf 380}, 363
 (1996).

 \bi{FOCUS} FOCUS Collaboration, J.M. Link {\it et al.,} Phys. Lett. B {\bf 544}, 89
 (2002).

 \bi{FOCUSDs} FOCUS Collaboration, J.M. Link {\it et al.,} hep-ex/0401001.

 \bi{Melikhov} D. Melikhov and B. Stech, \pr D {\bf 62}, 014006
 (2000).

 \bi{Ball91} P. Ball, V.M. Braun, and H. Dosch, \pr D {\bf 44}, 3567 (1991);
             P.~Ball,
             Phys.\ Rev.\ D {\bf 48}, 3190 (1993) [arXiv:hep-ph/9305267].

  \bi{LCSR} P. Ball and V.M. Braun, \pr D {\bf 58}, 094016 (1998); P.
 Ball, { J. High Energy Phys.} {\bf 9809}, 005 (1998)
 [hep-ph/9802394].

 \bi{Deandrea} A. Deandrea, R. Gatto, G. Nardulli, and A.D. Polosa,
 \pr D {\bf 59}, 074012 (1999).

 \bi{Aliev} T.M. Aliev and M. Savci, \pl B {\bf 456}, 256 (1999).


 \bi{BelleDs-2} Belle Collaboration, Y. Mikami {\it et al.,} KEK
 Preprint 2003-69.

 \bi{BaBarDs1} BaBar Collaboration, B. Aubert {\it et al.,} hep-ex/0310050.

 \bi{GodfreyDs} S. Godfrey, \pl B {\bf 568}, 254 (2003).

 \bi{BaBarD} BaBar Collaboration, B. Aubert {\it et al.,}  hep-ex/0308026.

\bi{Galik} CELO Collaboration, R.S. Galik, \np {\bf A663}, 647
(2000); S. Anderson {\it et al.,} CLEO-CONF-99-6 (1999).

 \bi{Neubert} M. Neubert, \pl {\bf B418}, 173 (1998).


 \bibitem{Georgi90} H. Georgi, Phys. Lett. B {\bf 240}, 447 (1990);
            E. Eichten and B. Hill, Phys. Lett. B {\bf 234}, 511
                (1990); {\bf 243}, 427 (1990).


\bibitem{Yan92} T. M. Yan, H. Y. Cheng, C. Y. Cheung, G. L. Lin, Y. C.
        Lin, and H. L. Yu, Phys. Rev. D {\bf 46}, 1148 (1992);
        H. Y. Cheng, C. Y. Cheung, G. L. Lin, Y. C. Lin, T. M.
        Yan, and H. L. Yu, {\it ibid.} D {\bf 46}, 5060 (1992);
        {\it ibid}. D {\bf 47}, 1030 (1993); M. B. Wise, Phys.
        Rev. D {\bf 45}, R2188 (1992); G. Burdman and J. Donoghue,
        Phys. Lett. B {\bf 280}, 287 (1992).


 \bibitem{De97} N. B. Demchuk, P. Yu. Kulikov, I. M. Narodetskii, and
        P. J. O'Donnell, Phys. Atom. Nucl. {\bf 60}, 1292 (1997).

 \bi{Morenas} V. Morenas, A. Le Yaouanc, L. Oliver, O P\'ene, and
 J.C. Raynal, \pr D {\bf 56}, 5668 (1997).

 \bi{CCCN} P. Cea, P. Colangelo, L. Cosmai, and G. Nardulli, \pl B
 {\bf 206}, 691 (1988).

 \bi{Deandrea98} A. Deandrea, N. Di Bartolomeo, R. Gatto, G.
 Nardulli, and A.D. Polosa, \pr D {\bf 58}, 034004 (1998).

 \bi{Veseli} S. Veseli and I. Dunietz, \pr D {\bf 54}, 6803 (1996).

 \bi{Godfrey} S. Godfrey and N. Isgur, \pr D {\bf 32}, 189 (1985).

 \bi{Colangelo} P. Colangelo, F. De Fazio, and N. Paver, \pr D
 {\bf 58}, 116005 (1998).

 \bi{Wambach} A. Wambach, \np B {\bf 343}, 647 (1995).

 \bi{Huang} M.Q. Huang and Y.B. Dai, \pr D {\bf 59}, 034018 (1999).

 \bibitem{deAraujo:1999cr}
            W.~R.~de Araujo, M.~Beyer, T.~Frederico, and H.~J.~Weber,
            J.\ Phys.\ G {\bf 25}, 1589 (1999). 

 \bibitem{DCJ03}
        M.~A.~DeWitt, H.~M.~Choi, and C.~R.~Ji,
        \pr D {\bf 68}, 054026 (2003).

 \bi{HFAG} Heavy Flavor Averaging Group,\\
 http://www.slac.stanford.edu/xorg/hfag/semi/summer03-eps/summer03.shtml.

 \bi{Boyd} C.G. Boyd, Z. Ligeti, I.Z. Rothstein, and M.B. Wise, \pr D {\bf 55}, 3027 (1997).

 \bi{Dorsten} M.P. Dorsten, hep-ph/0310025.

\end{thebibliography}
\end{document}